\documentclass[pre,showkeys,reprint,superscriptaddress,amsmath,amssymb,aps,longbibliography]{revtex4-1}

\usepackage{graphicx}
\usepackage{bm}
\usepackage{tabularx}

\usepackage{siunitx}
\usepackage{mathrsfs}
\usepackage{rotating}
\usepackage[bbgreekl]{mathbbol}
\usepackage{mathbbol}
\usepackage{overpic}
\usepackage{ragged2e}  
\DeclareSIUnit\px{px}
\newcommand{\ve}[1]{\ensuremath{\mbox{\boldmath$#1$}}}
\newcommand{\ma}[1]{\ensuremath{\mathbb{#1}}}

\newcommand{\vp}{\ve v}
\newcommand{\up}{{\hat{v}_p}}
\newcommand{\uq}{{\hat{v}_q}}
\newcommand{\un}{{\hat{v}_n}}
\newcommand{\rep}{{\rm Re}_{\rm p}}

\usepackage{xcolor}\usepackage{color}

\begin{document}
\title{Fluid-inertia torques from particle-shape symmetry}

\author{L. Sundberg}
\affiliation{Department of Physics, Gothenburg University, Gothenburg, SE-40530 Sweden}

\author{F. Candelier}
\affiliation{Aix-Marseille Univ.,  CNRS, IUSTI, Marseille, France}

\author{N. Fintzi}
\affiliation{IFP Energies Nouvelles, Rond-Point de l’echangeur de Solaize, 69360, Solaize, France}

\author{G. Voth}
\affiliation{Department of Physics, Wesleyan University, Middletown, CT 06459, USA}

\author{J. L. Pierson}
\affiliation{IFP Energies Nouvelles, Rond-Point de l’echangeur de Solaize, 69360, Solaize, France}

\author{K. Gustavsson}
\affiliation{Department of Physics, Gothenburg University, Gothenburg, SE-40530 Sweden}
\author{B. Mehlig}
\affiliation{Department of Physics, Gothenburg University, Gothenburg, SE-40530 Sweden}


\begin{abstract}
Numerical simulation of particle motion in fluids at low particle Reynolds numbers is often based on empirical force and torque models obtained by fitting force and torque from {\em ab-initio} computations for simple particle shapes such as spheres, spheroids, or cylindrical disks and fibres. To do the same for more complex particles shapes, one needs to first know how particle shape constrains the dependence of force and torque on flow velocity, its gradient, and on particle orientation. Here we use symmetry analysis and perturbation theory to determine the form of the hydrodynamic torque on a particle settling in a quiescent fluid at low but non-zero particle Reynolds numbers, for particle shapes with different point-group symmetries. The symmetry conclusions are verified  by comparing with
explicit calculations for nearly spherical particles.
\end{abstract}


\maketitle

\newpage
\makeatletter
\onecolumngrid

\section{Introduction}
Consider a small particle moving in a fluid. How does particle shape constrain the hydrodynamic force and torque experienced by the particle? This is an  important question, because 
numerical studies of particle motion in complex flows tend to rely on models for force and torque, which allow for efficient computer simulation. A common way to derive such models is to use parameterisations of force and torque obtained by ab-initio simulations \cite{Sanjeevi18,Ouchene15,Ouchene16,Ouchene20,Holzer09,Zastawny12,Froehlich20}, requiring an {\em ansatz} for how force and torque depend on the flow velocity and its gradients. Any such ansatz, in turn, depends on particle shape. 
In Stokes flow, it is known how particle-shape symmetries determine force and torque on small particles in a viscous fluid \cite{Bretherton:1962,brenner1964stokes,Happel_Brenner_1983,fries2017angular,witten2020review,collins2021lord,miara2024dynamics,huseby2024helical}.
In many applications, however, the particles are too large or move too rapidly for the Stokes approximation to apply,
because it neglects corrections to hydrodynamic force and torque that arise at non-zero particle Reynolds number 
$\rep$. 
Examples include ice crystals in cold clouds \cite{Heymsfield_1973,Sassen_1980,Klett_1995,Noel_Sassen_2005,Gustavsson_2021}, and other atmospheric particles~\cite{Newsom_Bruce_1994,newsom1998orientational,pierson2023inertial,bhowmick2024inertia,candelier2024torques}.
Atmospheric particles are frequently modelled using highly symmetric shapes
such as spheroids \cite{bhowmick2024inertia},
but often atmospheric particles are asymmetric, meaning that certain rotation and reflection symmetries are broken~\cite{Heymsfield_1973,cai2020origin}.

At small but non-zero particle Reynolds number $\rep$, hydrodynamic forces and torques on axisymmetric particles (such as straight fibres and flat disks) settling in a quiescent fluid have been calculated in perturbation theory
\cite{brenner1961oseen,Khayat_Cox_1989,Dabade_2015}, 
{\em ab-initio} computer simulations~\cite{Jiang_2021}, and verified
in laboratory experiments
~\cite{Lopez_Guazzelli_2017,roy2023orientation,Cabrera_2022}. This has led to reliable parameterisations of inertial forces and torques for such particles.  
\citet{brenner1963resistance} and \citet{cox1965steady} derived general expressions for the steady force on 
a particle of arbitrary shape settling
in a quiescent fluid, in terms of 
of tensors and pseudo-tensors of ranks two and three. They also  considered how shape symmetries (reflections and $\pi/2$ as well as $\pi$ rotations) constrain certain contributions to force and torque to vanish. 
Here we analyse how particle-shape symmetries constrain the general form of the non-zero contributions to the hydrodynamic torque on a particle with arbitrary shape settling in a quiescent fluid, and determine in particular how many parameters are needed to fully parameterise the torque to order $\rep$  (the general conclusion is: the more shape symmetries are broken, the more parameters are needed).
 We verify our symmetry considerations by comparing with explicit calculations for nearly spherical particles \cite{candelier2015role_a,collis2024unsteady} with different point-group symmetries, and discuss the implications of our results for steady settling of small asymmetrical particles in a quiescent fluid. Last but not least, we generalise the results of \citet{cox1965steady} to unsteady motion of small asymmetrical particles at non-zero $\rep$, indicating 
how history effects may change the transient angular dynamics for such particles. 

\begin{figure}
\begin{overpic}[width=5cm]{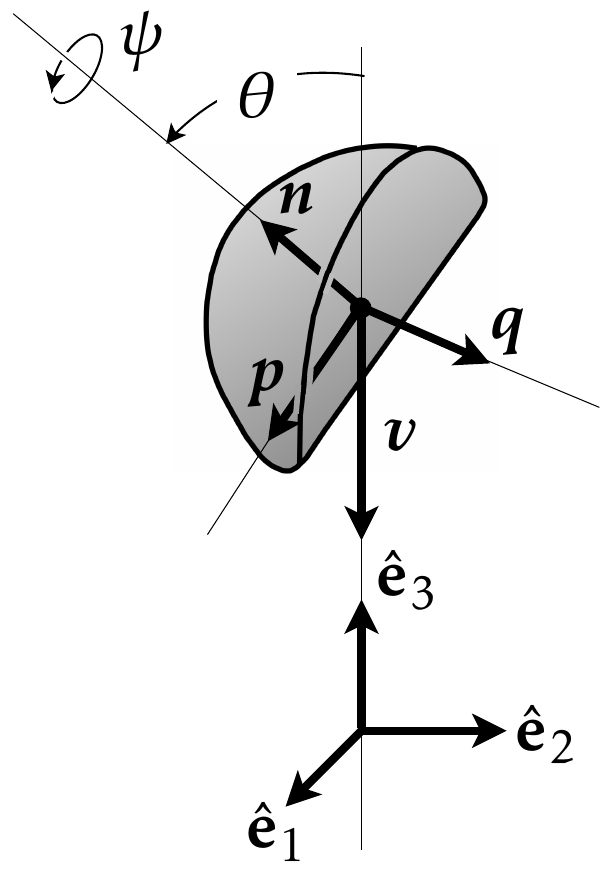}
\end{overpic}
\caption{\label{fig:cs} Particle with $C_{2v}$ shape symmetry moving through  a quiescent fluid with velocity $\ve v $.
The particle has the shape
of a bent disk \cite{miara2024dynamics}.
It has one $C_2$-rotation axis $\ve n$ and two reflection symmetries $\ve p \to -\ve p$ and $\ve q \to -\ve q$. Here $\ve p, \ve q, \ve n$ are the basis vectors of a coordinate system that rotates with the particle. The basis vectors of the laboratory coordinate system 
are denoted by $\hat{\bf e}_1,\hat{\bf e}_2, \hat{\bf e}_3$, and the velocity is $\ve v = -v \hat{\bf e}_3$. Also shown 
are the Euler angles $\theta$, $\psi$ that specify particle orientation. We use the $Z_\phi Y_\theta Z_\psi$ convention, as described in the supplemental material of Ref.~\cite{candelier2024torques}. }
\end{figure}

Our calculations are motivated in part by recent experiments, numerical simulations, and calculations
that show how breaking of fore-aft symmetry changes the
fluid-inertia torque \cite{Candelier_Mehlig_2016,Roy_2019,ravichandran2023orientation}.
Another motivation comes from recent experimental results for the fluid-inertia torque on curved atmospheric fibres settling in a quiescent fluid \cite{tatsii2024shape,candelier2024torques} which indicate that breaking a reflection symmetry can have a significant effect on how the fluid-inertia torque depends on the orientation of an asymmetric
particle (Figure~\ref{fig:cs})
that settles in a quiescent fluid at small but non-zero particle Reynolds numbers.

The remainder of this article is organised as follows.
In Section~\ref{sec:background}, we first recall results in the Stokes limit \cite{Bretherton:1962,Happel_Brenner_1983}.
Then we describe the general form of 
fluid-inertia corrections to Stokes torque for small 
but non-zero values of $\rep$, for particles of arbitrary shape in a uniform flow. 
In Section~\ref{sec:symmetries} we derive the form of the fluid-inertia torque for the particle shapes illustrated in  Figure~\ref{fig:shapes}, to order $\rep$. In Section \ref{sec:ns} we summarise explicit results for nearly spherical yet asymmetrical particles that confirm the symmetry analysis. We discuss our findings in Section~\ref{sec:discussion}, and relate them to known results for nearly spherical \cite{cox1965steady}, slender \cite{Khayat_Cox_1989}, and ellipsoidal \cite{bhowmick2024twist} particles. Section~\ref{sec:conc} contains our conclusions and a brief discussion of open questions, as well as possible applications.

\section{Background}
\label{sec:background}
Consider a particle moving with velocity $\ve v$ and
angular velocity $\ve \omega$ in a quiescent fluid (Figure~\ref{fig:cs}). Hydrodynamic force and torque upon a particle in a fluid are determined by integrals of components of the stress tensor over
the particle surface~$\mathscr{S}$:
\begin{align}\label{eq:torque}
    {\ve  F}&=  \int_{\mathscr{S}}\!\!\bbsigma\cdot {\rm d}\ve s \quad\mbox{and}\quad 
    {\ve  T}=  \int_{\mathscr{S}}\!\! \ve r\wedge(\bbsigma\cdot {\rm d}\ve s) \,.
\end{align}
Here $\ve r$ is the vector from the chosen reference point to a point on the particle surface (we take the centre-of-mass as the reference).
 The elements of the stress tensor $\bbsigma$ are given by $\sigma_{mn}= -p\delta_{mn} + 2 \mu S_{mn}$, where $\mu$ denotes the viscosity, and $S_{mn}$ are the elements of the strain-rate matrix, the symmetric part of the matrix of disturbance-velocity gradients $\partial w_i/\partial x_j$. In the frame translating with the particle, the disturbance flow $\ve w$ satisfies the Navier-Stokes equations:
\begin{equation}\label{eq:eom}
    {{\rm Re}_{p}\, {\rm Sl}\,\frac{\partial \ve w}{\partial t}}		-{\rm Re}_{p} {\ve v} \cdot \boldsymbol{\nabla} \ve{w}+{\rm Re}_{\rm{p}} \: \ve{w} \cdot \boldsymbol{\nabla} \ve{w} = - \boldsymbol{\nabla} p + \boldsymbol{\triangle} \ve{w} \,,\quad\ve \nabla \cdot \ve w = 0\,.
\end{equation}
We impose no-slip boundary conditions on the particle surface, $\ve w = \ve v + \ve \omega \wedge \ve r$,  and require that $\ve{w} \to \ve{0}$ as $|\ve r|\to \infty$.
Equation~(\ref{eq:eom}) was non-dimensionalised using an estimate $v_0$ of the particle velocity, the particle size $a$, the time scale $\tau_0$ at which the particle boundary conditions change, the angular-velocity scale $\omega_0=v_0/a$, and the pressure scale $\mu v_0/a$. Force and torque are non-dimensionalised as
\begin{align}
    \ve F /(\mu a v_0) \to \ve F \quad\mbox{and}\quad
    \ve T /(\mu a^2 v_0) \to \ve T \,.
\end{align} 
From here on, we quote all expressions in non-dimensional variables. 
The non-dimensional numbers in Eq.(\ref{eq:eom}) are the particle Reynolds number $\rep$ and  the Strouhal number Sl (a non-dimensional measure
of the time scale at which the particle boundary conditions change):
\begin{align}
    \rep =  a v_0/\nu\,,\quad {\rm Sl} = a/(v_0 \tau_0)\,.
\end{align}

For $\rep=0$, the left-hand side of Eq.~(\ref{eq:eom}) vanishes, and one obtains the steady Stokes approximation to the disturbance flow. Integrating the resulting $\bbsigma^{(0)}$ over the particle surface as specified in Eqs.~(\ref{eq:torque}) yields Stokes force $\ve F^{(0)}$ and torque $\ve T^{(0)}$. Both are linearly related to the particle velocity $\ve v$ and its angular velocity $\ve \omega$. In a quiescent fluid, this relation is written as \cite{Kim_Karrila_1991}
\begin{align}\label{eq:resistance_tensors}
    \begin{bmatrix}
        \ve F^{(0)} \\
        \ve T^{(0)}
    \end{bmatrix}
    = -
    \begin{bmatrix}
        \ma A & {\ma B}^{\sf T}  \\
        \ma B & \ma C 
    \end{bmatrix}
    \begin{bmatrix}
        {\ve v} \\
        \ve \omega
    \end{bmatrix}\,.
\end{align}
Here $\ma A, \ma B,$ and $\ma C$ are Stokes resistance tensors which depend on particle orientation, on particle shape, but not on particle size $a$. The superscript $\sf T$ denotes the transpose.

\citet{Happel_Brenner_1983} analysed how the resistance tensors in Eq.~(\ref{eq:resistance_tensors}) depend on particle shape (for corresponding results in shear flow, see Refs.~\cite{Bretherton:1962,brenner1963stokes,fries2017angular,ishimoto2020jeffery}).
Using the fact that the energy dissipation is a scalar, they determined how the Stokes resistance tensors transform under orthogonal coordinate transformations, and concluded the following: if the boundary conditions imposed by the particle upon the flow are invariant
under an orthogonal coordinate transformation $\ma O$, this constrains
the elements of the Stokes tensors as follows \cite{Happel_Brenner_1983}:
\begin{subequations}
\label{eq:stokes}
\begin{align}
    A_{ij} & = O_{mi} A_{mn}O_{nj} \,,\\
    B_{ij} & = \det[\ma O]O_{mi} B_{mn}O_{nj}\,,\label{eq:Bconstraint}\\
    C_{ij} &= O_{mi} C_{mn}O_{nj}\,.
\end{align} 
\end{subequations}
Alternatively, these transformation rules can be derived from the covariance of Eq.~\eqref{eq:resistance_tensors}. Transform $\ve F(\ve v,\ve \omega)$ and $\ve T(\ve v,\ve \omega)$ as $\ve{F}'=\ma O \ve{F}$ and $\ve{T}' = \det[\ma O] \ma O \ve{T}$, where $\ve{F}'=\ve F(\ma O \ve{v},\det[O] \ma O \ve\omega)$ and $\ve{T}'=\ve T(\ma O \ve{v},\det[O] \ma O \ve\omega)$, and $\ma O$ is a particle symmetry~\cite{fries2017angular}.
Eq.~(\ref{eq:Bconstraint}) shows that the translation-rotation coupling tensor $\ma B$ transforms in a different way from $\ma A$ and~$\ma C$. 
The factor $\det [\ma O]$ ensures that $\ma B=0$ for particles with sufficient numbers of reflection symmetries (spheroid, ellipsoid).
 
For non-zero particle Reynolds number $\rep$, the left side of Eq.~(\ref{eq:eom}) matters. To first order in $\rep$, general expressions for the steady hydrodynamic force and torque (${\rm Sl}=0$) were derived in Refs.~\cite{brenner1961oseen,brenner1963resistance,cox1965steady}. 
\citet{lovalenti1993hydrodynamic} derived 
the first-order force corrections in $\rep$ for unsteady flow. Here 
we obtain the corresponding expressions
for the unsteady torque, 
for particles of arbitrary shapes. In the steady limit, our results simplify to those in Refs.~\cite{brenner1961oseen,brenner1963resistance,cox1965steady}.

To compute the effects of weak fluid inertia upon hydrodynamic 
force and torque, one must account for the fact that the perturbation
problem in $\rep$ is singular, due to the slow spatial decay of the disturbance flow \cite{childress1964slow,saffman1965lift}. We use matched asymptotic expansions \cite{childress1964slow,saffman1965lift} 
to deal with this problem.
In this approach, an inner solution valid near the particle is matched to an outer solution valid far from the particle. In the following, it is assumed that $1 \gg \rep \gg \rep {\rm Sl}$.
Far from the particle, its boundary condition is replaced by a point-source forcing, so that Eq.~(\ref{eq:eom}) becomes
\begin{equation}\label{eq:outer}
{\rm Re}_{p}\,{\rm Sl}\,\frac{\partial \ve w_{\mbox{\scriptsize out}}}{\partial t}
- {\rm Re}_{p}\, \ve v \cdot \boldsymbol{\nabla} \ve w_{\mbox{\scriptsize out}}
+ {\rm Re}_{p}\, \ve w_{\mbox{\scriptsize out}} \cdot \boldsymbol{\nabla} \ve w_{\mbox{\scriptsize out}}
= - \boldsymbol{\nabla} p_{\rm out}
+ \boldsymbol{\triangle}\ve w_{\mbox{\scriptsize out}}
+ \ve f^{(0)}\,\delta(\ve r)\:.
\end{equation}
Here $\ve f^{(0)} = -\ve F^{(0)}$ denotes the force exerted by the particle on the fluid in the Stokes limit. 
This outer problem
can be solved by Fourier transform.
The outer solution in Fourier space can be expanded as a series in the small parameter $\sqrt{{\rm Re}_{p}\,{\rm Sl}}$, keeping in mind that this expansion must be carried out in the sense of distributions \citep{meibohm2016angular,redaelli2023hydrodynamic}.
In configuration space, the corresponding series expansion takes the form
\begin{align}\label{W_outer}
	{\ve w}_{\mbox{\scriptsize out}} &
	 = \frac{1}{8\pi}\left(\frac{\ma{1}}{r}  + \frac{\ve r\otimes \ve r}{r^3}\right)  \ve f^{(0)}
	 -{\rm Re}_{\rm{p}} \vp\cdot \boldsymbol{\nabla}\left[\frac{3 r}{32\pi} \left( \ma{1} - \frac{1}{3} \frac{\ve r \otimes \ve r}{r^2}\right)  \ve f^{(0)} \right]
    +\sqrt{\rep {\rm Sl}}\,\,\ve {\mathcal U}\,.
\end{align}
The term $\ve{\mathcal U}$ corresponds to a uniform flow and originates from a singular contribution to the expansion of the outer solution in Fourier space, namely, a term proportional to the Dirac distribution $\delta(\ve k)$.  
This uniform contribution takes the form
\begin{align}
\label{eq:U}
\sqrt{\rep {\rm Sl}}\,\ve{\mathcal U}
= \sqrt{\rep {\rm Sl}}
\int_0^t \!{\rm d}\tau\; {\ma K}^{(1)}(t,\tau)\cdot\frac{{\rm d}\ve f^{(0)}(\tau)}{{\rm d}\tau}
+ \rep
\int_0^t \!{\rm d}\tau\; {\ma K}^{(2)}(t,\tau)\cdot \ve f^{(0)}(\tau)\,.
\end{align}
The kernels ${\ma K}^{(1)}(t,\tau)$ and ${\ma K}^{(2)}(t,\tau)$ are given in Appendix~\ref{app:A}. They are equivalent to the kernel involved in the $\rep$-correction to the force in unsteady flow, derived by \citet{lovalenti1993hydrodynamic}. The kernels also appear
in the work of \citet{redaelli2023hydrodynamic}, except that there the expressions were simplified using the fact that the force $\ve f^{(0)}$ and the velocity $\ve v$ are collinear for a spherical particle. 

In the inner region, disturbance velocity and pressure are expanded as
\begin{equation}
{\ve w}_{\mbox{\scriptsize in}} 
= {\ve w}_{\mbox{\scriptsize in}}^{(0)}
+ \sqrt{\rep {\rm Sl}}\;{\ve w}_{\mbox{\scriptsize in}}^{(1)}
+ \ldots,
\qquad
p_{\mbox{\scriptsize in}}
= p_{\mbox{\scriptsize in}}^{(0)}
+ \sqrt{\rep {\rm Sl}}\; p_{\mbox{\scriptsize in}}^{(1)}
+ \ldots\,.
\end{equation}
These inner fields are then matched order by order with their counterparts in the outer solution. The leading-order term ${\ve w}_{\mbox{\scriptsize in}}^{(0)}$ is the solution of the steady Stokes problem. The term $\rep {\rm Sl} \,\,\partial_t{\ve w}^{(0)}_{\mbox{\scriptsize in}}$ can be neglected here because it is of higher order in the regular inner expansion.
This leading-order solution matches the first term of the outer expansion (\ref{W_outer}).  
The next correction ${\ve w}_{\mbox{\scriptsize in}}^{(1)}$ 
satisfies the inhomogeneous Stokes problem~\cite{redaelli2023hydrodynamic}
\begin{equation}\label{W_inner}
\sqrt{\frac{{\rm Re}_{p}}{{\rm Sl}}}
\left[
-\, \ve v \cdot \boldsymbol{\nabla} \ve w_{\mbox{\scriptsize in}}^{(0)}
+ \ve w_{\mbox{\scriptsize in}}^{(0)} \cdot \boldsymbol{\nabla} \ve w_{\mbox{\scriptsize in}}^{(0)}
\right]
= - \boldsymbol{\nabla} p^{(1)}
+ \boldsymbol{\triangle} \ve w_{\mbox{\scriptsize in}}^{(1)} \:.
\end{equation}
The function  ${\ve w}_{\mbox{\scriptsize in}}^{(1)}$ satisfies no-slip condition on the particle surface (i.e., $\ve w_{\mbox{\scriptsize in}}^{(1)} = \ve 0$) and matches the last two terms of the outer expansion (\ref{W_outer}).  
The solution of the inhomogeneous problem Eq.~(\ref{W_inner}) is sought in the form
$\ve w_{\mbox{\scriptsize in}}^{(1)} = (\ve w_{\rm p} + \ve{\mathcal U}) + \ve w_h$,
that is as the sum of a particular (forced) solution $\ve w_{\rm p} + \ve{\mathcal U}$, and a homogeneous solution $\ve w_h$. Here, $\ve w_{\rm p}$ satisfies (\ref{W_inner}), while $\ve w_h$ satisfies the Stokes equations together with the boundary condition
$\ve w_h = - (\ve w_{\rm p} + \ve{\mathcal U})$ on the particle surface,
ensuring that $\ve w_{\mbox{\scriptsize in}}^{(1)}$ fulfills the no-slip condition.  The particular solution $\ve w_{\rm p}$ depends quadratically on $\ve v$ and $\ve \omega$, due to the quadratic convective term in (\ref{W_inner}).   Similarly, the component of $\ve w_h$ linked to $\ve w_{\rm p}$ inherits this quadratic dependence because of the boundary condition imposed on the homogeneous part.  Consequently, the total force and torque at order $\rep$ are
\begin{equation}
\label{eq:FTsum}
\ve F = \ve F^{(0)} + \rep \,\ve F^{(1)}_{\rm reg} + \sqrt{\rep {\rm Sl}} \,\ve F^{(1)}_{\rm sing}\,,
\qquad
\ve T = \ve T^{(0)} + \rep \,\ve T^{(1)}_{\rm reg} + \sqrt{\rep {\rm Sl}} \,\ve T^{(1)}_{\rm sing}\,,
\end{equation}
where the so-called 'singular' contributions come from 
the singular part of the outer expansion,
\begin{align} 
\label{eq:sing}
\ve F^{(1)}_{\rm sing} = \ma A \cdot \ve{\mathcal U}, \quad 
\ve T^{(1)}_{\rm sing} = \ma B \cdot \ve{\mathcal U}\,,
\end{align}
and the 'regular' contributions take the form
\begin{subequations}
\label{eq:FTreg0}
\begin{align}
\ve F^{(1)}_{\rm reg} &= 
\ma M^{(vv)} : {\ve v} \otimes {\ve v}
+ \ma M^{(\omega\omega)} : {\ve \omega} \otimes {\ve \omega}
+ \ma M^{(v\omega)} : {\ve v} \otimes {\ve \omega}, \\
\ve T^{(1)}_{\rm reg} &= 
\ma N^{(vv)} : {\ve v} \otimes {\ve v}
+ \ma N^{(\omega\omega)} : {\ve \omega} \otimes {\ve \omega}
+ \ma N^{(v\omega)} : {\ve v} \otimes {\ve \omega}.
\end{align}
\end{subequations}
Here the
  colon denotes double contraction 
(the elements of $\ma M:\ma S$ with the second
order tensor $\ma S$ are $\sum_{jk}M_{ijk} S_{jk}$). 
We remark that terms of the form
$\ma M^{(\omega v)}:\ve \omega \otimes \ve v$
and 
$\ma N^{(\omega v)}:\ve \omega \otimes \ve v$ can be omitted since they have the same structure and transformation rules as $\ma M^{(v\omega)}:\ve v \otimes \ve \omega$
and  $\ma N^{(v\omega)}:\ve v \otimes \ve \omega$.
Because the contributions (\ref{eq:FTreg0}) are determined by the inner solution, they are valid instantaneously, also for non-zero angular velocities. Such quadratic terms were obtained  in Refs.~\cite{brenner1961oseen,brenner1963resistance,cox1965steady}  using a reformulation of the reciprocal theorem.

In the steady limit, Eq.~(\ref{eq:U}) simplifies. 
In order to obtain a steady disturbance flow,
$\ve v$, $\ve \omega$, and $f^{(0)}$ must be independent of time (or at least vary very slowly). 
Equation~(\ref{eq:resistance_tensors}) relates
these three vectors: 
$\ve f^{(0)} = -\ve F^{(0)} = \ma A \ve v+\ma B^{\sf T} \ve \omega$.
Accounting for a possible time dependence of $\ma A$, one finds:
\begin{equation}
\label{eq:extra}
\frac{{\rm d} \ve f^{(0)}}{{\rm d}t} 
= \ve \omega \wedge \ve f^{(0)} 
- \ma A \cdot (\ve \omega \wedge \ve v) 
+ \ma A \cdot \frac{{\rm d} \ve v}{{\rm d}t} 
+ \ma B^{\sf T} \cdot \frac{{\rm d} \ve \omega}{{\rm d}t}\:.
\end{equation}
We conclude that it is not possible in general  to simultaneously satisfy  
$\tfrac{{\rm d} }{{\rm d}t} \ve v= 
\tfrac{{\rm d} }{{\rm d}t} \ve \omega= 
\tfrac{{\rm d} }{{\rm d}t} \ve f^{(0)}= \ve 0$, at least when $\ma B \neq \ma 0$. An exception to
this rule is the following example: an axisymmetric pear-shaped particle that rotates around its symmetry axis
with constant angular velocity $\ve \omega$, so that
$\ve f^{(0)}$, $\ve \omega$, and $\ve v$ are collinear. 

A second example is a particle moving
at constant velocity with zero angular velocity. In this case, Eq.~(\ref{eq:extra}) 
allows for both $\ve f^{(0)}$ and $\ve v$
to be constant. When $\ve f^{(0)}$ varies slowly (so that its time scale of variation is much larger than $a^2/\nu$, or equivalently ${\rm Sl}\to 0$), Eq.~(\ref{eq:U}) simplifies.  The first integral on the right-hand side of Eq.~(\ref{eq:U}) vanishes, and $\ve f^{(0)}$ can be pulled out from the second integral. Upon performing the remaining integral, one obtains
\begin{align}
\label{eq:Usteady}
\ve {\mathcal U} 
= -\sqrt{\frac{\rep}{{\rm Sl}}}\,
\frac{3v}{32\pi}
\left( 
\ma I - \frac{1}{3}\, \hat{\ve v}\otimes \hat{\ve v}
\right)
\cdot \ve f^{(0)}\,,
\end{align}
where $\hat{\ve v}$ denotes the unit vector in the direction of $\ve v$,
and $\ma I$ is the identity matrix.
Since $\ve{\mathcal U}$  corresponds to a uniform flow at infinity, 
the singular contributions simplify to 
\begin{align} 
\ve F^{(1)}_{\rm sing} &=  -\sqrt{\frac{\rep}{{\rm Sl}}}\frac{3 v}{32 \pi}
\ma A \cdot \big[\ma A \cdot{\ve v}
 -\tfrac{1}{3} (\hat{\ve v}\otimes {\hat{\ve v}} )\cdot\ma A\cdot {\ve v} 
\big]\,,\,\quad
\ve T^{(1)}_{{\rm sing}} = -\sqrt{\frac{\rep}{{\rm Sl}}}\frac{3 v}{32 \pi}
{\ma B}\cdot\big[ \ma A\cdot {\ve v} 
 -\tfrac{1}{3} (\hat{\ve v}\otimes {\hat{\ve v}}) \cdot\ma A\cdot {\ve v} 
\big]\,.
\label{eq:Tsing_steady}
\end{align}  
These contributions to the steady force and torque  approximation were first obtained in Refs.~\cite{brenner1961oseen,brenner1963resistance,cox1965steady}  using the reformulation of the reciprocal theorem mentioned above. 
\citet{cox1965steady} includes terms with $\ve \omega \neq 0$ in his $\rep$-contributions for force and torque. As pointed out above, this is consistent in certain special cases, but not in general.

Asymptotic matching, employed here, offers  additional insights. First, it gives a clear interpretation of  where the different terms to the fluid-inertia forces and torques come from. The contributions (\ref{eq:Tsing_steady}) originate from the far field and may contain transient, time-dependent terms, while
the terms (\ref{eq:FTreg0}) are instantaneous contributions deriving from the inner solution. Second, asymptotic matching allows to compute the transient, time-dependent terms (\ref{eq:U}), so-called history forces and torques  that affect the transient dynamics of the particle. Third, the derivation of Eqs.~(\ref{eq:Tsing_steady})  from the unsteady $\mathcal U$ highlights that Eqs.~(\ref{eq:Tsing_steady}) assume that the particle velocity is steady. Upon a change of centre of reference at non-zero $\ve \omega$ this assumption may fail. Therefore one needs to go back to (\ref{eq:U}) and (\ref{eq:sing}) in order to see that the particle dynamics does not depend on the arbitrary choice of the centre-of-reference. 

Now consider how particle-shape symmetries constrain force and torque to order $\rep$. The tensors $\ma A$
and $\ma B$  in Eq.~(\ref{eq:sing})
are defined in the Stokes limit, and their components must obey  Eqs.~(\ref{eq:stokes}). This in turn implies the same for $\ve F^{(1)}_{\rm reg}$ and $\ve T^{(1)}_{\rm reg}$, 
which constrains the components of the tensors $\ma M$
and $\ma N$ as follows:
\begin{subequations}
\label{eq:3MN}
\begin{align}
N_{ijk}^{(vv)}   &= \det[\ma O]\,O_{mi} N_{mnl}^{(vv)}
O_{nj} O_{lk}\quad\,\mbox{and}\quad 
M_{ijk}^{(vv)} =  O_{mi} M_{mnl}^{(vv)} 
O_{nj} O_{lk} \,,\\
N_{ijk}^{(\omega\omega)}  & = \det[\ma O]\,O_{mi} N_{mnl}^{(\omega\omega)}
O_{nj} O_{lk}\quad\mbox{and}\quad 
M_{ijk}^{(\omega\omega)}  = O_{mi} M_{mnl}^{(\omega\omega)}
O_{nj} O_{lk}
\,,\\
N_{ijk}^{(v\omega)}   &= O_{mi} N_{mnl}^{(v\omega)}
O_{nj} O_{lk}\,,\hspace*{8mm}\quad\mbox{and}\quad
M_{ijk}^{(v\omega)}   = \det[\ma O]\,O_{mi} M_{mnl}^{(v\omega)}
O_{nj} O_{lk}
 \,.
\end{align}
\end{subequations}
Note that the rank-3 tensors $\ma M$ and $\ma N$ arise in the first
order of a systematic expansion in $\rep$. So their elements are determined by Stokes solutions, but the tensors depend on particle shape in a more complex way than solely through the rank-two Stokes resistance tensors $\ma A$, $\ma B$, and $\ma C$. Equation~(\ref{eq:3MN}) implies that 
$\ma N^{(vv)}$, $\ma N^{(\omega\omega)}$, and $\ma M^{(v\omega)}$ transform as the Stokes tensor $\ma H$ that
couples $\ve T^{(0)}$ to the
strain $\ma S$~\cite{Bretherton:1962,brenner1963stokes,fries2017angular,ishimoto2020jeffery}.
Likewise, $\ma M^{(vv)}$, $\ma M^{(\omega\omega)}$, and $\ma N^{(v\omega)}$ have the same transformation properties.

 \citet{brenner1963resistance} 
 and \citet{cox1965steady}  determined
 under which circumstances
 the components of the vectors $\ma M^{(vv)} :\ve v \otimes \ve v$ and $\ma N^{(vv)} :\ve v\otimes \ve v$ (and the corresponding contributions for the other rank-3 tensors) vanish, by considering 
$\tfrac{\pi}{2}$- and $\pi$-rotations as well as reflections. In general, these contributions do not vanish. Here we use Eqs.~(\ref{eq:stokes}), (\ref{eq:FTreg0}), (\ref{eq:Tsing_steady}), and (\ref{eq:3MN}) to derive constraints
on the non-zero components of the fluid-inertia torque for asymmetric particles, obtained from symmetric shapes by breaking rotation and/or reflection symmetries. This allows to understand similarities and differences in the dynamics of asymmetric particles settling in a quiescent fluid \cite{Candelier_Mehlig_2016,Roy_2019,candelier2024torques,miara2024dynamics,flapper2025settling}.

\section{Symmetry constraints on torque}\label{sec:symmetries}
\begin{figure}
    \begin{overpic}[width=15cm]{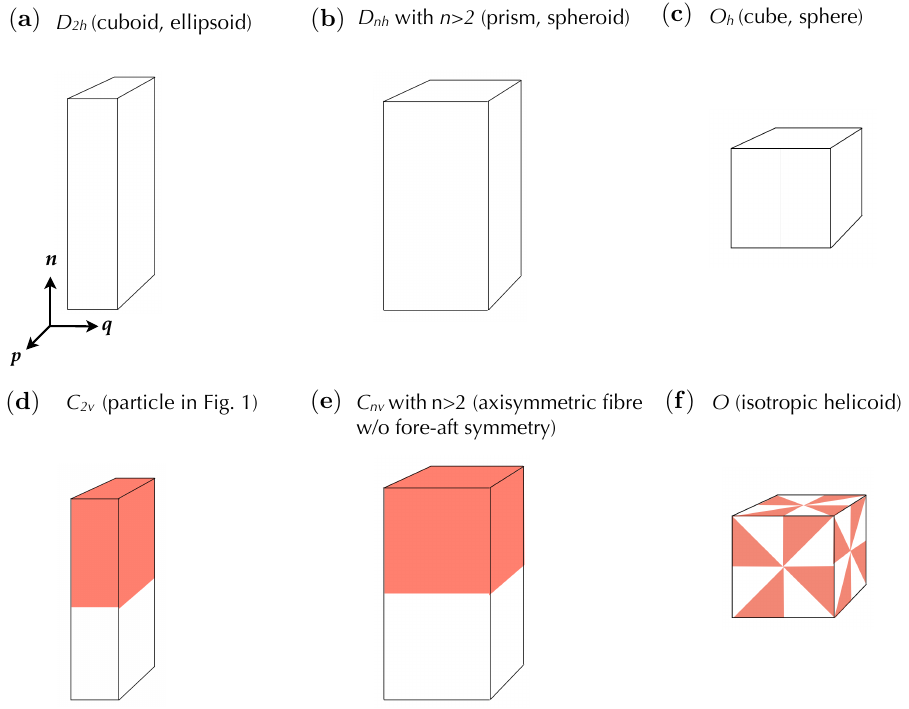}\end{overpic}\caption{\label{fig:shapes}
    Particle-shape symmetries considered in Section~\ref{sec:symmetries} (see that Section for the names of the six point groups). In panel~({\bf a}), the  orientation of the body-centered coordinate system $\ve p$, $\ve q$, $\ve n$ is shown (same for  other panels). Point-group symmetries represented as in Figs.~7.2 and 7.3 in Ref.~\cite{Ashcroft}.}
\end{figure}

Figure~\ref{fig:shapes} illustrates
particles with different point-group symmetries. 
The first row shows particles with reflection and discrete rotation symmetries, such as cuboid, prism, and cube. In the limit where the rotation symmetry becomes continuous, these generalize to ellipsoid, spheroid, and sphere. 
The second row shows shapes with broken reflection symmetries, namely
broken fore-aft symmetry (point-group symmetry $C_{2v}$), and the
isotropic helicoid \cite{kelvin1871hydrokinetic} (point-group symmetry $O$).

\subsection{Prismatic symmetry $D_{2h}$}\label{sec:D2h}
The cuboid shown in Figure~\ref{fig:shapes}({\bf a}) has three reflection symmetries in the planes orthogonal to the three principal axes (point group symmetry $D_{2h}$).
The three orthogonal reflections constrain the three Stokes resistance tensors to be $\ma A = {\rm diag}[A_{pp}, A_{qq}, A_{nn}]$, $\ma C = {\rm diag}[C_{pp}, C_{qq}, C_{nn}]$, and 
$\ma B = 0$. Further, the reflections constrain the inertial rank-3 tensors to $\ma N^{(v\omega)}=0$, and
\begin{align}\label{eq:Nellipsoid} 
    & {\ma N}_{p::}^{(vv)} = \begin{bmatrix}
    0 & 0 & 0 \\
    0 & 0 & {N}_{pqn}^{(vv)} \\
    0 & {N}_{pnq}^{(vv)} & 0
    \end{bmatrix}\,,\quad
    {\ma N}_{q::}^{(vv)} = \begin{bmatrix}
    0 & 0 & {N}_{qpn}^{(vv)} \\
    0 & 0 & 0 \\
    {N}_{qnp}^{(vv)} & 0 & 0
    \end{bmatrix}\,,\quad
    {\ma N}_{n::}^{(vv)} = \begin{bmatrix}
    0 & N_{npq}^{(vv)} & 0\\
    N_{nqp}^{(vv)} & 0 & 0 \\
    0 & 0 & 0
    \end{bmatrix}\,.
\end{align}
It follows from the discussion in Section~\ref{sec:background} that the tensors $\ma N^{(\omega\omega)}$ and $\ma M^{(v\omega)}$ have the same form as $\ma N^{(vv)}$, but with independent parameters. The tensors $\ma M^{(vv)}$ and $\ma M^{(\omega\omega)}$ vanish. The above constraints yield the following expressions for the $\rep$-corrections to the torque
\begin{subequations}\label{eq:ellipsoid}
\begin{align}\label{eq:Fp_ellipsoid_sym}
T_p^{(1)}&=
2N_{p(nq)}^{(vv)}  v_n v_q+2N_{p(nq)}^{(\omega\omega)}  \omega_n \omega_q \,,\\
T_q^{(1)}&= 
2N_{q(pn)}^{(vv)}  v_pv_n+2N_{q(pn)}^{(\omega\omega)}  \omega_p\omega_n \,,\\
T_n^{(1)}&=
2N_{n(pq)}^{(vv)} v_p v_q+2N_{n(pq)}^{(\omega\omega)} \omega_p \omega_q\,
\end{align}
\end{subequations}
in the  basis $\ve p, \ve q, \ve n$  (Figure~\ref{fig:shapes}). 
Here, $N_{p(nq)}^{(vv)}= \tfrac{1}{2} (N_{pnq}^{(vv)}+N_{pqn}^{(vv)})$,\ldots  are symmetrised coupling-coefficients.
Since $\ma B =0$, the singular contributions to the inertial torque, Eq.~(\ref{eq:Tsing_steady}), vanish. Thus, only the instantaneous terms from the inner solution contribute, meaning that Eqs.~(\ref{eq:ellipsoid}) are valid in the unsteady case, in particular for non-zero values of $\ve \omega$.

\subsection{Prismatic symmetry ($D_{nh}$ for $n>2$)}
Straight columns with a regular $n$-gon cross sections with $n>2$ ({\em prismatic symmetry $D_{nh}$} with $n>2$) possess discrete $n$-fold rotational symmetry ($C_n$) about the $\ve n$-axis. 
The case $n=4$ is illustrated in Fig.~\ref{fig:shapes}({\bf b}). Another example is cloud-ice crystals, often possessing hexagonal ($n=6$) cross sections~\cite{nakaya1954snow,furukawa2007snow}. Imposing $C_n$-rotation symmetry with $n>2$ 
around $\ve n$ imposes additional constraints on the non-zero coefficients in Eq.~\eqref{eq:ellipsoid} compared with $n=2$:
\begin{subequations}
    \label{eq:NspheroidAll}
    \begin{align}\label{eq:Nspheroid}
        A_{pp} = A_{qq} = A_\perp, \quad A_{nn} = A_\parallel,\quad C_{pp} = C_{qq} = C_\perp, \quad C_{nn} = C_\parallel,\\
        N_{pqn}^{(vv)} = -N_{qpn}^{(vv)} = N_1^{(vv)}\,,
        N_{pnq}^{(vv)} = -N_{pnq}^{(vv)} = N_2^{(vv)}\,, 
        N_{npq}^{(vv)} = -N_{nqp}^{(vv)} = N_3^{(vv)}\,.
    \end{align}
\end{subequations}
The elements of $\ma N^{(\omega\omega)}$ are constrained in the same way. This leaves six undetermined parameters in the $\ma N$-tensors. The $\rep$-contributions to the torque simplify to:
\begin{subequations}\label{eq:spheroid}
    \begin{align}
        T_p^{(1)}&=(N_1^{(vv)}+N_2^{(vv)}) v_q v_n +(N_1^{(\omega\omega)}+N_2^{(\omega\omega)}) \omega_q \omega_n\,,\\
        T_q^{(1)}&=-(N_1^{(vv)}+N_2^{(vv)}) v_p v_n -(N_1^{(\omega\omega)}+N_2^{(\omega\omega)}) \omega_p \omega_n\,,\\
        T_n^{(1)}&=0\label{eq:Tspheroid}
    \end{align}
\end{subequations}
in the basis $\ve p, \ve q, \ve n$ (Figure~\ref{fig:shapes}). Similar to Eqs.~(\ref{eq:ellipsoid}), these expressions are valid in the unsteady case. 
The rotational symmetry around $\ve n$ causes the torque $T_n$ to vanish. This follows because ${\ma N}_{n::}^{(vv)}$ and ${\ma N}_{n::}^{(\omega\omega)}$ are anti-symmetric, and contracted with the symmetric $\bm{v}\otimes\bm{v}$ and $\bm{\omega}\otimes\bm{\omega}$, respectively.
If $\ve\omega=0$, the torque is orthogonal to $\ve n$, and proportional to the velocity component $v_n$: $\ve T = (N_{1}^{(vv)}+N_{2}^{(vv)}) (\ve v\cdot\ve n)(\ve v\wedge \ve n)$, consistent with Refs.~\cite{Khayat_Cox_1989,Dabade_2015,Jiang_2021}.

\subsection{Full octahedral symmetry ($O_h$)}\label{sec:Oh}
Now consider particles with {\em full octahedral symmetry} $O_h$  [(Figure~\ref{fig:shapes}({\bf c})]. An example is the cube, but also the
sphere with continuous rotation symmetry.
For both examples, rotation and reflection symmetries further constrain Eqs.~(\ref{eq:NspheroidAll}). The  Stokes tensors $\ma A$ and $\ma C$ become proportional to the identity matrix, and the tensors $\ma N^{(vv)}$ and $\ma N^{(\omega\omega)}$ become proportional to the anti-symmetric Levi-Civita tensor
$\ve \epsilon$: $A_\perp = A_\parallel = A$, $C_\perp = C_\parallel = C$, $\ma N^{(vv)}=N^{(vv)} \ve \epsilon$, and $\ma N^{(\omega\omega)}=N^{(\omega\omega)} \ve \epsilon$. 
The latter conditions cause the torque (\ref{eq:spheroid}) to vanish.

As discussed in Section~\ref{sec:background}, $\ma M^{(v\omega)}$ is constrained in the same way as $\ma N^{(vv)}$, so $\ma M^{(v\omega)} = M^{(v\omega)} \ve \epsilon$, yielding a contribution proportional to $\ve v \wedge \ve \omega$, the Rubinow-Keller lift force for a spherical particle (Eq.~(71) in Ref.~\cite{rubinow1961transverse}), and its generalisation to particles with $O_h$ symmetry. See Appendix~\ref{app:NS} for details. 

Using that $\ma A=A\ma I$ and that $\hat{\ve v}$ is a unit vector, the singular contribution to the force (\ref{eq:Tsing_steady}) is isotropic, proportional to $\tfrac{2}{3} A^2 v^2 \hat {\ve v}$ \cite{brenner1963resistance,cox1965steady}. The theory to linear order in $\rep$ is isotropic, but does not rule out the existence of anisotropic terms at higher orders in in the Reynolds number. 
It is instructive to compare this conclusion with a more general symmetry argument that assumes only that force and torque to order $\rep$ are third-order polynomials in the components of $\hat{\ve v}$ with coefficients that are constrained by symmetry. Requiring invariance of force and torque under particle-shape symmetry transformations yields results consistent with the above conclusions, but they constrain force and torque less because the relation between force and torque encoded in Eq.~(\ref{eq:sing}) is not accounted for.
 
\subsection{Orthorhombic pyramidal symmetry ($C_{2v}$)}
Figure~\ref{fig:shapes}({\bf d}) shows a particle that has the same symmetry as $D_{2h}$, but with broken reflection symmetry in the $\ve n$-direction
(orthorhombic pyramidal symmetry  $C_{2v}$). Examples of such particles are curved planar fibres with constant radius of curvature \cite{candelier2024torques}, and symmetrically bent disks \cite{miara2024dynamics}
(Fig.~\ref{fig:cs}). The  tensors $\ma A$, $\ma C$, $\ma N^{(vv)}$, and $\ma N^{(\omega\omega)}$ of particles with $C_{2v}$ symmetry have the 
same form as for particles with $D_{2h}$ symmetry (discussed above). 
But the broken reflection symmetry results in a non-zero Stokes translation-rotation coupling:
\begin{equation}
\label{eq:bc2v}
    \ma B = \begin{bmatrix}
    0 & B_{pq} & 0\\
    B_{qp} & 0 & 0\\
    0 & 0 & 0
\end{bmatrix}\,,
\end{equation}
as well as a non-zero $\ma N^{(v\omega)}$:
\begin{equation}
    {\ma N}_{p::}^{(v\omega)} = 
    \begin{bmatrix}
    0 & 0 & {N}_{ppn}^{(v\omega)}\\
    0 & 0 & 0 \\
    {N}_{pnp}^{(v\omega)} & 0 & 0
    \end{bmatrix}\, \quad
    {\ma N}_{q::}^{(v\omega)} = \begin{bmatrix}
    0 & 0 & 0 \\
    0 & 0 & {N}_{qqn}^{(v\omega)} \\
    0 & {N}_{qnq}^{(v\omega)} & 0
    \end{bmatrix}\,,\quad
    {\ma N}_{n::}^{(v\omega)} = \begin{bmatrix}
    {N}_{npp}^{(v\omega)} & 0 & 0 \\
    0 & {N}_{nqq}^{(v\omega)} & 0 \\
    0 & 0 & {N}_{nnn}^{(v\omega)}
    \end{bmatrix}\,.
\end{equation}
Now consider the corresponding
expression for the fluid-inertia correction to the hydrodynamic torque. 
Since the general symmetry analysis
of the unsteady contribution $\mathcal{U}$
is complicated, we only consider
the steady limit, which requires 
$\ve \omega=0$, as discussed in the previous Section. For the steady $\rep$-contributions to the torque we find:
\begin{subequations}\label{eq:C_2v}
\begin{align}
    T_p^{(1)} &= -\tfrac{1}{32\pi}vB_{pq}[3A_{qq} - ( A_{pp} \up^2+A_{qq}\uq^2 + A_{nn}\un^2)]v_q + 2N_{p(nq)}^{(vv)} v_nv_q \,,\\\label{eq:TqFibre}
    T_q^{(1)} &= -\tfrac{1}{32\pi}vB_{qp}[3A_{pp} - ( A_{pp} \up^2+A_{qq}\uq^2 + A_{nn}\un^2)]v_p
+ 2N_{q(np)}^{(vv)} v_nv_p \,,\\
    T_n^{(1)} &= 2N_{n(qp)}^{(vv)} v_qv_p 
     \end{align}
\end{subequations}
in the  basis $\ve p, \ve q, \ve n$  (Figure~\ref{fig:shapes}). 
Comparing with Eq.~(\ref{eq:ellipsoid}) for the ellipsoid, we see that the form of the quadratic terms is unaffected by the reflection symmetry for $\ve \omega=0$.  Broken reflection symmetry still changes the fluid-inertia torque, due to the non-zero translation-rotation coupling $\ma B$ in the Stokes limit, yielding a non-zero singular contribution from Eq.~(\ref{eq:Tsing_steady}).

\subsection{Pyramidal symmetry ($C_{nv}$ with $n>2$)}\label{sec:cnv}
Particles with $n$-fold axisymmetry without fore-aft symmetry have {\em $C_{nv}$}-symmetry, $n>2$ [Figure~\ref{fig:shapes}({\bf e})]. Other examples invariant under $C_{nv}$ are dumbbells with spheres of different radii \cite{Candelier_Mehlig_2016,maches2024settling}, and straight fibres with continuous rotation symmetry but varying cross section \cite{Roy_2019}. For particles invariant under this symmetry, we find the following non-zero coefficients
\begin{subequations}\label{eq:C_4v}
\begin{equation}
    A_{pp} = A_{qq} = A_\perp, \quad A_{nn} = A_\parallel, \quad C_{pp} = C_{qq} = C_\perp, \quad C_{nn} = C_\parallel,\quad B_{pq} = -B_{qp} = B.
\end{equation}
as well as 
\begin{align}\label{eq:Mforeaftbreak}
    {N}_{ppn}^{(v\omega)}& = {N}_{qqn}^{(v\omega)} = N_1^{(v\omega)}, \quad {N}_{pnp}^{(v\omega)} = {N}_{qnq}^{(v\omega)} = N_2^{(v\omega)}, \quad
    {N}_{npp}^{(v\omega)} = {N}_{nqq}^{(v\omega)} = N_3^{(v\omega)}, \quad {N}_{nnn}^{(v\omega)} = N_4^{(v\omega)}\,,\\
\label{eq:Nforeaftbreak} 
    N_{pqn}^{(vv)} &= -N_{qpn}^{(vv)} = N_1^{(vv)}, \quad
    N_{pnq}^{(vv)} = -N_{pnq}^{(vv)} = N_2^{(vv)}, \quad
    N_{npq}^{(vv)} = -N_{nqp}^{(vv)} = N_3^{(vv)}\,.
\end{align}
\end{subequations}
The tensor $\ma N^{(\omega\omega)}$ has the same form as $\ma N^{(vv)}$. The steady $\rep$-contributions to the torque components are
\begin{subequations}
\label{eq:Tforeaftbroken}
\begin{align}
    T_p^{(1)} &=\tfrac{-1}{32\pi}vB[A_\perp(3-(\up^2+\uq^2)) - A_\parallel \un^2]v_q + (N_1^{(vv)} + N_2^{(vv)})v_qv_n \,,\\
    T_q^{(1)} &= \tfrac{1}{32\pi}vB[A_\perp(3-(\up^2+\uq^2)) - A_\parallel \un^2]v_p- (N_1^{(vv)} + N_2^{(vv)})v_pv_n \,,\\
    T_n^{(1)} &=0 \,.
\end{align}
\end{subequations}
Comparing with Eq.~\eqref{eq:spheroid}, we see that breaking of fore-aft symmetry introduces additional terms in the steady $\rep$-corrections to the torque, depending on the translation-rotation coupling $B$ in the Stokes limit. The 
form of the quadratic terms in the steady torque does not change, as was mentioned for the fibres. The steady fluid-inertia torque (\ref{eq:Tforeaftbroken}) can be written as~\cite{Jiang_2021}
\begin{equation}\label{eq:tsph}
    \ve T^{(1)} = v^2 g({\un})({\hat{\ve v}}\wedge \hat{\ve n})\,,
\end{equation}
with $g(\un) = -B[2A_\perp + ( A_\perp - {A_\parallel}) \un^2]/({32\pi}) + (N_1^{(vv)}+N_2^{(vv)})\un$. The general form of Eq.~(\ref{eq:tsph}) is consistent with Eq.~(2.11) in \citet{Roy_2019} who considered particles without fore-aft symmetry made out of non-slender rods with changing cross section or inhomogeneous mass distribution. However, Eq.~(2.11) in Ref.~\cite{Roy_2019} implies that $g$ is proportional to $\hat v_n$. This is explained by the fact that Eq.~(2.11) was derived in the slender-body approximation including terms up to $1/(\log\kappa)^2$, whereas the first two terms in the above expression for $g$ are at least of order $1/(\log\kappa)^3$ (see  Section~\ref{sec:discussion}). 
\citet{Candelier_Mehlig_2016} computed the torque on  an asymmetric dumbbell settling in a quiescent fluid. In this case, $g$ is a linear function of $\hat{v}_n$. We do not know why the coefficient of the quadratic term vanishes. It may be a consequence of the fact that the authors considered the special case of a slender and weakly asymmetric dumbbell,
constraining the coefficients
in Eq.~(\ref{eq:Tforeaftbroken}) further. 

\subsection{Chiral octahedral symmetry $O$}
Lord Kelvin's isotropic helicoid \cite{kelvin1871hydrokinetic} is a particle which has {\em chiral octahedral symmetry} $O$. That is, it has the same rotational symmetries as $O_h$ but no reflection symmetries [Figure~\ref{fig:shapes}({\bf f})]. The rotation symmetries constrains $\ma A, \ma B$ and $\ma C$ to be isotropic: $\ma A = A\ma I,\;\ma B = B\ma I, \ma C = C\ma I$. It thus has the same structure as a cube/sphere but with an isotropic translation-rotation coupling \cite{kelvin1871hydrokinetic,collins2021lord}.
 Here we conclude in addition that all $\ma N$-tensors are proportional to $\ve \epsilon$. For the sphere we found 
 $\ma N^{(v\omega)}=0$, but here 
 $\ma N^{(v\omega)}=\ve\epsilon$ because there are no reflection symmetries, so $\det[\ma O]\equiv 1$. As a consequence, the steady translation-rotation coupling remains isotropic to order $\rep$,
\begin{equation}\label{eq:chiral}
\ve T^{(1)}= -\frac{3v}{32\pi} B\, (\ma I - \tfrac{1}{3} \ve{\hat{v}} \otimes \ve{\hat{v}})\cdot(A \ve v +B \ve \omega)
+ N^{(v\omega)} \,\ve v \wedge \ve \omega\,.
\end{equation}
Note that $\ve f^{(0)}$ is constant for an isotropic helicoid that
translates with constant velocity $\ve v$ and rotates with constant angular velocity $\ve \omega$, regardless of how this vectors align. Therefore we quoted the steady result for non-zero frequency above. The last term on the right-hand side
is the analogue of the Rubinow-Keller lift force on a translating and rotating sphere.

\begin{table*}[p]
\centering
\caption{\label{tab:nearlyspherical}Summary of geometric constraints and contributions to the force and torque for different particle symmetries. The form of
the tensor $\ma C$ is the same as $\ma A$ (with independent elements). The forms
of $\ma M^{(vv)}$ and $\ma M^{(\omega \omega)}$
 are the same as the form 
of $\ma N^{(v \omega)}$, and the forms of $\ma M^{(v \omega)}$ and $\ma N^{(\omega \omega)}$ are
the same as that of 
$\ma N^{(vv)}$. 
}
\renewcommand{\arraystretch}{1.25}
\begin{tabularx}{\textwidth}{ l  >{\raggedright\arraybackslash}X  >{\centering\arraybackslash}p{7cm} }
\hline\hline &&\\[-1mm]
{Symmetry} &  {Constraints on} $\ma A,\ma B,\ma C, \ma N$ 
&{Examples (nearly spherical particles)}  \\[3mm]
\hline
$\begin{array}{l}
\mbox{Full octahedral 
 ($\mathrm{O}_h$) }
 \end{array}$ 
 &
$\begin{array}{l l}
 &  \ma A = A\ma I\,,\\
& \ma B = \ma 0\,, \\
&  \ma N^{(v\omega)} = \ma 0\,,\\
&\ma N^{(v v)} = N^{(vv)} 
\ve \epsilon \quad (\ve \epsilon \equiv \mbox{Levi-Civita tensor}).\\ 
\end{array}$ 
&
\begin{minipage}{\linewidth}\centering
\includegraphics[width=0.4\linewidth]{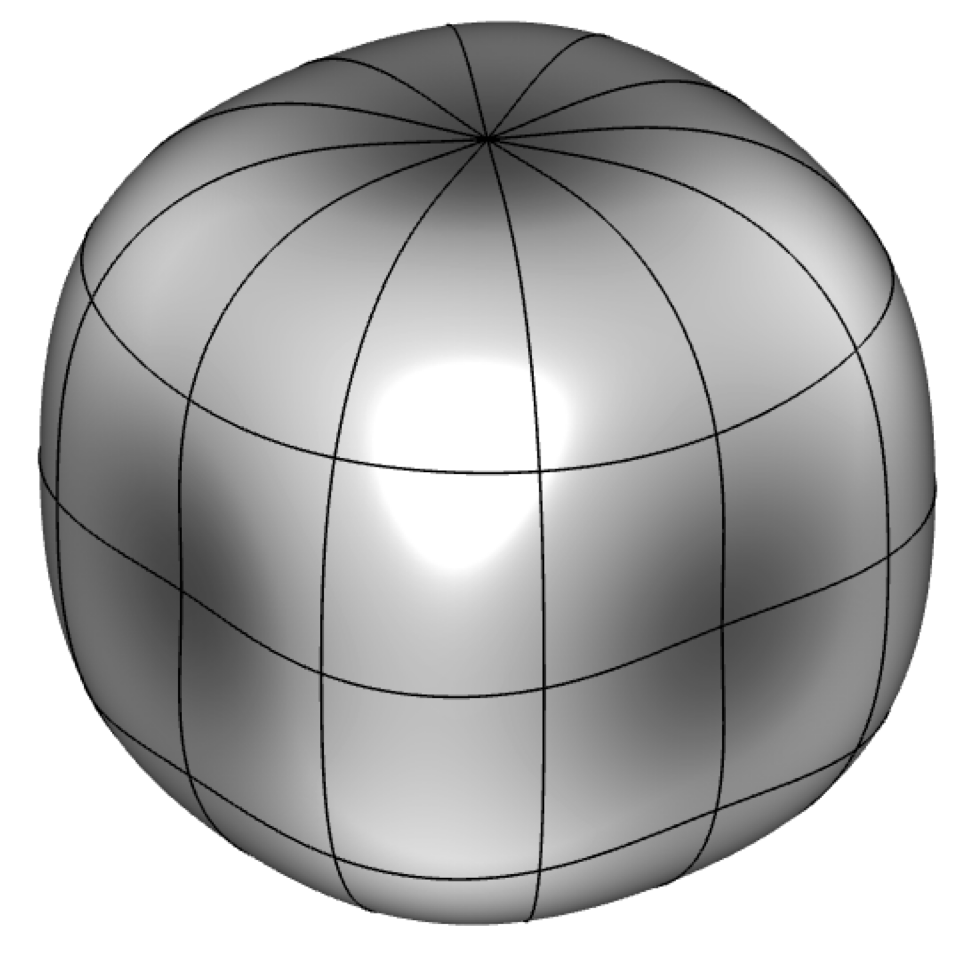}\\[4pt]
  $r = 1 - \tfrac{1}{2}\varepsilon(\sin^4\vartheta\cos^4\varphi + \sin^4\vartheta\sin^4\varphi + \cos^4\vartheta)$\\
  ($\varepsilon=0.5$)
\end{minipage}
\vspace{11pt}
 \\
\hline
\hline
$\begin{array}{l}
\mbox{Prismatic} \,\,
 (\mathrm{D}_{2h})\end{array}$ &
$\begin{array}{l l}
 &  \ma A = {\rm diag}[A_{pp}, A_{qq}, A_{nn}]\,,\\
& \ma B = \ma 0\,,\\
 &  \ma N^{(v\omega)} =\ma 0 \,,\\
& \ma N^{(vv)} = N_{pqn}^{(vv)}\,\ve p \otimes \ve q \otimes \ve n + N_{pnq}^{(vv)}\,\ve p \otimes \ve n \otimes \ve q \\
& \qquad \quad \!+  N_{qpn}^{(vv)}\,\ve q \otimes \ve p \otimes \ve n + N_{qnp}^{(vv)}\,\ve q \otimes \ve n \otimes \ve p \\
& \qquad \quad \! + N_{npq}^{(vv)}\,\ve n \otimes \ve p \otimes \ve q + N_{nqp}^{(vv)}\,\ve n \otimes \ve q \otimes \ve p\,.
\end{array}$ 
&
\begin{minipage}{\linewidth}\centering
\includegraphics[width=0.4\linewidth]{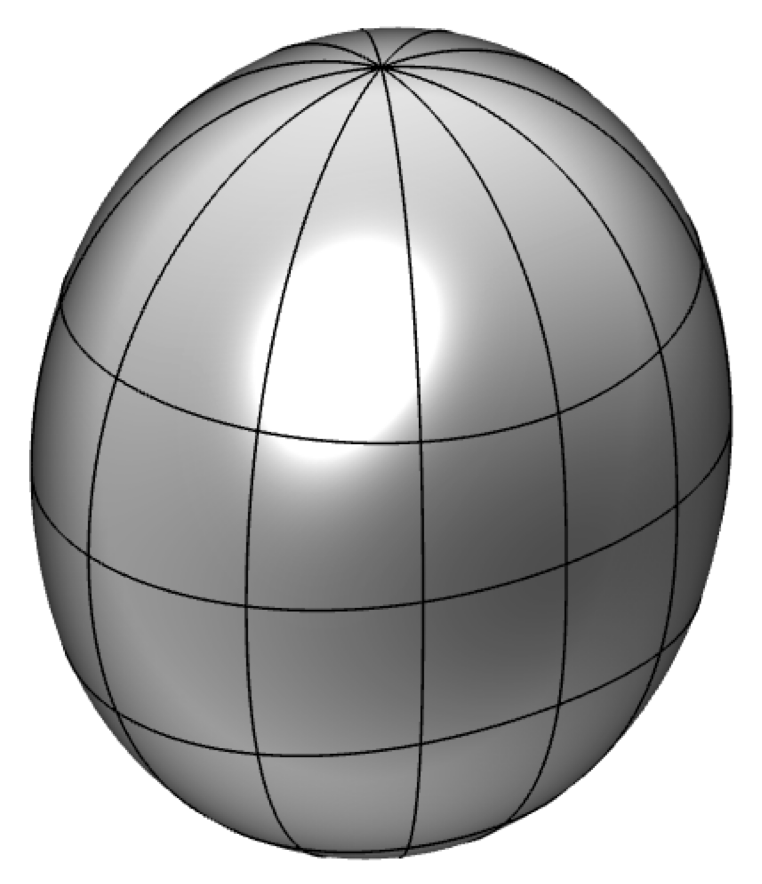}\\[4pt]
  $r=1-\varepsilon \sin^2\vartheta (\alpha - \alpha \sin^2\varphi+\sin^2\varphi)$\\($\varepsilon=0.3\,,\alpha=0.5$)
\end{minipage}
 \\
\hline
\hline
$\begin{array}{l}
\mbox{Pyramidal}\,\,
 (\mathrm{C}_{\infty v})
\end{array}$   &
$\begin{array}{l l}
&  \ma A = {\rm diag}[A_{\bot}, A_{\bot}, A_{\parallel}]\,,\\&\ma B =  B\, (\ve p \otimes \ve q - \ve q \otimes \ve p)\,,\\
 &  \ma N^{(v\omega)} = \:N_1^{(v\omega)} (\ve p \otimes \ve p \otimes \ve n + \ve q \otimes \ve q \otimes \ve n ) ,\\
& \qquad \quad  \!+N_2^{(v\omega)} (\ve p \otimes \ve n \otimes \ve p + \ve q \otimes \ve n \otimes \ve q )\\
& \qquad \quad \!+ N_3^{(v\omega)} (\ve n \otimes \ve p \otimes \ve p + \ve n \otimes \ve q \otimes \ve q )\\
& \qquad \quad \!+  N_4^{(v\omega)} (\ve n \otimes \ve n \otimes \ve n)\,,\\
& \\
&  \ma N^{(vv)} = \:N_1^{(vv)} (\ve p \otimes \ve q \otimes \ve n - \ve q \otimes \ve p \otimes \ve n ) \\
& \qquad \quad \!+ N_2^{(vv)} (\ve p \otimes \ve n \otimes \ve q - \ve q \otimes \ve n \otimes \ve p )\\
& \qquad \quad \!+ N_3^{(vv)} (\ve n \otimes \ve p \otimes \ve q - \ve n \otimes \ve q \otimes \ve p )\,.\\
&
\end{array}$ 
&
\begin{minipage}{\linewidth}\centering
  \includegraphics[width=0.4\linewidth]{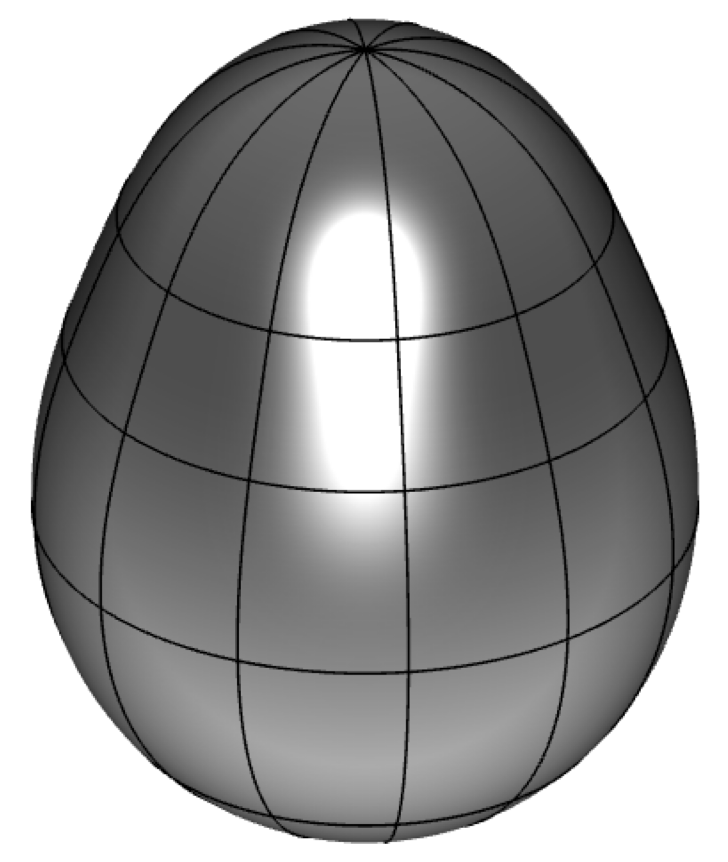}\\[4pt]
  $r=1-\varepsilon (\sin^2\vartheta+ \tfrac{\gamma}{2} \cos\vartheta)$ \\
  ($\varepsilon=0.3\,,\gamma =1.5$)
  \end{minipage}
 \\
\hline
\hline
$\begin{array}{l}
\mbox{Orthorhombic}\\
\mbox{pyramidal}\,\,(\mathrm{C}_{2v})
\end{array}$  &
$\begin{array}{l l}
 &  \ma A = {\rm diag}[A_{pp}, A_{qq}, A_{nn}]\,,\\&\ma B =  B_{pq}\ve p \otimes \ve q + B_{qp} \ve q \otimes \ve p\,,\\
 &  \ma N^{(v\omega)} = \:N_{ppn}^{(v\omega)} \ve p \otimes \ve p \otimes \ve n + N_{pnp}^{(v\omega)} \ve p \otimes \ve n \otimes \ve p  \\
& \qquad \quad\!+ \:N_{qqn}^{(v\omega)} \ve q \otimes \ve q \otimes \ve n + N_{qnq}^{(v\omega)} \ve q \otimes \ve n \otimes \ve q  \\
& \qquad \quad\!+ \:N_{npp}^{(v\omega)} \ve n \otimes \ve p \otimes \ve p + N_{nqq}^{(v\omega)} \ve n \otimes \ve q \otimes \ve q \,,\\
&\hspace{3.5cm} +N_{nnn}^{(v\omega)} \ve n \otimes \ve n \otimes \ve n \,,\\
&\\
&  \ma N^{(vv)} = \:N_{pqn}^{(vv)} \ve p \otimes \ve q \otimes \ve n + N_{pnq}^{(vv)} \ve p \otimes \ve n \otimes \ve q  \\
& \qquad \quad \!+N_{qpn}^{(vv)} \ve q \otimes \ve p\otimes \ve n+ N_{qnp}^{(vv)} \ve q \otimes \ve n \otimes \ve p  \\
& \qquad \quad\!+ \:N_{npq}^{(vv)} \ve n \otimes \ve p \otimes \ve q + N_{nqp}^{(vv)} \ve n \otimes \ve q \otimes \ve p\,.\\
&
\end{array}$ 
&
\begin{minipage}{\linewidth}\centering
  \includegraphics[width=0.7\linewidth]{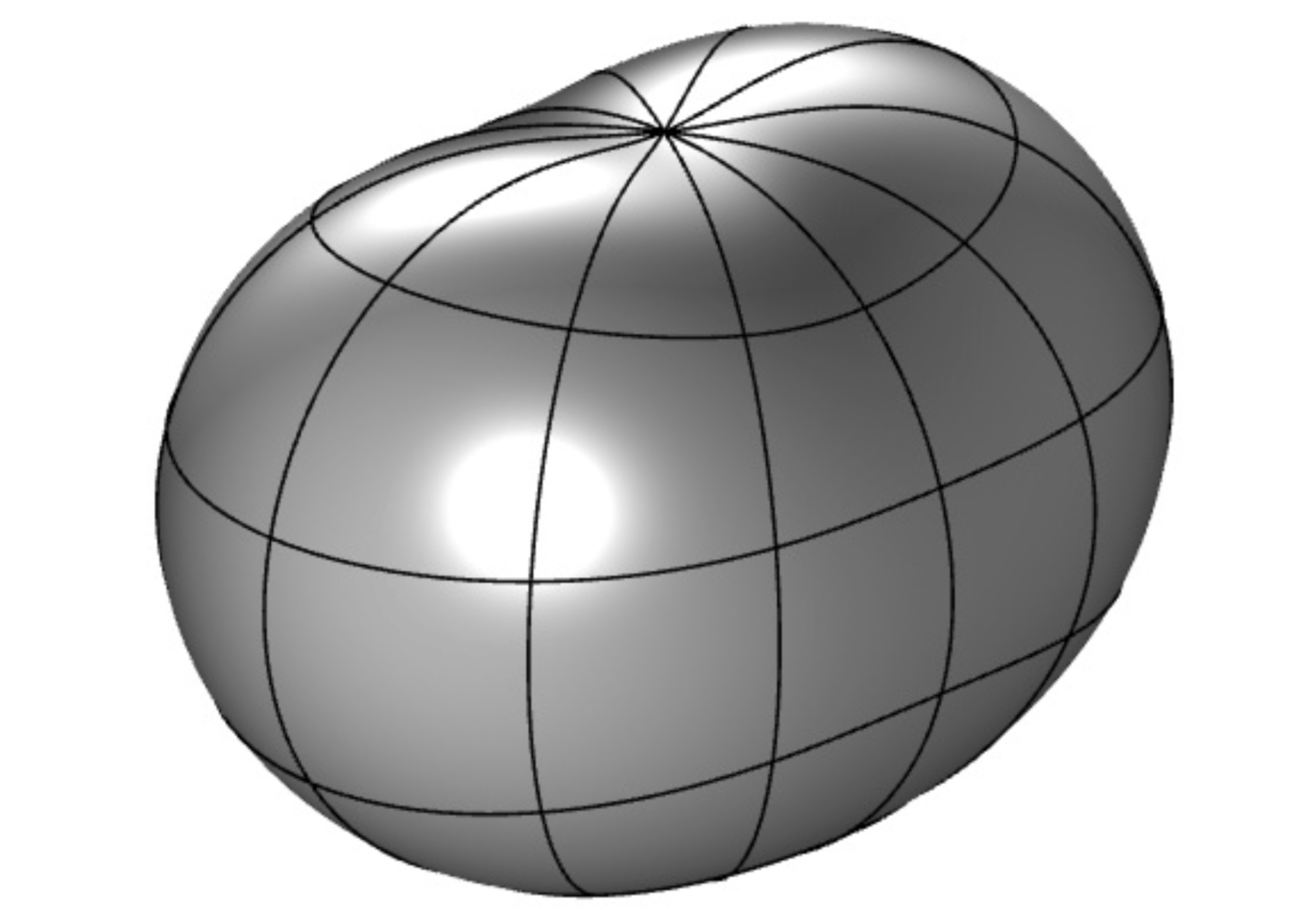}\\[4pt]
 $r=1-\varepsilon (1+\tfrac{\beta}{2} \cos \vartheta   - \cos^2 \varphi \sin^2\vartheta)$
 \\($\varepsilon=0.3\,,\beta=1.5$)
 \end{minipage}
 \\
\hline
\end{tabularx}
\end{table*}

\section{Nearly spherical particles}
\label{sec:ns}
Table~\ref{tab:nearlyspherical} summarises the structure of the coupling tensors derived in the previous Section for the symmetries $O_{\rm h}$, $D_{\rm 2h}$, $C_{\infty \rm v}$, and $C_{\rm 2v}$. 
 We tested these results for nearly spherical particles with the shape symmetries listed in Figure~\ref{fig:shapes}.
To this end, we parameterised the particle boundary
by the radius $r(\varphi,\vartheta)$. Here $\varphi$ and $\vartheta$ are azimuthal and polar angles, that is
 $x=r(\varphi,\vartheta) \sin\vartheta\cos\varphi$, $y=r(\varphi,\vartheta) \sin\vartheta\sin\varphi$, $z=r(\varphi,\vartheta)\cos\vartheta$.
 Table~\ref{tab:nearlyspherical} lists
parameterisations for four particle shapes, 
namely $O_{\rm h}$, $D_{\rm 2h}$, $C_{\infty \rm v}$, and $C_{\rm 2v}$. 
Using the theory described in Section \ref{sec:background}, we computed hydrodynamic force and torque as a joint perturbation expansion
in $\rep$ and $\varepsilon$ \cite{candelier2015role_a,collis2024unsteady}. 
The former is a singular expansion, the latter a regular one.
This allows to extract the elements of the tensors discussed in the previous Section, or at least of certain sums of their elements.
Our results for the particles in Table~\ref{tab:nearlyspherical} confirm the symmetry analysis in Section~\ref{sec:symmetries} (Appendix~\ref{app:NS}). We remark, however, that some coefficients vanish although they are allowed to be non-zero by 
particle-shape symmetries (Section~\ref{sec:discussion}). 

For example, for the special case $\alpha=1$ in the second row of Table~\ref{tab:nearlyspherical}, one obtains a parameterisation of 
the  hydrodynamic torque on a nearly spherical spheroid.
In this case, the torque component $T_n$ vanishes, confirming the results of the symmetry analysis, Eq.~(\ref{eq:Nspheroid}). The case of a nearly spherical spheroid was also considered by Cox \cite{cox1965steady}. He parameterised
the major axis as $1+2\epsilon$ and the minor axis as
$1-\epsilon$, in terms of a small  parameter $\epsilon$. Comparing with our parameterisation ($c=1$ and $a=b=1-\varepsilon$), we see that
 $\varepsilon = 3\epsilon$. Accounting for this factor of  three, as well as a factor of $6\pi$ in
the non-dimensionalisation of the torque in Ref~\cite{cox1965steady}, one concludes that our results agree with those of \citet{cox1965steady}.

\section{Discussion}
\label{sec:discussion}
The results summarised in Section~\ref{sec:symmetries} show how particle-shape symmetries constrain the hydrodynamic torque, including fluid-inertia contributions to order $\rep$, extending earlier work by Brenner and Cox on corresponding constraints on the hydrodynamic force~\cite{brenner1963resistance}. They determined under which circumstances the contributions from the $\ma M$ tensor to the hydrodynamic force vanish, by analysing how 
reflections as well as $\tfrac{\pi}{2}$- and $\pi$-rotations constrain $\ma M^{(vv)}:\ve v \otimes \ve v$.
In a similar fashion, \citet{cox1965steady} considered
under which conditions
the components 
of the vectors $\ma N^{(vv)} :\ve v\otimes \ve v$,
$\ma N^{(v\omega)} :\ve v\otimes \ve \omega $ and $\ma N^{(\omega\omega)} :\ve \omega\otimes \ve \omega$ to vanish, because in that case steady force and torque are solely determined by the Stokes resistance tensors $\ma A$ (drag)
and $\ma B$ (translation-rotation coupling).

For general shapes, these contributions do not vanish. In this case, the question is: how many parameters are needed to parameterise the dependence of the torque (and the force) upon particle velocity and angular velocity? Here we used symmetry analysis
to answer this question. We determined how particle-shape symmetry
constrains the non-zero elements of the coupling tensors. This determines the form of the torque, and in particular how many parameters are needed in parameterisations. 
  The general structure of the torque becomes more complex (involves more parameters to be determined) 
as more shape symmetries are broken. In particular, any given shape symmetry implies that the resistance tensors  $\ma A$ and $\ma C$ commute with the generator of the corresponding symmetry transformation.
Full octahedral symmetry $O_{\rm h}$ requires that $\ma A, \ma B$, and $\ma C$ commute with all rotational generators. This implies that
these tensors are proportional to the unit matrix (isotropy). Now consider how rotations constrain the rank-3 tensors: the symmetries in the point-group $O_{\rm h}$ 
require that
  all rank-3 tensors are proportional to the Levi-Civita tensor $\ve \epsilon$, the only antisymmetric rank-3 tensor. 
  Reflection symmetries constrain the rank-3
  tensors further, namely $\ma N^{(v\omega)}$ 
  as well as $\ma M^{(vv)}$ and $\ma M^{(\omega\omega)}$ must vanish. Reflection symmetries also cause $\ma B$ to vanish.
  The last two statements apply to all particles in the top row of~Fig.~\ref{fig:shapes}. 

For nearly spherical particles with different shape symmetries (Section~\ref{sec:ns}), we found some cases where the symmetry analysis
from Section~\ref{sec:symmetries} allows a non-zero coupling, but the coupling nevertheless vanishes to a certain order in perturbation theory. For $C_{\infty v}$-symmetry for example,
the translation-rotation coupling $\ma B$ vanishes to order $\varepsilon^2$, although allowed by symmetry. In general, one expects that all coupling allowed by symmetry are non-zero unless there are additional symmetries not considered. The elements of $\ma B$ vanish to order $\varepsilon^2$ 
because the parameterisation $r=1-\varepsilon(\sin^2\vartheta + \tfrac{\gamma}{2}\cos\vartheta)$ corresponds to a shifted spheroid. Expressing the torque w.r.t. the centre of mass cancels any potential contributions to the torque up to order $\varepsilon^2$. 
This happens also for $C_{2v}$-symmetry. In the latter case, we also find that $N_{n(pq)}^{(vv)}$ and $N_{n(pq)}^{(\omega\omega)} $ evaluate to zero to order $\varepsilon$. We expect that higher orders in $\varepsilon$ are non-zero. 

\citet{candelier2024torques} used the slender-body approximation to calculate the steady hydrodynamic force and torque on curved atmospheric fibres with $C_{2v}$ symmetry, settling in quiescent air.
In slender-body theory \cite{Khayat_Cox_1989},  force and torque are expressed as
\begin{align}
\ve F &= \int\!\! {\rm d}s\, \ve f(s)\,,\quad
\ve T = \int \!\!{\rm d}s \,\ve r(s) \wedge \ve f(s)\,,
\label{eq:t}
\end{align}
replacing Eqs.~(\ref{eq:torque}).
Here $\ve f(s)$ is 
the force on a small segment of a slender body. 
It is obtained by asymptotic matching, as an expansion in $1/\log \kappa$ where $\kappa$ is the (large) aspect ratio of the fibre. 
The results of \citet{Khayat_Cox_1989}
are obtained including the order $1/(\log \kappa)^2$ and assuming that the particle Reynolds number w.r.t the (small) radius $a$ of the fibre is small (the particle Reynolds number w.r.t. to the fibre length can be of order unity). 
It has been shown that the predictions of \citet{Khayat_Cox_1989} are accurate for aspect ratios greater than 30 and for small Reynolds numbers based on the fibre radius, \textit{i.e.} $v_0 a/\nu\leq 0.1$.
To order $\rep=v_0 a/\nu$, 
the resulting hydrodynamic force and torque contain 
at most third powers of $\hat{\ve v}$, just as Eqs.~(\ref{eq:sing}) and (\ref{eq:FTreg0}).
Let us compare the slender-body results for the torque on curved fibres with $C_{2v}$ symmetry [Eq.~(3) in \citet{candelier2024torques}] with the present symmetry analysis,
Eq.~(\ref{eq:C_2v}). The two expressions are
consistent. The differences arise from a difference in the orientation of the basis. 
Here, the reflection symmetry $\ve n \to -\ve n$
is broken, while in Ref.~\cite{candelier2024torques}, the symmetry $\ve p \to -\ve p $ is broken. 
However, 
 the slender-body approximation has fewer terms than allowed by the symmetry analysis, 
 similar to the nearly-spherical particles discussed above, 
 although different terms are zero in that case.
 The allowed terms that vanish in the slender-body approximation are  proportional to $B_{qn} A_{nn}$, while allowed terms
 proportional to $B_{nq} A_{qq}$ appear. This can be understood as follows. Computing the Stokes resistance tensors for planar fibres with constant radius of curvature using $\ve f(s)$
 from Ref.~\cite{Khayat_Cox_1989}, one concludes that $A_{qq}$ and $A_{nn}$
 are of order $1/\log\kappa$, and so is $B_{nq}$ \cite{cox1970motion}. The coefficient $B_{qn}$, by contrast, is proportional to $ 1/(\log \kappa)^2$. This means that the corresponding contribution to the torque is of order $1/(\log \kappa)^3$, and thus not contained in the slender-body theory of \citet{Khayat_Cox_1989}, to the order quoted there. 
 The above dependencies on $\kappa$
 are confirmed by
  numerical results for $B_{nq}/B_{qn}$ for a bead model
  obtained by modeling the fibres as chains of hydrodynamically interacting, identical  beads
 (Table~S1 in supplemental material of Ref.~\cite{candelier2024torques}): the numerical results are  consistent with $B_{nq}/B_{qn}\sim \log\kappa$. The accuracy is  approximately 2\% when comparing $\kappa=20$ with $\kappa=40$.
  
 This shortcoming of slender-body theory stems from approximations for the inner solution, namely that hydrodynamic interactions
 between different segments of the fibre are neglected (one has to go to higher orders in 
 $1/(\log \kappa)$ to account for these interactions \cite{keller1976slender}). It is instructive to compare these conclusions with results 
 for the bead model
 \cite{candelier2024torques}. In that case, $\ma B=0$ without hydrodynamic interactions in the centre-of-mass of the chain. This is no longer true if the beads have different sizes,  or in the
 slender-body approximation where the fibre segments have different orientations.
 
 In slender-body theory, higher orders in $\rep=av_0/\nu$ result in higher powers of $\hat {\ve v}$ in the expressions for hydrodynamic force and torque.  To obtain reasonable agreement between theory and experiment, \citet{candelier2024torques} found it necessary to include higher orders in $\rep$, not contained in Eq.~(\ref{eq:C_2v}) in Section~\ref{sec:symmetries}
 (although for spheroids, \citet{bhowmick2024inertia} found that the linear approximation in $\rep$ works well if the prefactor is adjusted 
 by an $\rep$-dependence).
It is likely that
 higher-order singular contributions to force and torque are entirely determined by the Stokes tensors $\ma A$, $\ma B$, and $\ma C$. If this is the case, any other symmetry-breaking effects at larger particle Reynolds numbers must come from the inner solution. 

 At small but finite $\rep$, fluid inertia torques tend to align highly symmetric particles with the slip velocity. This is important in the atmospheric sciences, where the alignment of symmetric ice crystals ($D_{nh}$ symmetry with $n>2$) settling in the quiescent or weakly turbulent atmosphere~\cite{Heymsfield_1973,Sassen_1980,Klett_1995,Noel_Sassen_2005,Gustavsson_2021,bhowmick2024inertia}. Calculations of torque and force for nearly spherical spheroids and slender fibres with circular cross sections \cite{Khayat_Cox_1989} show that such particles -- with $D_{nh}$ symmetry with $n>2$ -- align so that they settle in quiescent air with their broad sides down. The fixed points of angular dynamics are dictated by rotational and fore-aft symmetry. Their stability
is determined by the sign of the coefficient $N_1^{(vv)}+ N_2^{(vv)}$ in Eq.~(\ref{eq:spheroid}) which depends on the aspect ratio of the spheroid. The coefficient changes sign at aspect ratio $\lambda=1$ (sphere).  \citet{Dabade_2015} calculated how  $N_1^{(vv)}+ N_2^{(vv)}$ depends on the particle aspect ratio, although their result differs from that of \citet{cox1965steady} in the nearly spherical limit.

\citet{Andersen2005-2} used an empirical torque model to describe fluttering and tumbling of platelets settling in a quiescent fluid at $\rep$ of order $10^3$. For planar motion,
they parameterise the out-of-plane component
of the torque as $T_3 = c_1\omega + c_2\omega^2$
[their Eq.~(3.9)]. The first term is 
consistent with Stokes torque (\ref{eq:resistance_tensors}). However, our symmetry analysis (Section \ref{sec:symmetries}) 
does not produce the suggested $\omega^2$ correction. Instead we find   $(N_1^{(\omega\omega)}+N_2^{(\omega\omega)}) \omega^2 {\rm sin}2\theta$ for a spheroid to order $\rep$. In the spherical limit, the quadratic correction to the torque vanishes to this order in $\rep$, as observed by \citet{rubinow1961transverse}.  
Analysis of ab-initio data for force and torque on a thin disk (aspect ratio $\lambda = 0.2$) in 
unsteady motion in a quiescent fluid at $\rep\sim 2$ shows that there is a small $\omega^2$ contribution to the torque  consistent with $(N_1^{(\omega\omega)}+N_2^{(\omega\omega)})\omega^2 {\rm sin}2\theta$. 
Any possible contribution of the form $c_2 \omega^2$ is small. This is consistent with the direct numerical simulations of \citet{pierson2021hydrodynamic}  who considered the torque on a cylinder rotating perpendicular to its symmetry axis (with $\theta=0$). They observe an $\omega^2$-contribution, but only at large Reynolds numbers, not attainable from a
first-order expansion in $\rep$.

For particles with prismatic symmetry (Fig.~\ref{fig:shapes}({\bf a})), the symmetry analysis predicts that
all three torque components are non-zero, in general. For specific orientations
the torque may vanish. For example, if 
$\ve \omega=0$, and $\hat {\ve v} = -\hat {\bf e}_3$,  we have $v_p=\sin\theta\cos\psi,  v_q=-\sin\theta\sin\psi$, and $v_n=-\cos\theta$. We see that there are six orientations where the torque vanishes, each corresponding to one side of the particle in Figure~\ref{fig:shapes}{({\bf a})} pointing down ($\theta=0,\pm\tfrac{\pi}{2}$ and $\psi =0,\pm\tfrac{\pi}{2}$).
 These are 
 potential steady states of the angular dynamics at small $\rep$.

The laboratory experiments of \citet{bhowmick2024twist} show that ellipsoidal particles settling in quiescent air approach a steady orientation with their broadest side down. 
Eq.~(\ref{eq:ellipsoid}) explain this observation
for nearly spherical ellipsoids, as the linear-stability analysis described in Appendix~\ref{app:stability} shows. 
More generally, the symmetry analysis of Section~\ref{sec:symmetries} explains that
particles with broken reflection symmetries ($C_{\infty v}$ and $C_{2v}$, for example) 
align differently, compared 
more symmetric particles discussed in the previous paragraph, such as spheroids and ellipsoids. The reason is  that both Stokes and fluid-inertia torque are non-zero and can balance for certain orientations, provided that $\rep$ is not too large~\cite{Candelier_Mehlig_2016,maches2024settling,Roy_2019,ravichandran2023orientation}. 
Particles that 
align at a non-zero tilt angle as they settle in quiescent air  move horizontally. This can affect the transport of such particles in the atmosphere, but standard models for the transport of atmospheric particles assume symmetrical shapes (spheroids)~\cite{tatsii2024shape}.

The previous paragraph discussed the steady-state problem, the orientation of a steadily settling particle. Now consider how the particle orientation changes as the steady state is approached. As explained in Section~\ref{sec:background}, the transient is
affected by history forces and torques arising from inertial fluid acceleration (the
term $\partial_t \ve w$ in the unsteady Stokes equation. 
In this limit,  the history force is well understood for a spherical particle settling in a quiescent fluid. The history force arises from the outer solution and decays slowly, as $t^{-1/2}$, after a sudden change of the particle dynamics~\cite{landau1987hydrodynamics}. Fluid-inertia corrections at small $\rep$ modify the power-law decay in a non-universal fashion \cite{lovalenti1993hydrodynamic}, but the decay remains slow at small Reynolds numbers. The history torque, by contrast, decays rapidly for a sphere. Its kernel is essentially a Dirac delta function \cite{candelier2023second}. Since the torque is determined by the inner solution, it is expected that small-$\rep$  corrections do not change this fact. Eqs.~(\ref{eq:U}) and (\ref{eq:sing}) show that the situation is quite
different for particles
with non-zero translation–rotation coupling $\ma B$, because $\ma B$ results
in a slow decay of the history torque as well. This history torque is expected to affect the transient dynamics of asymmetric particles settling at non-zero Reynolds numbers \cite{candelier2024torques}. 
 For spheroids ($\ma B=0$) the relaxation of the orientational dynamics is well described \cite{bhowmick2024inertia}, but as yet there is no reliable theory for the relaxation of particles with non-zero translation-rotation coupling
 outside the Stokes limit \cite{candelier2024torques}. The main complication are the above-mentioned slowly decaying history torques.

 Note that we computed the torque w.r.t. the centre-of-mass of the particle, and assumed a homogeneous mass-density distribution. An inhomogeneous distribution of the particle-mass density can shift the centre-of-mass, breaking further shape symmetries. Adjusting its centre-of-mass allows gliders settling in a quiescent fluid to navigate to land on a certain spot~\cite{jiang2025smart}. 

Our symmetry analysis is based on the theory for hydrodynamic forces and torques 
described in Section~\ref{sec:background}, which
is valid at non-zero  but small
$\rep$. How small the Reynolds number must be for the theory to work qualitatively or quantitatively depends on particle shape.
For a spherical particle, the left of Eqs.~(\ref{eq:Tsing_steady})
simplifies to the Oseen force (Appendix~{\ref{app:NS}}),
and it is well known that this works up
to $\rep \sim 0.5$ \citep{chester1969flow}.
The Rubinow-Keller lift force for a spherical particle [Eq.~(\ref{eq:RK})] is in good qualitative
agreement with experiments up to $\rep \sim 5$ \cite{oesterle1998experiments}. For fibres, \citet{fintzi2023inertial} demonstrated that it is necessary to include the full functional form of the singular contribution at moderate Reynolds numbers ($\rep \geq 0.5$), especially for large aspect ratios, even though this range of Reynolds numbers probably lies beyond the validity of the present asymptotic analysis. 

\citet{bhowmick2024inertia}
 used empirical parameterisations of the fluid-inertia torque for spheroidal particles based on Eq.~(\ref{eq:spheroid}), with
$\rep$-dependent coefficients. This works well up to 
$\rep\sim 20$, although the functional dependence of the coefficients has to be determined by comparison with particle-resolving simulations. In a similar way, the simulation results of \citet{Jiang_2021}
and the experiments of \citet{Cabrera_2022} indicate that the functional form (\ref{eq:spheroid}) is correct for fibres and spheroids, see also Refs.~\cite{ouchene2016new,sanjeevi2017orientational,fintzi2023inertial}. As mentioned above, slender-body theory \cite{Khayat_Cox_1989} allows to include higher-order corrections in $\rep$, albeit only to order $1/(\log\kappa)^2$. The ab-initio simulations of 
\citet{fintzi2023inertial} (see also Ref.~\citet{kharrouba2021flow}) indicate that these refined predictions fail quantitatively for $\kappa =30$ and $\rep \geq 0.5$ (with Reynolds number based on the radius of the fibre), because  the large-aspect-ratio limit required by the theory of \citet{Khayat_Cox_1989} was most likely not attained in the simulations. \citet{candelier2024torques}, on the other hand, found qualitative agreement between the full slender-body theory for the torque on curved atmospheric fibres and experiments. 
\citet{Roy_2019} compared theory and experiment
for the settling of straight slender fibres with broken fore-aft symmetry. They found good agreement between measurements and slender-body theory (including high-order $\rep$ corrections). These results indicate that the results presented here help to constrain fluid-inertia torques on particles of general shape. 
In the past, some authors have attempted to constrain the functional form of the fluid-inertia force by comparing with the potential-flow limit. 
The results of \citet{candelier2023second} indicate, however, that this approach may be fallacious, because the expressions for the hydrodynamic force on a sphere in a linear flow are qualitatively different in the two limits.
Whether it may help to constrain fluid-inertia torques remains an open question. In summary, 
only direct numerical simulations and dedicated experiments can serve as the final benchmark to test the range of validity of the theory present here. 

We determined how shape symmetry constrains the hydrodynamical torque
on an isotropic helicoid, an isotropic particle with non-zero translation-rotation coupling~\cite{kelvin1871hydrokinetic}. Our analysis shows that the coupling remains isotropic to order $\rep$, just as in the potential-flow limit analysed by Lord Kelvin. 
We also saw that symmetry allows for a non-zero contribution $\ma N^{(v\omega)} : \ve v \otimes \ve \omega\propto \ve v \wedge \ve \omega$ to the fluid-inertia torque,
analogous to the Rubinow-Keller lift force on
a rotating sphere translating in a fluid at
rest. 
In the Stokes limit, helical ribbons exhibit an infinite sequence of bifurcations as the translation-rotation coupling tensor passes through axisymmetry \cite{huseby2024helical}. 
The theory described here can be used to determine fluid-inertia torques (and forces) on helical ribbons. This allows to determine how fluid-inertia corrections change the phase portraits of the angular dynamics. 

The results described up to this point ignore the unsteady term of the disturbance equation in the inner solution, as one must in the stated limit
$1 \gg \rep \gg \rep{\rm Sl}$. To conclude the discussion,
we briefly consider the opposite limit
$1  \gg \rep{\rm Sl}\gg \rep$ (details in Appendix~\ref{app:iu}).
In this limit, there are 
additional contributions to
the regular parts of  hydrodynamic force and torque
\begin{subequations}
\label{eq:iu}
\begin{align}
\ve F^{(2)}_{\rm u}
&=\rep {\rm Sl}(\ma P \dot {\ve v} + \ma Q \dot {\ve \omega})\,,\\
\ve T^{(2)}_{\rm u}&=\rep {\rm Sl}(\ma Q^{\sf T}\dot  {\ve v} + \ma R\dot  {\ve \omega})\,.
    \end{align}
\end{subequations}
Here dots denote time derivatives, and 
$\ma P$, $\ma Q$, and $\ma R$ are second-rank tensors that transform as Eqs.~(\ref{eq:stokes}). Therefore they obey the same constraints as $\ma A$, $\ma B$, and $\ma C$. Eqs.~(\ref{eq:iu}) were first derived by \citet{gavze1990accelerated} who obtained a general solution for the unsteady Stokes equation for a particle of arbitrary shape. 
We note that the unsteady solution also contains history terms. In Appendix~\ref{app:iu} we show that 
these history terms agree with those derived in Section \ref{sec:background} to order $\sqrt{\rep {\rm Sl}}$.
\citet{collis2024unsteady} solved the unsteady Stokes equation for nearly spherical particles with several different shapes in the frequency domain. One must obtain the kernels derived in Appendix~\ref{app:iu} by integration over frequency.

\section{Conclusions}
\label{sec:conc}
In this paper we used symmetry analysis to constrain the hydrodynamic torque on particles moving in a quiescent fluid at small but non-zero particle Reynolds number. For highly symmetrical particles, such as slender fibres with circular cross section and spheroids, fluid inertia torques were well understood~\cite{brenner1961oseen,Khayat_Cox_1989,Dabade_2015}, and known to describe how highly symmetric particles align as they settle in quiescent or weakly turbulent fluids~\cite{bhowmick2024inertia}. 

The present study was motivated by recent interest in the angular dynamics of asymmetrical atmospheric particles~\cite{Candelier_Mehlig_2016,Roy_2019,ravichandran2023orientation,maches2024settling,candelier2024torques}, such as curved fibres, asymmetric ice crystals and ash particles. Here we show how breaking of shape symmetry allows new terms to appear in the hydrodynamic torque (and force), that can significantly change how particles align as they settle.
The analysis shows that the bent disks of \citet{miara2024dynamics}, and planar curved
atmospheric fibres with constant radius of curvature \cite{candelier2024torques} obey the same equations of motion, albeit with different parameters. The form of the equation
of motion determines the possible dynamical
behaviours by constraining the bifurcations that may occur as the parameters are varied. For example, in the Stokes limit the transition from alignment at small $\rep$ (curved fibre) to oscillations (bent disk)
occurs when the product $B_{pq} B_{qp}$ of the non-zero translation-rotation coupling coefficients in Eq.~(\ref{eq:bc2v}) changes sign \cite{candelier2024torques}. The results summarised here allow to determine how the bifurcations change for larger ribbons, as fluid-inertia corrections become important.

Regarding the hydrodynamic torque, our results generalise the theory of \cite{cox1965steady} to include unsteady effects, leading to slowly decaying history torques for asymmetrical particles with translation-rotation coupling (for symmetrical particles, history torques decay rapidly). Such unsteady effects influence the transient angular dynamics. In order to understand the transient angular dynamics for asymmetrical, curved atmospheric fibres \cite{bhowmick2024inertia}, we expect history torques to play a role. Similar conclusions are expected for bent disks \cite{miara2024dynamics}, which have the same point-group shape symmetry as planar fibres with constant radius of curvature. 

Even in the absence of history torques, one expects complex settling trajectories even for symmetrical particles settling in quiescent air.  For example, our theory shows that the transient dynamics
of ellipsoids is governed by three couplings (the six non-zero elements of the rank-3 tensor
$\ma N^{(vv)}$ which take different values depending on the aspect ratios of the ellipsoidal particle. This -- together with the fact that all three components of the hydrodynamical torque are non-zero -- explains the rich variety of settling behaviours observed in the experiment \cite{bhowmick2024twist}, but
a detailed computation of the transient dynamics remains a question for the future. Regarding the steady limit, our calculations show that nearly spherical ellipsoids settle with their broadest side down, as observed in the experiment \cite{bhowmick2024twist}. To confirm that this conclusion is valid for ellipsoidal particles with arbitrary aspect ratios, one needs to compute the three symmetrised coefficients in Eq.~(\ref{eq:ellipsoid}) for the torque by direct numerical simulation. 

The results discussed here were obtained in the limit of small Reynolds numbers, $\rep\ll 1$. An important question is how the particle dynamics changes as the Reynolds number increases. For spheroids in shear flow, bifurcations that occur at larger shear Reynolds numbers Re$_s$ have been mapped out, and it turns out that the small-Re theory remains qualitatively valid for Re$_s$ up to order unity. For spheroids settling in a quiescent fluid, one can obtain quantitative agreement with experiments using empirical torque models similar to those discussed here \cite{bhowmick2024inertia}. For particles with broken shape symmetry \cite{candelier2024torques}, the 
expressions derived here yield qualitative but not quantitative agreement with the experiments. How the accuracy of small-Re approximations breaks down for highly asymmetric particles as the Reynolds number increases remains an open question. It may be instructive to compare the results of the present analysis with the potential-flow limit, the limit considered by Lord Kelvin in his analysis of the isotropic helicoid~\cite{kelvin1871hydrokinetic}.
Corresponding expressions for hydrodynamic force and torque can be obtained for particles of arbitrary shape using either the Lagrangian formalism \citep{lamb1924hydrodynamics}, or by direct integration of the pressure on the body \citep{Howe_1995}. However, the results of \citet{candelier2023second} indicate that even for a spherical particle, hydrodynamic force and torque can be quite different in the two limits. As a consequence it may be challenging to find convincing interpolation formulae between the two limits. 

Last but not least, symmetry analysis as described here can be used to constrain hydrodynamical forces and torques on motile micro-organisms. The standard model assumes that the organisms are spherical, but this is an idealisation. In the future we will analyse how the interaction of microswimmers with the surrounding flow changes when shape symmetries are broken. To this end, one needs to generalise the present analysis to account for strain and vorticity of the undisturbed flow. For special cases this is known \cite{Bretherton:1962,brenner1963stokes,fries2017angular,ishimoto2020jeffery}, but not in general. 

 \acknowledgments{LS and BM were supported by VR grant no.  2021-4452.
 This work was completed while BM held a Gauss professorship in G\"ottingen, financed by the G\"ottingen Academy of Sciences and Humanity in Lower Saxony. KG was supported by VR grant no. 2023-03617. }

\vfill\eject\newpage
\appendix

\section{Fluid-inertia kernels}
\label{app:A}
We use asymptotic matching \cite{childress1964slow,saffman1965lift} to find a solution to the disturbance equation (\ref{eq:eom})
assuming that $1\gg \rep \gg \rep {\rm Sl}$. The calculation is very similar to
the one in Ref.~\cite{redaelli2023hydrodynamic}, except for two differences. First, here we consider a passive particle which simplifies the expression for $\ve f_0$. Second, here we consider non-spherical particles and allow force and particle velocity to point in different directions. For spherical particles, hydrodynamic force \cite{legendre1997note} and torque \cite{meibohm2016angular} are
entirely determined by the outer solution. For a particle of general shape the inner solution depends on particle shape and can contribute to force and torque. In the outer solution, particle shape enters only as the amplitude of a source term in the disturbance equation. The outer solution is obtained as a convolution of this shaped-dependent amplitude with a  universal kernel (equivalent to that obtained by \citet{lovalenti1993hydrodynamic})
that depends on the history of the particle dynamics. 
The kernels in Eq.~(\ref{eq:U}) in the main text have elements:
\begin{equation}
\begin{split}
[{\ma{K}^{(1)}}(t,\tau)]_{ij}= &  - \frac{1}{8\pi} \left( \frac{\exp(-A(t, \tau)^2)}{ \sqrt{\pi} A(t, \tau)^2}  + \frac{\mbox{erf}(A(t, \tau)) (A(t, \tau)^2 - \frac{1}{2}))}{A(t, \tau)^3} \right)\frac{\delta_{ij}}{\sqrt{t-\tau}}\\
&+ \frac{1}{8 \pi}\left(\frac{3 \exp(-A(t, \tau)^2)}{2 \sqrt{\pi} A(t, \tau)^2} + \frac{\mbox{erf}(A(t, \tau))(A(t, \tau)^2 - \frac{3}{2})}{2  A(t, \tau)^3}\right) \frac{\delta_{ij} - 
\hat{a}_i(t,\tau) \hat{a}_j(t,\tau)}{\sqrt{t-\tau}}\:.
\label{K1}
\end{split}
\end{equation}
and
\begin{equation}
\begin{split}
[{\ma{K}^{(2)}}(t,\tau)]_{ij} = &   \frac{1}{8 \pi} \frac{1}{4 A(t,\tau)}\left(\frac{3 \exp(-A(t, \tau)^2)}{ \sqrt{\pi} A(T, \tau)^2 } + \frac{\mbox{erf}(A(t, \tau)) (A(t, \tau)^2 - \frac{3}{2})}{A(t, \tau)^3} \right) \\
& \hspace{5cm}\times \frac{v_i(\tau) \hat{a}_j(t,\tau) + \hat{a}_i(t,\tau)v_j(\tau) - 4 [{\ve v}(\tau)\cdot \hat{{\ve a}}(t,\tau)] \hat{a}_i(t,\tau) \hat{a}_j(t,\tau)}{t - \tau}\\
&
- \frac{1}{8 \pi} \frac{1}{4 A(t,\tau)} \left( \frac{(-4 A(t, \tau)^2 - 3 ) \exp(-A(t, \tau)^2))}{ \sqrt{\pi} A(t, \tau)^2} +\frac{\mbox{erf}(A(t, \tau)) (A(t, \tau)^2 + \frac{3}{2})}{ A(t, \tau)^3} \right)[{\ve v}(\tau)\cdot \hat{{\ve a}}(t,\tau)] \\
& \hspace{5cm} \times \frac{ \delta_{ij} - \hat{a}_i(t,\tau) \hat{a}_j(t,\tau) }{t-\tau}\:.
\label{K2}
\end{split}
\end{equation}
In these expressions, $\ve f^{(0)}$ is the force exerted by the particle upon the fluid in the Stokes limit, $A(t, \tau)$ is the norm of the pseudo-displacement vector $\ve a(t,\tau)$ initially introduced by \citet{lovalenti1993hydrodynamic}
\begin{equation}
    \label{eq:a}
\ve a(t,\tau) =\frac{1}{2}\sqrt{\frac{\rep}{{\rm Sl}}}\frac{1}{\sqrt{t-\tau}} \int_\tau^t \mbox{d}t' \ve v(t')\:,
\end{equation}
and $\hat{\ve a} = \ve a / A(t,\tau)$. 
In the steady limit, Eq.~(\ref{eq:a}) implies
$\ve a =\tfrac{1}{2}\sqrt{\frac{\rep}{{\rm Sl}}}
\sqrt{t-\tau} \: \ve v$ and therefore $\hat {\ve a} = \hat{\ve v}$. Using these results, we see that Eq.~(\ref{eq:U}) simplifies to Eq.~(\ref{eq:Usteady}) in the steady limit. 
These kernels  are equivalent to those obtained by \citet{lovalenti1993hydrodynamic}. But the details differ because the decomposition into two integrals, one involving ${\ve f}^{(0)}$ and the other involving ${{\rm d}{\ve f}^{(0)}}/{{\rm d} \tau}$ is not unique. 
The above equations are more general than those derived in Ref.~\cite{redaelli2023hydrodynamic}, where it was assumed that ${\ve f}^{(0)}$ and $\hat{\ve v}$ are  collinear.

\section{Coupling tensors for nearly spherical particles}
\label{app:NS}
In this Appendix, we quote results for the elements of the coupling tensors obtained for the  nearly spherical particles listed in Table~\ref{tab:nearlyspherical}. The results were obtained as follows. To each order in the asymptotic-matching calculation one needs to solve a Stokes problem, as described in Ref.~\cite{redaelli2023hydrodynamic}.
These Stokes problems are solved for nearly spherical particles by expanding in the small non-sphericity parameter
$\varepsilon$ as described in the Appendix of Ref.~\cite{candelier2015role_a}.

\subsection{$O_h$-symmetry}
The results for $O_h$ symmetry are well known
\cite{Happel_Brenner_1983}. The tensors $\ma A$ and $\ma C$ are isotropic:
\begin{subequations} 
\begin{align}
A_{pp}&=A_{qq}=A_{nn} = 6 \pi  - \frac{9}{5} \pi \varepsilon  + \frac{116}{525} \pi \varepsilon^2 \,,
\quad \mbox{and}\quad  C_{pp}=C_{qq}=C_{nn} = 8 \pi - \frac{36}{5} \varepsilon \pi + \frac{1678}{525} \varepsilon^2 \pi\,,
\end{align}
The tensor $\ma B$ vanishes. For $\varepsilon=0$, Eq.~(\ref{eq:Tsing_steady})
yields the Oseen drag on a spherical particle
\begin{equation}
    \label{eq:Oseen}
    \ve F = 6\pi (1-\tfrac{3}{8} \rep v) \ve v\,.
\end{equation}

The following rank-3 tensors  
are non-zero:
$\ma M^{(v \omega)} = M^{(v \omega)} \ve \epsilon$,
$\ma N^{(v v)} = N^{(v v)} \ve \epsilon$, and 
$\ma N^{(\omega \omega)} = N^{(\omega \omega)} \ve \epsilon$. 
The tensors $\ma N^{(vv)}$
and $\ma N^{(\omega\omega)}$
do not contribute because
they are doubly contracted 
with a symmetrical tensor.
For $M^{(v\omega)}$  we find to order $\varepsilon$
\begin{equation}
\label{eq:Mvw}
M^{(v \omega)} = - \big(\pi - \tfrac{9}{10} \pi \varepsilon\big)\,.
\end{equation}
\end{subequations}
Eq.~(\ref{eq:Mvw}) yields the Keller-Rubinow lift force 
for a nearly spherical particle with $O_h$ symmetry. In dimensional form, we obtain
\begin{align}
    \label{eq:RK}
\ve F = -\varrho_{\rm f} \nu a v_0 \rep \pi(1-\tfrac{9}{10}\varepsilon) \ve v \wedge \ve \omega /(v_0 \omega_0)=\varrho_{\rm f} a^3 \pi(1-\tfrac{9}{10}\varepsilon) \ve \omega \wedge \ve v\,.
\end{align}
For $\varepsilon=0$, this is Eq.~(71) in Ref.~\cite{rubinow1961transverse}.
 
\subsection{$D_{2h}$-symmetry}
For $D_{2h}$ symmetry, we find
the following non-zero coefficients, 
\label{Res_D_inf_h}
\begin{align}
A_{pp} &=   6 \pi +   \left(-\tfrac{12}{5} \pi - \tfrac{6}{5} \alpha \pi \right) \varepsilon \,,
\quad A_{qq}  = 6 \pi + \left(-\tfrac{6}{5}\pi - \tfrac{12}{5}\alpha \pi \right)\varepsilon \,,
\quad A_{nn} =6 \pi + \left(-\tfrac{12}{5}\alpha \pi - \tfrac{12}{5}\pi \right)\varepsilon \,,
\\
C_{pp} &=8 \pi + \left(-\tfrac{48}{5}\pi - \tfrac{24}{5}\alpha \pi \right)\varepsilon\,,
\quad C_{qq} = 8 \pi + \left(-\tfrac{24}{5}\pi - \tfrac{48}{5}\alpha \pi \right)\varepsilon \,,
\quad C_{nn} =8 \pi + \left(-\tfrac{48}{5}\pi - \tfrac{48}{5}\alpha \pi \right)\varepsilon \,,\nonumber\\
2N_{p(nq)}^{(\omega \omega)}& =\tfrac{1}{15}\pi \varepsilon \,,\quad 2N_{q(np)}^{(\omega \omega)} = -\tfrac{1}{15}\pi\alpha \varepsilon\,,\quad 2N_{n(pq)}^{(\omega \omega)} =  \tfrac{1}{15}\pi(\alpha-1) \varepsilon\,,\nonumber\\
2N_{p(nq)}^{(vv)}& =\tfrac{29}{20}\pi \varepsilon \,,\quad 2N_{q(np)}^{(vv)} = -\tfrac{29}{20}\pi\alpha \varepsilon\,,\quad 2N_{n(pq)}^{(vv)} =  \tfrac{29}{20}\pi(\alpha-1) \varepsilon\,,\nonumber\\
M_{pqn}^{(v\omega)}&=-\pi + \left(\tfrac{2}{5} \alpha \pi + \tfrac{37}{20}\pi
\right) \varepsilon\,,\quad 
M_{pnq}^{(v\omega)}=\pi - \left(\tfrac{2}{5} \alpha \pi + \tfrac{3}{4}\pi\right) \varepsilon\,, \nonumber\\
M_{qnp}^{(v\omega)}&=\pi - \left(\tfrac{2}{5} \alpha \pi + \tfrac{3}{4}\pi\right) \varepsilon\,,\quad 
M_{qpn}^{(v\omega)}=\pi - \left(\tfrac{2}{5} \pi + \tfrac{37}{20}\alpha \pi\right) \varepsilon\,, \nonumber\\
M_{nqp}^{(v\omega)}&=\pi - \left(\tfrac{3}{4} \alpha \pi + \tfrac{37}{20}\pi
\right) \varepsilon\,,\quad 
M_{npq}^{(v\omega)}=-\pi + \left(\tfrac{3}{4} \pi + \tfrac{37}{20}\alpha \pi\right) \varepsilon\,,\nonumber
\end{align}
 confirming the symmetry analysis in 
Section~\ref{sec:symmetries} [Eq.~(\ref{eq:Nellipsoid})]. For $\alpha=1$
the particle shape is spheroidal ($C_{\infty h}$). As explained in the main text, this constrains more coefficients to vanish [Eq.~(\ref{eq:spheroid})].

\subsection{$C_{\infty v}$-symmetry}
For a particle with
broken fore-aft symmetry (third row in Table~\ref{tab:nearlyspherical}), we
find the following non-zero coupling coefficients 
\begin{align}
\label{Res_C_inf_v}
&A_{\perp} =   6 \pi -\tfrac{18}{5} \pi  \varepsilon + \left(\tfrac{198}{175} \pi +\tfrac{9}{20} \gamma^2 \pi\right) \varepsilon^2\,,\quad
 A_{\parallel} =6 \pi -\tfrac{24}{5} \pi  \varepsilon + \left(\tfrac{192}{175} \pi +\tfrac{3}{5} \gamma^2 \pi\right) \varepsilon^2\,,\\
&C_{\perp} = 8 \pi -\tfrac{72}{5}\pi\varepsilon + \left( \tfrac{2736}{175} \pi + \tfrac{9}{5}\gamma^2 \pi \right)\varepsilon^2 \,,\quad C_{\parallel} =8 \pi -\tfrac{96}{5}\pi\varepsilon + \left( \tfrac{3264}{175} \pi + \tfrac{12}{5}\gamma^2 \pi \right)\varepsilon^2,\nonumber \\
&N_{1}^{(vv)}+N_{2}^{(vv)} =\tfrac{29}{20}\pi \varepsilon\quad\mbox{and}\quad N_{1}^{(\omega\omega)}+N_{2}^{(\omega\omega)}= \tfrac{\pi}{15}\,,\nonumber \\
&N_1^{(v\omega)} =\pi \gamma \varepsilon\,,\quad  
\mbox{and}\quad N_{3}^{(v\omega)} = -\pi \gamma \varepsilon\,,\nonumber \\
&M_{1}^{(\omega \omega)}+M_{2}^{(\omega \omega)} =-\tfrac{1}{6}\pi \gamma \varepsilon\,,\quad\mbox{and}\quad M_{3}^{(\omega \omega)}= \tfrac{1}{6}\pi \gamma \varepsilon\,,\nonumber\\
&M_{1}^{(v \omega)} =\pi -\tfrac{23}{20}\pi \varepsilon\,,\quad M_{2}^{(v \omega)}= -\pi+\tfrac{9}{4} \pi \varepsilon\,, \quad \mbox{and} \quad M_{3}^{(v \omega)}= -\pi+\tfrac{13}{5} \pi \varepsilon\,.\nonumber
\end{align}
 The tensors $\ma B$, $\ma N^{(vv)}$, and 
 $\ma N^{(\omega\omega)}$
 were computed with respect to the centre-of-mass $\ve x_{\rm com} = - \tfrac{\gamma}{2} \varepsilon \hat{\ve e}_3$.
The coupling coefficients 
 $M_4^{(vv)}$ and $M_4^{(\omega\omega)}$
 vanish to order $\varepsilon$
 although they are allowed to be non-zero by symmetry.
  The tensor $\ma B$ vanishes to order $\varepsilon^2$, although allowed to be non-zero by the symmetry analysis. 
 We quote the elements of $\ma A$ and $\ma C$ to second order
 in $\varepsilon$, to show their $\gamma$-dependence which occurs only at this order. 

\subsection{$C_{2v}$-symmetry}
 For a nearly spherical particle with $C_{2v}$-symmetry (fourth row in Table~\ref{tab:nearlyspherical}),
we find the following non-zero coupling coefficients
\begin{align}
A_{pp} &\!=\!6 \pi - \tfrac{24\pi}{5} \varepsilon 
\!+ \!\left(\tfrac{192\pi}{175} \! + \! \tfrac{9\pi}{20} \beta^2  \right)\varepsilon^2,
A_{qq}\! = \!6 \pi \!- \!\tfrac{18\pi}{5}  \varepsilon 
\!+ \!\left(\tfrac{198\pi}{175} \! + \!\tfrac{9\pi}{20} \beta^2  \right)\varepsilon^2\,,
A_{nn} \!=\!6 \pi - \tfrac{18\pi}{5} \varepsilon 
\!+ \!\left(\tfrac{198\pi}{175} \! + \! \tfrac{3\pi}{5} \beta^2  \right)\varepsilon^2,\\
C_{pp} 
\!&=\!= \!8 \pi - \tfrac{96}{5} \pi \varepsilon   \!+ \! \left(\tfrac{3264\pi}{175} + \tfrac{9\pi}{5} \beta^2 \right)\varepsilon^2 \,,
C_{qq}   \!= \! 8 \pi  \!- \! \tfrac{72\pi}{5} \, \varepsilon   \!+ \! \left(\tfrac{2736 \pi}{175} + \tfrac{9\pi}{5} \beta^2 \right)\varepsilon^2\,,
C_{nn} \!=\!8 \pi\! -\! \tfrac{72}{5} \pi \, \varepsilon \! + \! \left(\tfrac{2736\pi}{175} \!+ \!\tfrac{12\pi}{5} \beta^2 \right)\varepsilon^2, \nonumber\\
&\hspace*{-6mm}2\,N_{q(pn)}^{(vv)} =\tfrac{29}{20}\pi \varepsilon \,,\quad 2\,N_{n(pq)}^{(vv)} = -\tfrac{29}{20}\pi \varepsilon\,,
\nonumber\\
&\hspace*{-6mm}2\,N_{q(pn)}^{(\omega \omega)} =\tfrac{1}{15}\pi \varepsilon \,,\quad 2\,N_{n(pq)}^{(\omega \omega)} = -\tfrac{1}{15}\pi \varepsilon\,,
\nonumber\\
&\hspace*{-6mm}N_{pnp}^{(v\omega)}=N_{qnq}^{(v\omega)}= \beta \pi \varepsilon\,,\quad N_{npp}^{v \omega} =N_{nqq}^{v \omega} = - \beta \pi \varepsilon\,\nonumber\,,\\
&\hspace*{-6mm}2\,M_{p(pn)}^{(\omega\omega)}=2\,M_{q(np)}^{(\omega\omega)}= -\tfrac{1}{6}\pi \beta \varepsilon \,,\quad 
2\,M_{n(pp)}^{(\omega\omega)}=2\,M_{n(qq)}^{(\omega\omega)}= \tfrac{1}{6}\pi \beta \varepsilon \,,\nonumber\\
&\hspace*{-6mm}M_{pqn}^{(v\omega)}=-M_{pnq}^{(v\omega)}= - \pi+ \tfrac{13}{5}\pi  \varepsilon \,,\quad M_{qpn}^{(v\omega)}=-M_{npq}^{(v\omega)}=\pi-\tfrac{23}{20}\pi  \varepsilon\,,\quad  M_{qnp}^{(v\omega)}=-M_{nqp}^{(v\omega)}=-\pi + \tfrac{9}{4}\pi  \varepsilon\,,\quad \nonumber
\end{align}
The elements of $\ma B$ $\ma C$, and $\ma N$ were computed with respect to the centre-of-mass $\ve x_{\rm com} = - \tfrac{\beta}{2} \varepsilon \hat{\ve e}_1$. 
The tensor $\ma B$ vanishes to order $\varepsilon^2$, 
and the combinations
$N_{n(qp)}^{(vv)}$ and $N_{n(qp)}^{(\omega\omega)}$ vanish to order $\varepsilon$, although allowed to be non-zero by the symmetry analysis. 

\section{Stability of a settling ellipsoid}
\label{app:stability}
We consider the sedimentation of an ellipsoidal particle in a quiescent fluid. 
Such particles have $D_{2h}$ symmetry. The force and torque are given by
\begin{subequations}
\begin{align}
    \bm F &= - \mathbb{A} \cdot \bm v + {\rep}\ve F^{(1)} - G\hat{\bf e}_3 \,, \\
    \bm T &= - \mathbb{C} \cdot \bm \omega + {\rep}\ve T^{(1)}\,.
\end{align}
\end{subequations}
Here $\ve F^{(1)}=\ve F^{(1)}_{\rm reg} + \ve F^{(1)}_{\rm sing}$, where $\ve F^{(1)}_{\rm reg}$ takes the form in Eq.~(\ref{eq:FTreg0}) with $M^{(vv)}=M^{(\omega\omega)}=0$ and $M^{(\omega v)}$ being of the same form as $N^{(vv)}$ due to the symmetry constraints posed by Eqs.~(\ref{eq:3MN}), and $\ve F^{(1)}_{\rm sing}$ is given in Eq.~(\ref{eq:Tsing_steady}).
Moreover, $\ve T^{(1)}=\ve T^{(1)}_{\rm reg}$ is given in Eq.~(\ref{eq:ellipsoid}) [$\ve T^{(1)}_{\rm sing}=0$ since $\ma B=0$ for $D_{2h}$ symmetry].
Finally, $-G\hat{\ve e}_3$ is the buoyancy force, with $G=(\rho_{\rm p}-\rho_{\rm f})V_{\rm p}g/(\mu v_0a)$, where $V_{\rm p}$ is the particle volume, and $\rho_{\rm p}$ and $\rho_{\rm f}$ are the densities of the particle and fluid, respectively.
We express the dynamics in the coordinate system $\ve p, \ve q, \ve n$ that rotates with the particle.
In this system, $\ma A$ and $\ma C$ are diagonal for $D_{2h}$. Neglecting the fluid-inertia contribution to the hydrodynamic force, 
we obtain
\begin{subequations}
\label{eq:EllipsoidDynamics}
\begin{align}
\dot v_p&=-A_{pp}v_p+F_p^{(1)}+G\cos\psi\sin\theta+\omega_n v_q - \omega_q v_n\\
\dot v_q&=-A_{qq}v_q+F_q^{(1)}-G\sin\psi\sin\theta+\omega_p v_n - \omega_n v_p\\
\dot v_n&=-A_{nn}v_n+F_n^{(1)}-G\cos\theta+\omega_q v_p - \omega_p v_q\\
\dot\theta&=\omega_q\cos\psi+\omega_p\sin\psi\,,\quad
\dot\psi=\omega_n+\cot\theta[\omega_p\cos\psi-\omega_q\sin\psi]\,,\\
\dot \omega_p&=\frac{1}{I_{pp}}[-C_{pp}\omega_p+{\rep}T_p^{(1)}+\omega_n\omega_q(I_{qq}-I_{nn})]\\
\dot \omega_q&=\frac{1}{I_{qq}}[-C_{qq}\omega_q+{\rep}T_q^{(1)}+\omega_p\omega_n(I_{nn}-I_{pp})]\\
\dot \omega_n&=\frac{1}{I_{nn}}[-C_{nn}\omega_n+{\rep}T_n^{(1)}+\omega_q\omega_p(I_{pp}-I_{qq})]\,.
\end{align}
\end{subequations}
Here, $I_{pp},\ldots$ are the elements of the inertia
tensor of the ellipsoidal particle. The $\omega_iv_j$ and $\omega_i\omega_j$ terms result from the time derivative of the particle orientation.
We consider the dynamics in the subsystem $\ve V=[v_p,v_q,v_n,\theta,\psi,\omega_p,\omega_q,\omega_n]^{\sf T}$, not including $\ve x$ and $\phi$, since these coordinates do not affect $\ve V$.

We search for a rotation-free steady state, $\ve\omega=0$.
By solving $\dot{\ve V}=0$ under this condition, the system (\ref{eq:EllipsoidDynamics}) has six fixed points corresponding to the particle settling with either $\ve p$, $\ve q$, $\ve n$ aligned with or against gravity.
The settling speed is obtained as a solution to a second-order polynomial. For small Reynolds numbers, it takes the form $v_g^*=G/A^{(g)}$, where $A^{(g)}$ is the component of the resistance tensor in the direction that aligns/anti-aligns with gravity.

The $8\times 8$ stability matrix $\ma J$ has elements $J_{ij}=\partial_{V_j}\dot V_i$.
Evaluating $\ma J$ at either of the fixed points and suitably permuting the rows and columns, we
find that it is block-diagonal with two 1-blocks and two 3-blocks. The eigenvalues of the 1-blocks are the negative resistance elements $\lambda_1=-A^{(g)}$ and $\lambda_2=-C^{(g)}/I^{(g)}$.
The eigenvalues of the 3-blocks each satisfy a third-order polynomial being equal to zero, giving the stability exponents $\lambda_{3-5}$ and $\lambda_{6-8}$, respectively.
For particles settling with $\ve n$ pointing downwards, these become (neglecting contributions of order ${\rep^2}$)
\begin{align}
0&=\lambda^3+P_2\lambda^2+P_1\lambda+P_0\mbox{ and }0=\lambda^3+Q_2\lambda^2+Q_1\lambda+Q_0 \mbox{ with }
P_0=-\frac{2N^{(vv)}_{p(qn)}}{I_{pp}} \frac{G^2{\rep}}{A_{nn}}
\,,\;
Q_0=\frac{2N^{(vv)}_{q(np)}}{I_{qq}}\frac{G^2{\rep}}{A_{nn}}
\,,\nonumber\\
P_1&=\frac{A_{qq}C_{pp}}{I_{pp}}(1-\frac{G{\rep}}{32\pi}\frac{A_{nn}-3A_{qq}}{A_{nn}})
-2N^{(vv)}_{p(qn)}\frac{G^2{\rep}}{A_{nn}^2I_{pp}}
\,,\;
Q_1=\frac{A_{pp}C_{qq}}{I_{qq}}(1-\frac{G{\rep}}{32\pi}\frac{A_{nn}-3A_{pp}}{A_{nn}})
+2N^{(vv)}_{q(np)}\frac{G^2{\rep}}{A_{nn}^2I_{qq}}
\,,\nonumber\\
P_2&=A_{qq}+\frac{C_{pp}}{I_{pp}}-\frac{G{\rep}}{32\pi}\frac{A_{qq}}{A_{nn}}(A_{nn}-3A_{qq})\,,\;
Q_2=A_{pp}+\frac{C_{qq}}{I_{qq}}-\frac{G{\rep}}{32\pi}\frac{A_{pp}}{A_{nn}}(A_{nn}-3A_{pp})\,.
\end{align}
A necessary condition for these polynomials to only have negative solutions is that all polynomial coefficients are positive, $P_i>0$ and $Q_i>0$, which requires 
\begin{align}
N^{(vv)}_{p(qn)}<0 \mbox{ and }N^{(vv)}_{q(np)}>0\,.
\label{eq:EllipsoidSufficient}
\end{align}
A sufficient condition (Routh-Hurwitz criterion) is that additionally the product of the coefficients of $\lambda$ and $\lambda^2$ is larger than the constant coefficient:
\begin{align}
P_1P_2>P_0 \mbox{ and }Q_1Q_2>Q_0\,.
\label{eq:EllipsoidAdditionalCondition}
\end{align}
Alignment with $\ve n$ pointing against gravity (instead of along) gives the same stability exponents. The stability exponents for the remaining four orientations are obtained by permutations of $\ve p$, $\ve q$, and $\ve n$. The conditions corresponding to Eq.~(\ref{eq:EllipsoidSufficient}) when settling along or against $\ve p$ and $\ve q$ are 
\begin{subequations}
\begin{align}
N^{(vv)}_{n(pq)}<0 \mbox{ and }N^{(vv)}_{p(qn)}>0\mbox{ for }\ve q\parallel\ve {e}_3\\
N^{(vv)}_{q(np)}<0 \mbox{ and }N^{(vv)}_{n(pq)}>0\mbox{ for }\ve p\parallel\ve {e}_3
\end{align}
\label{eq:EllipsoidSufficientpq}
\end{subequations}The three conditions (\ref{eq:EllipsoidSufficient}) and (\ref{eq:EllipsoidSufficientpq}) are mutually exclusive.
This means that at most one orientation can be stable.
In the overdamped limit (large components of the resistance tensor), Eqs.~(\ref{eq:EllipsoidAdditionalCondition}) are satisfied and the ellipsoid obtain a stable orientation provided that at least two of the components $N^{(vv)}_{n(pq)}$, $N^{(vv)}_{q(np)}$, and $N^{(vv)}_{n(pq)}$ have opposite signs.
This orientation may become unstable for less inertial particles, for which Eqs.~(\ref{eq:EllipsoidAdditionalCondition}) are no longer satisfied.

For nearly spherical ellipsoids, the shape parameters are given in Eq.~(\ref{Res_D_inf_h}). Substituting the results into 
Eqs.~(\ref{eq:EllipsoidSufficient}), (\ref{eq:EllipsoidAdditionalCondition}), and (\ref{eq:EllipsoidSufficientpq}) shows that, as long as $|\epsilon|$ is not too large, the particle aligns $\ve n$ with gravity if $\varepsilon<0$ and $\alpha>0$ and $\alpha>0$, aligns $\ve q$ if $\varepsilon>0$ and $\alpha<1$, and aligns $\ve p$ if $\varepsilon>0$ and $\alpha>1$ or if $\varepsilon<0$ and $\alpha<0$.
All these cases correspond to the particle aligning the broadest cross section perpendicular to gravity.

\section{Asymptotic matching for large Strouhal numbers}
\label{app:iu}
Here we show how to obtain Eqs.~(\ref{eq:iu}). 
When the Strouhal number is large, in the limit $1 \gg \rep \rm{Sl} \gg \rep$, the equations for the outer flow may be simplified by neglecting the convective terms. 
Force and torque on the particle can still be obtained using matched-asymptotic expansions, but extended to the second order, following
\citet{candelier2023second}, but setting the undisturbed flow to $0$. 
In short, the idea is to seek the disturbance flow 
in the form 
\begin{equation}
	{\ve{w}}_{\mbox{\scriptsize in}} = {\ve{w}}_{\mbox{\scriptsize in}}^{(0)} + \sqrt{\rep {\rm Sl}} \: {\ve{w}}_{\mbox{\scriptsize in}}^{(1)} + \rep {\rm Sl} \: {\ve{w}}_{\mbox{\scriptsize in}}^{(2)} +  \ldots \quad  \mbox{and}\quad 
	{p}_{\mbox{\scriptsize in}} = {p}_{\mbox{\scriptsize in}}^{(0)} + \sqrt{\rep {\rm Sl}} \: {p}_{\mbox{\scriptsize in}}^{(1)} +   
    \rep  {\rm Sl} \: {p}_{\mbox{\scriptsize in}}^{(1)} +   \ldots \,.
\end{equation}
To second order, the outer flow equations 
require two source terms, $f^{(0)}$ and $f^{(1)}$,
\begin{equation}\label{eq:eom_2}
{{\rm Re}_{p}\, {\rm Sl}\,\frac{\partial \ve w}{\partial t}}= - \boldsymbol{\nabla} p + \boldsymbol{\triangle} \ve{w} + \left(\ve f^{(0)} + \sqrt{\rep {\rm Sl}} \,\ve f^{(1)}\right) \delta(\ve r) \:.
	\end{equation}
The solution of this outer equation takes the form
\begin{equation}
\begin{split}
{\ve w}_{\mbox{\scriptsize out}} = &  \frac{1}{8\pi}\left(\frac{\ma{1}}{r}  + \frac{\ve r\otimes \ve r}{r^3}\right)\cdot \left(\ve f^{(0)} + \sqrt{\rep {\rm Sl}} \,\ve f^{(1)}\right) + \sqrt{{\rm Re}_p {\rm Sl}}\, \ve  {\mathcal U}_1 \\
& +{\rm Re}_p {\rm Sl} \, \left(\frac{3 r}{32\pi}\left(\ma{I}  -\frac{1}{3} \frac{\ve r\otimes \ve r}{r^2}\right)  \frac{{\rm d} \ve f^{(0)}}{{\rm d} t}+\,\ve  {\mathcal U}_2  \right)
\end{split}
\end{equation}
where 
\begin{equation}
\label{eq:UU1}
\ve  {\mathcal U}_1  = - \int_0^t \frac{{\rm d } \tau}{6 \pi} \frac{1}{\sqrt{\pi(t-\tau)}} \frac{{\rm{d}} \ve f^{(0)}}{{\rm{d}} \tau} \quad \mbox{and} 
\quad \ve  {\mathcal U}_2  = - \int_0^t \frac{{\rm d } \tau}{6 \pi} \frac{1}{\sqrt{\pi(t-\tau)}} \frac{{\rm{d}} \ve f^{(1)}}{{\rm{d}} \tau} \:.  
\end{equation}
The matching procedure leads us to $\ve f^{(1)} = - \ma A \cdot \ve  {\mathcal U}_1 $. 
The corresponding
singular contributions to force and torque read: 
$$
\ve F^{(1)}_{\rm sing} = \ma A \cdot \ve {\mathcal U}_1,\quad 
\ve F^{(2)}_{\rm sing} = \ma A \cdot \ve {\mathcal U}_2,
$$
$$
\ve T^{(1)}_{\rm sing} = \ma B \cdot \ve {\mathcal U}_1,\quad 
\ve T^{(2)}_{\rm sing} = \ma B \cdot \ve {\mathcal U}_2\:.
$$
In the limit ${\rm Sl}\gg 1$, the two kernels 
$\ma K^{(1)}$ and $\ma K^{(2)}$, defined in Eqs.~(\ref{K1})--(\ref{K2}), 
simplify as
\begin{align}
\label{eq:limit}
\ma K^{(1)} \longrightarrow \frac{1}{6\pi \sqrt{t-\tau}}\,,
\qquad
\ma K^{(2)} \longrightarrow \ma 0\:,
\end{align}
the same as the kernel $\mathcal{\ve U}_1$ in Eq.~(\ref{eq:UU1}). 
The derivation underlying Eqs.~(\ref{K1})--(\ref{K2})  assumed  ${\rm Sl}\ll 1$. Eq.~(\ref{eq:limit}) shows that the result remains uniformly valid at order $\sqrt{\rep\,{\rm Sl}}$.
To order ${\rep\,{\rm Sl}}$ there are additional contributions to the memory kernel in the form of a double convolution. For an isotropic particle where $\ma A \propto \ma I$, the double convolution simplifies to an instantaneous contribution connected to the added mass \cite{candelier2023second}. 

Now consider the regular contribution. The regular expansion for the disturbance velocity in the inner region is linearly
related to the particle velocity $\ve v$ and its angular velocity $\ve \omega$
\begin{align}
\ve w_{\rm in}^{(0)}
= \ma W_v(\ve X) \cdot \ve v
+ \ma W_\omega (\ve X) \cdot \ve \omega\,,
\end{align}
where $X_i$ are coordinates in the particle-fixed coordinate system.
 Eq.~(\ref{eq:eom}), however,  is written  in a coordinate system that translates with particle (but does not rotate with it). Evaluating the time
derivative of $\ve w_{\rm in}^{(0)}$ in the translating coordinate system yields
an expression that depends linearly on $\dot{\ve v}$ and $\dot{\ve \omega}$,
and there are also second-order terms proportional to
$\ve \omega  \otimes \ve v$
and  $\ve \omega \otimes \ve \omega$. Hydrodynamic force and torque must have the same tensorial structure. There are quadratic terms just like Eq.~(\ref{eq:FTreg0}). But there are also 
 unsteady contributions to force and torque of the form
\begin{subequations}
\label{eq:iu_appendix}
\begin{align}
\ve F^{(2)}_{\rm u}
&=\rep{\rm Sl}\,(\ma P \dot {\ve v} + \ma Q \dot {\ve \omega})\,,\\
\ve T^{(2)}_{\rm u}&={\rm Re}_p{\rm Sl}\,(\ma Q^{\sf T} \dot {\ve v} + \ma R \dot {\ve \omega})\,,
    \end{align}
\end{subequations}
distinct from the terms discussed in the main text. Here
$\ma P$, $\ma Q$, $\ma R$ are second-order tensors, constrained by symmetry in the same way as the second-order Stokes resistance tensors. The fact that
$\ma Q^{\sf T}$ appears in the expression for the torque follows from the reciprocal theorem.
The terms (\ref{eq:iu_appendix}) that look like added-mass contributions. They were first obtained by \citet{gavze1990accelerated}, who derived a general solution of the unsteady Stokes equation for a particle of arbitrary shape. More recently, \citet{collis2024unsteady} considered the same limit as \citet{gavze1990accelerated}.
They solved the unsteady Stokes equation for several
nearly spherical particles to obtain hydrodynamic force and torque in the unsteady limit, at large Strouhal numbers. No  history terms appear in their expressions, because they work in the frequency domain. Integration
over frequency must give these terms, in the same way as for a sphere oscillating in a fluid at rest~\cite{landau1987hydrodynamics}.

\vfill\eject

\begin{thebibliography}{74}%
\makeatletter
\providecommand \@ifxundefined [1]{%
 \@ifx{#1\undefined}
}%
\providecommand \@ifnum [1]{%
 \ifnum #1\expandafter \@firstoftwo
 \else \expandafter \@secondoftwo
 \fi
}%
\providecommand \@ifx [1]{%
 \ifx #1\expandafter \@firstoftwo
 \else \expandafter \@secondoftwo
 \fi
}%
\providecommand \natexlab [1]{#1}%
\providecommand \enquote  [1]{``#1''}%
\providecommand \bibnamefont  [1]{#1}%
\providecommand \bibfnamefont [1]{#1}%
\providecommand \citenamefont [1]{#1}%
\providecommand \href@noop [0]{\@secondoftwo}%
\providecommand \href [0]{\begingroup \@sanitize@url \@href}%
\providecommand \@href[1]{\@@startlink{#1}\@@href}%
\providecommand \@@href[1]{\endgroup#1\@@endlink}%
\providecommand \@sanitize@url [0]{\catcode `\\12\catcode `\$12\catcode
  `\&12\catcode `\#12\catcode `\^12\catcode `\_12\catcode `\%12\relax}%
\providecommand \@@startlink[1]{}%
\providecommand \@@endlink[0]{}%
\providecommand \url  [0]{\begingroup\@sanitize@url \@url }%
\providecommand \@url [1]{\endgroup\@href {#1}{\urlprefix }}%
\providecommand \urlprefix  [0]{URL }%
\providecommand \Eprint [0]{\href }%
\providecommand \doibase [0]{http://dx.doi.org/}%
\providecommand \selectlanguage [0]{\@gobble}%
\providecommand \bibinfo  [0]{\@secondoftwo}%
\providecommand \bibfield  [0]{\@secondoftwo}%
\providecommand \translation [1]{[#1]}%
\providecommand \BibitemOpen [0]{}%
\providecommand \bibitemStop [0]{}%
\providecommand \bibitemNoStop [0]{.\EOS\space}%
\providecommand \EOS [0]{\spacefactor3000\relax}%
\providecommand \BibitemShut  [1]{\csname bibitem#1\endcsname}%
\let\auto@bib@innerbib\@empty
\bibitem [{\citenamefont {Sanjeevi}\ \emph {et~al.}(2018)\citenamefont
  {Sanjeevi}, \citenamefont {Kuipers},\ and\ \citenamefont
  {Padding}}]{Sanjeevi18}%
  \BibitemOpen
  \bibfield  {author} {\bibinfo {author} {\bibfnamefont {S.K.P.}\ \bibnamefont
  {Sanjeevi}}, \bibinfo {author} {\bibfnamefont {J.A.M.}\ \bibnamefont
  {Kuipers}}, \ and\ \bibinfo {author} {\bibfnamefont {J.T.}\ \bibnamefont
  {Padding}},\ }\bibfield  {title} {\enquote {\bibinfo {title} {Drag, lift and
  torque correlations for non-spherical particles from {Stokes} limit to high
  {R}eynolds numbers},}\ }\href@noop {} {\bibfield  {journal} {\bibinfo
  {journal} {Int. J. Multiphase Flow}\ }\textbf {\bibinfo {volume} {106}},\
  \bibinfo {pages} {325--337} (\bibinfo {year} {2018})}\BibitemShut {NoStop}%
\bibitem [{\citenamefont {Ouchene}\ \emph {et~al.}(2015)\citenamefont
  {Ouchene}, \citenamefont {Khalij}, \citenamefont {{Tani{\`e}re}},\ and\
  \citenamefont {Arcen}}]{Ouchene15}%
  \BibitemOpen
  \bibfield  {author} {\bibinfo {author} {\bibfnamefont {R.}~\bibnamefont
  {Ouchene}}, \bibinfo {author} {\bibfnamefont {M.}~\bibnamefont {Khalij}},
  \bibinfo {author} {\bibfnamefont {A.}~\bibnamefont {{Tani{\`e}re}}}, \ and\
  \bibinfo {author} {\bibfnamefont {B.}~\bibnamefont {Arcen}},\ }\bibfield
  {title} {\enquote {\bibinfo {title} {Drag, lift and torque coefficients for
  ellipsoidal particles: from low to moderate particle {R}eynolds numbers},}\
  }\href@noop {} {\bibfield  {journal} {\bibinfo  {journal} {Computers \&
  Fluids}\ }\textbf {\bibinfo {volume} {113}},\ \bibinfo {pages} {53--64}
  (\bibinfo {year} {2015})}\BibitemShut {NoStop}%
\bibitem [{\citenamefont {Ouchene}\ \emph
  {et~al.}(2016{\natexlab{a}})\citenamefont {Ouchene}, \citenamefont {Khalij},
  \citenamefont {Arcen},\ and\ \citenamefont {{A. Tani{\`e}re}}}]{Ouchene16}%
  \BibitemOpen
  \bibfield  {author} {\bibinfo {author} {\bibfnamefont {R.}~\bibnamefont
  {Ouchene}}, \bibinfo {author} {\bibfnamefont {M.}~\bibnamefont {Khalij}},
  \bibinfo {author} {\bibfnamefont {B.}~\bibnamefont {Arcen}}, \ and\ \bibinfo
  {author} {\bibnamefont {{A. Tani{\`e}re}}},\ }\bibfield  {title} {\enquote
  {\bibinfo {title} {A new set of correlations of drag, lift and torque
  coefficients for non-spherical particles at large {R}eynolds numbers},}\
  }\href@noop {} {\bibfield  {journal} {\bibinfo  {journal} {Powder
  Technology}\ }\textbf {\bibinfo {volume} {303}},\ \bibinfo {pages} {33--43}
  (\bibinfo {year} {2016}{\natexlab{a}})}\BibitemShut {NoStop}%
\bibitem [{\citenamefont {Ouchene}(2020)}]{Ouchene20}%
  \BibitemOpen
  \bibfield  {author} {\bibinfo {author} {\bibfnamefont {R.}~\bibnamefont
  {Ouchene}},\ }\bibfield  {title} {\enquote {\bibinfo {title} {Numerical
  simulation and modeling of the hydrodynamic forces and torque acting on
  individual oblate spheroids},}\ }\href@noop {} {\bibfield  {journal}
  {\bibinfo  {journal} {Phys. Fluids}\ }\textbf {\bibinfo {volume} {32}},\
  \bibinfo {pages} {073303} (\bibinfo {year} {2020})}\BibitemShut {NoStop}%
\bibitem [{\citenamefont {{H{\"o}lzer}}\ and\ \citenamefont
  {Sommerfeld}(2009)}]{Holzer09}%
  \BibitemOpen
  \bibfield  {author} {\bibinfo {author} {\bibfnamefont {A.}~\bibnamefont
  {{H{\"o}lzer}}}\ and\ \bibinfo {author} {\bibfnamefont {M.}~\bibnamefont
  {Sommerfeld}},\ }\bibfield  {title} {\enquote {\bibinfo {title} {Lattice
  {Boltzmann} simulations to determine drag, lift and torque acting on
  non-spherical particles},}\ }\href@noop {} {\bibfield  {journal} {\bibinfo
  {journal} {Computers \& Fluids}\ }\textbf {\bibinfo {volume} {38}},\ \bibinfo
  {pages} {572--589} (\bibinfo {year} {2009})}\BibitemShut {NoStop}%
\bibitem [{\citenamefont {Zastawny}\ \emph {et~al.}(2012)\citenamefont
  {Zastawny}, \citenamefont {Mallouppas}, \citenamefont {Zhao},\ and\
  \citenamefont {van Wachem}}]{Zastawny12}%
  \BibitemOpen
  \bibfield  {author} {\bibinfo {author} {\bibfnamefont {M.}~\bibnamefont
  {Zastawny}}, \bibinfo {author} {\bibfnamefont {G.}~\bibnamefont
  {Mallouppas}}, \bibinfo {author} {\bibfnamefont {F.}~\bibnamefont {Zhao}}, \
  and\ \bibinfo {author} {\bibfnamefont {B.}~\bibnamefont {van Wachem}},\
  }\bibfield  {title} {\enquote {\bibinfo {title} {Derivation of drag and lift
  forces and torque coefficients for non-spherical particles in flow},}\
  }\href@noop {} {\bibfield  {journal} {\bibinfo  {journal} {International
  Journal of Multiphase Flow}\ }\textbf {\bibinfo {volume} {39}},\ \bibinfo
  {pages} {227--239} (\bibinfo {year} {2012})}\BibitemShut {NoStop}%
\bibitem [{\citenamefont {Fr{\"o}hlich}\ \emph {et~al.}(2020)\citenamefont
  {Fr{\"o}hlich}, \citenamefont {Meinke},\ and\ \citenamefont
  {Schr{\"o}der}}]{Froehlich20}%
  \BibitemOpen
  \bibfield  {author} {\bibinfo {author} {\bibfnamefont {K.}~\bibnamefont
  {Fr{\"o}hlich}}, \bibinfo {author} {\bibfnamefont {M.}~\bibnamefont
  {Meinke}}, \ and\ \bibinfo {author} {\bibfnamefont {W.}~\bibnamefont
  {Schr{\"o}der}},\ }\bibfield  {title} {\enquote {\bibinfo {title}
  {Correlations for inclide prolates based on highly resolved simulations},}\
  }\href@noop {} {\bibfield  {journal} {\bibinfo  {journal} {J. Fluid Mech.}\
  }\textbf {\bibinfo {volume} {901}},\ \bibinfo {pages} {A5} (\bibinfo {year}
  {2020})}\BibitemShut {NoStop}%
\bibitem [{\citenamefont {Bretherton}(1962)}]{Bretherton:1962}%
  \BibitemOpen
  \bibfield  {author} {\bibinfo {author} {\bibfnamefont {F.P.}\ \bibnamefont
  {Bretherton}},\ }\bibfield  {title} {\enquote {\bibinfo {title} {The motion
  of rigid particles in a shear flow at low {R}eynolds number},}\ }\href@noop
  {} {\bibfield  {journal} {\bibinfo  {journal} {J. Fluid Mech.}\ }\textbf
  {\bibinfo {volume} {14}},\ \bibinfo {pages} {284--304} (\bibinfo {year}
  {1962})}\BibitemShut {NoStop}%
\bibitem [{\citenamefont {Brenner}(1964{\natexlab{a}})}]{brenner1964stokes}%
  \BibitemOpen
  \bibfield  {author} {\bibinfo {author} {\bibfnamefont {H.}~\bibnamefont
  {Brenner}},\ }\bibfield  {title} {\enquote {\bibinfo {title} {The {S}tokes
  resistance of a slightly deformed sphere},}\ }\href@noop {} {\bibfield
  {journal} {\bibinfo  {journal} {Chem. Eng. Sci.}\ }\textbf {\bibinfo {volume}
  {19}},\ \bibinfo {pages} {519--539} (\bibinfo {year}
  {1964}{\natexlab{a}})}\BibitemShut {NoStop}%
\bibitem [{\citenamefont {Happel}\ and\ \citenamefont
  {Brenner}(1983)}]{Happel_Brenner_1983}%
  \BibitemOpen
  \bibfield  {author} {\bibinfo {author} {\bibfnamefont {J.}~\bibnamefont
  {Happel}}\ and\ \bibinfo {author} {\bibfnamefont {H.}~\bibnamefont
  {Brenner}},\ }\href {\doibase 10.1007/978-94-009-8352-6} {\emph {\bibinfo
  {title} {Low Reynolds Number Hydrodynamics}}}\ (\bibinfo  {publisher}
  {Martinus Nijhoff Publishers},\ \bibinfo {address} {Hague},\ \bibinfo {year}
  {1983})\BibitemShut {NoStop}%
\bibitem [{\citenamefont {Fries}\ \emph {et~al.}(2017)\citenamefont {Fries},
  \citenamefont {Einarsson},\ and\ \citenamefont {Mehlig}}]{fries2017angular}%
  \BibitemOpen
  \bibfield  {author} {\bibinfo {author} {\bibfnamefont {J.}~\bibnamefont
  {Fries}}, \bibinfo {author} {\bibfnamefont {J.}~\bibnamefont {Einarsson}}, \
  and\ \bibinfo {author} {\bibfnamefont {B.}~\bibnamefont {Mehlig}},\
  }\bibfield  {title} {\enquote {\bibinfo {title} {Angular dynamics of small
  crystals in viscous flow},}\ }\href@noop {} {\bibfield  {journal} {\bibinfo
  {journal} {Phys. Rev. Fluids}\ }\textbf {\bibinfo {volume} {2}},\ \bibinfo
  {pages} {014302} (\bibinfo {year} {2017})}\BibitemShut {NoStop}%
\bibitem [{\citenamefont {Witten}\ and\ \citenamefont
  {Diamant}(2020)}]{witten2020review}%
  \BibitemOpen
  \bibfield  {author} {\bibinfo {author} {\bibfnamefont {T.A.}\ \bibnamefont
  {Witten}}\ and\ \bibinfo {author} {\bibfnamefont {H.}~\bibnamefont
  {Diamant}},\ }\bibfield  {title} {\enquote {\bibinfo {title} {A review of
  shaped colloidal particles in fluids: anisotropy and chirality},}\
  }\href@noop {} {\bibfield  {journal} {\bibinfo  {journal} {Reports on
  progress in physics}\ }\textbf {\bibinfo {volume} {83}},\ \bibinfo {pages}
  {116601} (\bibinfo {year} {2020})}\BibitemShut {NoStop}%
\bibitem [{\citenamefont {Collins}\ \emph {et~al.}(2021)\citenamefont
  {Collins}, \citenamefont {Hamati}, \citenamefont {Candelier}, \citenamefont
  {Gustavsson}, \citenamefont {Mehlig},\ and\ \citenamefont
  {Voth}}]{collins2021lord}%
  \BibitemOpen
  \bibfield  {author} {\bibinfo {author} {\bibfnamefont {D.}~\bibnamefont
  {Collins}}, \bibinfo {author} {\bibfnamefont {R.J.}\ \bibnamefont {Hamati}},
  \bibinfo {author} {\bibfnamefont {F.}~\bibnamefont {Candelier}}, \bibinfo
  {author} {\bibfnamefont {K.}~\bibnamefont {Gustavsson}}, \bibinfo {author}
  {\bibfnamefont {B.}~\bibnamefont {Mehlig}}, \ and\ \bibinfo {author}
  {\bibfnamefont {G.A.}\ \bibnamefont {Voth}},\ }\bibfield  {title} {\enquote
  {\bibinfo {title} {Lord {Kelvin}'s isotropic helicoid},}\ }\href@noop {}
  {\bibfield  {journal} {\bibinfo  {journal} {Phys. Rev. Fluids}\ }\textbf
  {\bibinfo {volume} {6}},\ \bibinfo {pages} {074302} (\bibinfo {year}
  {2021})}\BibitemShut {NoStop}%
\bibitem [{\citenamefont {Miara}\ \emph {et~al.}(2024)\citenamefont {Miara},
  \citenamefont {Vaquero-Stainer}, \citenamefont {Pihler-Puzovi{\'c}},
  \citenamefont {Heil},\ and\ \citenamefont {Juel}}]{miara2024dynamics}%
  \BibitemOpen
  \bibfield  {author} {\bibinfo {author} {\bibfnamefont {T.}~\bibnamefont
  {Miara}}, \bibinfo {author} {\bibfnamefont {C.}~\bibnamefont
  {Vaquero-Stainer}}, \bibinfo {author} {\bibfnamefont {D.}~\bibnamefont
  {Pihler-Puzovi{\'c}}}, \bibinfo {author} {\bibfnamefont {M.}~\bibnamefont
  {Heil}}, \ and\ \bibinfo {author} {\bibfnamefont {A.}~\bibnamefont {Juel}},\
  }\bibfield  {title} {\enquote {\bibinfo {title} {Dynamics of inertialess
  sedimentation of a rigid u-shaped disk},}\ }\href@noop {} {\bibfield
  {journal} {\bibinfo  {journal} {Comm. Phys.}\ }\textbf {\bibinfo {volume}
  {7}},\ \bibinfo {pages} {47} (\bibinfo {year} {2024})}\BibitemShut {NoStop}%
\bibitem [{\citenamefont {Huseby}\ \emph {et~al.}(2025)\citenamefont {Huseby},
  \citenamefont {Gissinger}, \citenamefont {Candelier}, \citenamefont {Pujara},
  \citenamefont {Verhille}, \citenamefont {Mehlig},\ and\ \citenamefont
  {Voth}}]{huseby2024helical}%
  \BibitemOpen
  \bibfield  {author} {\bibinfo {author} {\bibfnamefont {E.}~\bibnamefont
  {Huseby}}, \bibinfo {author} {\bibfnamefont {J.}~\bibnamefont {Gissinger}},
  \bibinfo {author} {\bibfnamefont {F.}~\bibnamefont {Candelier}}, \bibinfo
  {author} {\bibfnamefont {N.}~\bibnamefont {Pujara}}, \bibinfo {author}
  {\bibfnamefont {G.}~\bibnamefont {Verhille}}, \bibinfo {author}
  {\bibfnamefont {B.}~\bibnamefont {Mehlig}}, \ and\ \bibinfo {author}
  {\bibfnamefont {G.}~\bibnamefont {Voth}},\ }\bibfield  {title} {\enquote
  {\bibinfo {title} {Helical ribbons: Simple chiral sedimentation},}\
  }\href@noop {} {\bibfield  {journal} {\bibinfo  {journal} {Phys. Rev.
  Fluids}\ }\textbf {\bibinfo {volume} {10}},\ \bibinfo {pages} {024101}
  (\bibinfo {year} {2025})}\BibitemShut {NoStop}%
\bibitem [{\citenamefont {Heymsfield}(1973)}]{Heymsfield_1973}%
  \BibitemOpen
  \bibfield  {author} {\bibinfo {author} {\bibfnamefont {A.~J.}\ \bibnamefont
  {Heymsfield}},\ }\bibfield  {title} {\enquote {\bibinfo {title} {Laboratory
  and field observations of the growth of columnar and plate crystals from
  frozen droplets},}\ }\href {\doibase
  10.1175/1520-0469(1973)030<1650:LAFOOT>2.0.CO;2} {\bibfield  {journal}
  {\bibinfo  {journal} {JAS}\ }\textbf {\bibinfo {volume} {30}},\ \bibinfo
  {pages} {1650--1656} (\bibinfo {year} {1973})}\BibitemShut {NoStop}%
\bibitem [{\citenamefont {Sassen}(1980)}]{Sassen_1980}%
  \BibitemOpen
  \bibfield  {author} {\bibinfo {author} {\bibfnamefont {K.}~\bibnamefont
  {Sassen}},\ }\bibfield  {title} {\enquote {\bibinfo {title} {Remote sensing
  of planar ice crystal fall attitudes},}\ }\href {\doibase
  10.2151/jmsj1965.58.5_422} {\bibfield  {journal} {\bibinfo  {journal} {J.
  Meteor. Soc. Japan. Ser. {II}}\ }\textbf {\bibinfo {volume} {58}},\ \bibinfo
  {pages} {422--429} (\bibinfo {year} {1980})}\BibitemShut {NoStop}%
\bibitem [{\citenamefont {Klett}(1995)}]{Klett_1995}%
  \BibitemOpen
  \bibfield  {author} {\bibinfo {author} {\bibfnamefont {J.~D.}\ \bibnamefont
  {Klett}},\ }\bibfield  {title} {\enquote {\bibinfo {title} {Orientation model
  for particles in turbulence},}\ }\href {\doibase
  10.1175/1520-0469(1995)052<2276:OMFPIT>2.0.CO;2} {\bibfield  {journal}
  {\bibinfo  {journal} {JAS}\ }\textbf {\bibinfo {volume} {52}},\ \bibinfo
  {pages} {2276 -- 2285} (\bibinfo {year} {1995})}\BibitemShut {NoStop}%
\bibitem [{\citenamefont {Noel}\ and\ \citenamefont
  {Sassen}(2005)}]{Noel_Sassen_2005}%
  \BibitemOpen
  \bibfield  {author} {\bibinfo {author} {\bibfnamefont {V.}~\bibnamefont
  {Noel}}\ and\ \bibinfo {author} {\bibfnamefont {K.}~\bibnamefont {Sassen}},\
  }\bibfield  {title} {\enquote {\bibinfo {title} {Study of planar ice crystal
  orientations in ice clouds from scanning polarization lidar observations},}\
  }\href {\doibase 10.1175/JAM2223.1} {\bibfield  {journal} {\bibinfo
  {journal} {J. Appl. Meteorol.}\ }\textbf {\bibinfo {volume} {44}},\ \bibinfo
  {pages} {653 -- 664} (\bibinfo {year} {2005})}\BibitemShut {NoStop}%
\bibitem [{\citenamefont {Gustavsson}\ \emph {et~al.}(2021)\citenamefont
  {Gustavsson}, \citenamefont {Sheikh}, \citenamefont {Naso}, \citenamefont
  {Pumir},\ and\ \citenamefont {Mehlig}}]{Gustavsson_2021}%
  \BibitemOpen
  \bibfield  {author} {\bibinfo {author} {\bibfnamefont {K.}~\bibnamefont
  {Gustavsson}}, \bibinfo {author} {\bibfnamefont {M.~Z.}\ \bibnamefont
  {Sheikh}}, \bibinfo {author} {\bibfnamefont {A.}~\bibnamefont {Naso}},
  \bibinfo {author} {\bibfnamefont {A.}~\bibnamefont {Pumir}}, \ and\ \bibinfo
  {author} {\bibfnamefont {B.}~\bibnamefont {Mehlig}},\ }\bibfield  {title}
  {\enquote {\bibinfo {title} {Effect of particle inertia on the alignment of
  small ice crystals in turbulent clouds},}\ }\href {\doibase
  10.1175/JAS-D-20-0221.1} {\bibfield  {journal} {\bibinfo  {journal} {JAS}\
  }\textbf {\bibinfo {volume} {78}},\ \bibinfo {pages} {2573 -- 2587} (\bibinfo
  {year} {2021})}\BibitemShut {NoStop}%
\bibitem [{\citenamefont {Newsom}\ and\ \citenamefont
  {Bruce}(1994)}]{Newsom_Bruce_1994}%
  \BibitemOpen
  \bibfield  {author} {\bibinfo {author} {\bibfnamefont {R.~K.}\ \bibnamefont
  {Newsom}}\ and\ \bibinfo {author} {\bibfnamefont {C.~W.}\ \bibnamefont
  {Bruce}},\ }\bibfield  {title} {\enquote {\bibinfo {title} {The dynamics of
  fibrous aerosols in a quiescent atmosphere},}\ }\href {\doibase
  10.1063/1.868347} {\bibfield  {journal} {\bibinfo  {journal} {Phys. Fluids}\
  }\textbf {\bibinfo {volume} {6}},\ \bibinfo {pages} {521--530} (\bibinfo
  {year} {1994})}\BibitemShut {NoStop}%
\bibitem [{\citenamefont {Newsom}\ and\ \citenamefont
  {Bruce}(1998)}]{newsom1998orientational}%
  \BibitemOpen
  \bibfield  {author} {\bibinfo {author} {\bibfnamefont {R.K.}\ \bibnamefont
  {Newsom}}\ and\ \bibinfo {author} {\bibfnamefont {C.W.}\ \bibnamefont
  {Bruce}},\ }\bibfield  {title} {\enquote {\bibinfo {title} {Orientational
  properties of fibrous aerosols in atmospheric turbulence},}\ }\href {\doibase
  10.1016/S0021-8502(97)10030-1} {\bibfield  {journal} {\bibinfo  {journal} {J.
  Aerosol Sci.}\ }\textbf {\bibinfo {volume} {29}},\ \bibinfo {pages}
  {773--797} (\bibinfo {year} {1998})}\BibitemShut {NoStop}%
\bibitem [{\citenamefont {Pierson}(2023)}]{pierson2023inertial}%
  \BibitemOpen
  \bibfield  {author} {\bibinfo {author} {\bibfnamefont {J.-L.}\ \bibnamefont
  {Pierson}},\ }\bibfield  {title} {\enquote {\bibinfo {title} {Inertial
  settling of an arbitrarily oriented cylinder in a quiescent flow: From
  short-time to quasisteady motion},}\ }\href@noop {} {\bibfield  {journal}
  {\bibinfo  {journal} {Phys. Rev. Fluids}\ }\textbf {\bibinfo {volume} {8}},\
  \bibinfo {pages} {104301} (\bibinfo {year} {2023})}\BibitemShut {NoStop}%
\bibitem [{\citenamefont {Bhowmick}\ \emph {et~al.}(2024)\citenamefont
  {Bhowmick}, \citenamefont {Seesing}, \citenamefont {Gustavsson},
  \citenamefont {Guettler}, \citenamefont {Wang}, \citenamefont {Pumir},
  \citenamefont {Mehlig},\ and\ \citenamefont {Bagheri}}]{bhowmick2024inertia}%
  \BibitemOpen
  \bibfield  {author} {\bibinfo {author} {\bibfnamefont {T.}~\bibnamefont
  {Bhowmick}}, \bibinfo {author} {\bibfnamefont {J.}~\bibnamefont {Seesing}},
  \bibinfo {author} {\bibfnamefont {K.}~\bibnamefont {Gustavsson}}, \bibinfo
  {author} {\bibfnamefont {J.}~\bibnamefont {Guettler}}, \bibinfo {author}
  {\bibfnamefont {Y.}~\bibnamefont {Wang}}, \bibinfo {author} {\bibfnamefont
  {A.}~\bibnamefont {Pumir}}, \bibinfo {author} {\bibfnamefont
  {B.}~\bibnamefont {Mehlig}}, \ and\ \bibinfo {author} {\bibfnamefont
  {G.}~\bibnamefont {Bagheri}},\ }\bibfield  {title} {\enquote {\bibinfo
  {title} {Inertia induces strong orientation fluctuations of nonspherical
  atmospheric particles},}\ }\href@noop {} {\bibfield  {journal} {\bibinfo
  {journal} {Phys. Rev. Lett.}\ }\textbf {\bibinfo {volume} {132}},\ \bibinfo
  {pages} {034101} (\bibinfo {year} {2024})}\BibitemShut {NoStop}%
\bibitem [{\citenamefont {Candelier}\ \emph {et~al.}(2024)\citenamefont
  {Candelier}, \citenamefont {Gustavsson}, \citenamefont {Sharma},
  \citenamefont {Sundberg}, \citenamefont {Pumir}, \citenamefont {Bagheri},\
  and\ \citenamefont {Mehlig}}]{candelier2024torques}%
  \BibitemOpen
  \bibfield  {author} {\bibinfo {author} {\bibfnamefont {F.}~\bibnamefont
  {Candelier}}, \bibinfo {author} {\bibfnamefont {K.}~\bibnamefont
  {Gustavsson}}, \bibinfo {author} {\bibfnamefont {P.}~\bibnamefont {Sharma}},
  \bibinfo {author} {\bibfnamefont {L.}~\bibnamefont {Sundberg}}, \bibinfo
  {author} {\bibfnamefont {A.}~\bibnamefont {Pumir}}, \bibinfo {author}
  {\bibfnamefont {G.}~\bibnamefont {Bagheri}}, \ and\ \bibinfo {author}
  {\bibfnamefont {B.}~\bibnamefont {Mehlig}},\ }\href@noop {} {\enquote
  {\bibinfo {title} {Torques on curved atmospheric fibres},}\ } (\bibinfo
  {year} {2024}),\ \Eprint {http://arxiv.org/abs/2409.19004} {arXiv:2409.19004}
  \BibitemShut {NoStop}%
\bibitem [{\citenamefont {Cai}\ \emph {et~al.}(2020)\citenamefont {Cai},
  \citenamefont {Mitrano}, \citenamefont {Heuberger}, \citenamefont
  {R.Hufenus},\ and\ \citenamefont {Nowack}}]{cai2020origin}%
  \BibitemOpen
  \bibfield  {author} {\bibinfo {author} {\bibfnamefont {Y.}~\bibnamefont
  {Cai}}, \bibinfo {author} {\bibfnamefont {D.M.}\ \bibnamefont {Mitrano}},
  \bibinfo {author} {\bibfnamefont {M.}~\bibnamefont {Heuberger}}, \bibinfo
  {author} {\bibnamefont {R.Hufenus}}, \ and\ \bibinfo {author} {\bibfnamefont
  {B.}~\bibnamefont {Nowack}},\ }\bibfield  {title} {\enquote {\bibinfo {title}
  {The origin of microplastic fiber in polyester textiles: The textile
  production process matters},}\ }\href@noop {} {\bibfield  {journal} {\bibinfo
   {journal} {Journal of Cleaner Production}\ }\textbf {\bibinfo {volume}
  {267}},\ \bibinfo {pages} {121970} (\bibinfo {year} {2020})}\BibitemShut
  {NoStop}%
\bibitem [{\citenamefont {Brenner}(1961)}]{brenner1961oseen}%
  \BibitemOpen
  \bibfield  {author} {\bibinfo {author} {\bibfnamefont {H.}~\bibnamefont
  {Brenner}},\ }\bibfield  {title} {\enquote {\bibinfo {title} {The {O}seen
  resistance of a particle of arbitrary shape},}\ }\href {\doibase
  10.1017/S0022112061000755} {\bibfield  {journal} {\bibinfo  {journal} {J.
  Fluid Mech.}\ }\textbf {\bibinfo {volume} {11}},\ \bibinfo {pages} {604--610}
  (\bibinfo {year} {1961})}\BibitemShut {NoStop}%
\bibitem [{\citenamefont {Khayat}\ and\ \citenamefont
  {Cox}(1989)}]{Khayat_Cox_1989}%
  \BibitemOpen
  \bibfield  {author} {\bibinfo {author} {\bibfnamefont {R.E.}\ \bibnamefont
  {Khayat}}\ and\ \bibinfo {author} {\bibfnamefont {R.G.}\ \bibnamefont
  {Cox}},\ }\bibfield  {title} {\enquote {\bibinfo {title} {Inertia effects on
  the motion of long slender bodies},}\ }\href {\doibase
  10.1017/S0022112089003174} {\bibfield  {journal} {\bibinfo  {journal} {J.
  Fluid Mech.}\ }\textbf {\bibinfo {volume} {209}},\ \bibinfo {pages}
  {435--462} (\bibinfo {year} {1989})}\BibitemShut {NoStop}%
\bibitem [{\citenamefont {Dabade}\ \emph {et~al.}(2015)\citenamefont {Dabade},
  \citenamefont {Marath},\ and\ \citenamefont {Subramanian}}]{Dabade_2015}%
  \BibitemOpen
  \bibfield  {author} {\bibinfo {author} {\bibfnamefont {V.}~\bibnamefont
  {Dabade}}, \bibinfo {author} {\bibfnamefont {N.~K.}\ \bibnamefont {Marath}},
  \ and\ \bibinfo {author} {\bibfnamefont {G.}~\bibnamefont {Subramanian}},\
  }\bibfield  {title} {\enquote {\bibinfo {title} {Effects of inertia and
  viscoelasticity on sedimenting anisotropic particles},}\ }\href {\doibase
  10.1017/jfm.2015.360} {\bibfield  {journal} {\bibinfo  {journal} {J.\ Fluid
  Mech.}\ }\textbf {\bibinfo {volume} {778}},\ \bibinfo {pages} {133--188}
  (\bibinfo {year} {2015})}\BibitemShut {NoStop}%
\bibitem [{\citenamefont {Jiang}\ \emph {et~al.}(2021)\citenamefont {Jiang},
  \citenamefont {Zhao}, \citenamefont {Andersson}, \citenamefont {Gustavsson},
  \citenamefont {Pumir},\ and\ \citenamefont {Mehlig}}]{Jiang_2021}%
  \BibitemOpen
  \bibfield  {author} {\bibinfo {author} {\bibfnamefont {F.}~\bibnamefont
  {Jiang}}, \bibinfo {author} {\bibfnamefont {L.}~\bibnamefont {Zhao}},
  \bibinfo {author} {\bibfnamefont {H.~I.}\ \bibnamefont {Andersson}}, \bibinfo
  {author} {\bibfnamefont {K.}~\bibnamefont {Gustavsson}}, \bibinfo {author}
  {\bibfnamefont {A.}~\bibnamefont {Pumir}}, \ and\ \bibinfo {author}
  {\bibfnamefont {B.}~\bibnamefont {Mehlig}},\ }\bibfield  {title} {\enquote
  {\bibinfo {title} {Inertial torque on a small spheroid in a stationary
  uniform flow},}\ }\href {\doibase 10.1103/PhysRevFluids.6.024302} {\bibfield
  {journal} {\bibinfo  {journal} {Phys. Rev. Fluids}\ }\textbf {\bibinfo
  {volume} {6}},\ \bibinfo {pages} {024302} (\bibinfo {year}
  {2021})}\BibitemShut {NoStop}%
\bibitem [{\citenamefont {Lopez}\ and\ \citenamefont
  {Guazzelli}(2017)}]{Lopez_Guazzelli_2017}%
  \BibitemOpen
  \bibfield  {author} {\bibinfo {author} {\bibfnamefont {D.}~\bibnamefont
  {Lopez}}\ and\ \bibinfo {author} {\bibfnamefont {E.}~\bibnamefont
  {Guazzelli}},\ }\bibfield  {title} {\enquote {\bibinfo {title} {Inertial
  effects on fibers settling in a vortical flow},}\ }\href {\doibase
  10.1103/PhysRevFluids.2.024306} {\bibfield  {journal} {\bibinfo  {journal}
  {Phys. Rev. Fluids}\ }\textbf {\bibinfo {volume} {2}},\ \bibinfo {pages}
  {024306} (\bibinfo {year} {2017})}\BibitemShut {NoStop}%
\bibitem [{\citenamefont {Roy}\ \emph {et~al.}(2023)\citenamefont {Roy},
  \citenamefont {Kramel}, \citenamefont {Menon}, \citenamefont {Voth},\ and\
  \citenamefont {Koch}}]{roy2023orientation}%
  \BibitemOpen
  \bibfield  {author} {\bibinfo {author} {\bibfnamefont {A.}~\bibnamefont
  {Roy}}, \bibinfo {author} {\bibfnamefont {S.}~\bibnamefont {Kramel}},
  \bibinfo {author} {\bibfnamefont {U.}~\bibnamefont {Menon}}, \bibinfo
  {author} {\bibfnamefont {G.A.}\ \bibnamefont {Voth}}, \ and\ \bibinfo
  {author} {\bibfnamefont {D.L.}\ \bibnamefont {Koch}},\ }\bibfield  {title}
  {\enquote {\bibinfo {title} {Orientation of finite {R}eynolds number
  anisotropic particles settling in turbulence},}\ }\href@noop {} {\bibfield
  {journal} {\bibinfo  {journal} {Journal of Non-Newtonian Fluid Mechanics}\
  }\textbf {\bibinfo {volume} {318}},\ \bibinfo {pages} {105048} (\bibinfo
  {year} {2023})}\BibitemShut {NoStop}%
\bibitem [{\citenamefont {Cabrera}\ \emph {et~al.}(2022)\citenamefont
  {Cabrera}, \citenamefont {Sheikh}, \citenamefont {Mehlig}, \citenamefont
  {Plihon}, \citenamefont {Bourgoin}, \citenamefont {Pumir},\ and\
  \citenamefont {Naso}}]{Cabrera_2022}%
  \BibitemOpen
  \bibfield  {author} {\bibinfo {author} {\bibfnamefont {F.}~\bibnamefont
  {Cabrera}}, \bibinfo {author} {\bibfnamefont {M.~Z.}\ \bibnamefont {Sheikh}},
  \bibinfo {author} {\bibfnamefont {B.}~\bibnamefont {Mehlig}}, \bibinfo
  {author} {\bibfnamefont {N.}~\bibnamefont {Plihon}}, \bibinfo {author}
  {\bibfnamefont {M.}~\bibnamefont {Bourgoin}}, \bibinfo {author}
  {\bibfnamefont {A.}~\bibnamefont {Pumir}}, \ and\ \bibinfo {author}
  {\bibfnamefont {A.}~\bibnamefont {Naso}},\ }\bibfield  {title} {\enquote
  {\bibinfo {title} {Experimental validation of fluid inertia models for a
  cylinder settling in a quiescent flow},}\ }\href {\doibase
  10.1103/PhysRevFluids.7.024301} {\bibfield  {journal} {\bibinfo  {journal}
  {Phys. Rev. Fluids}\ }\textbf {\bibinfo {volume} {7}},\ \bibinfo {pages}
  {024301} (\bibinfo {year} {2022})}\BibitemShut {NoStop}%
\bibitem [{\citenamefont {Brenner}\ and\ \citenamefont
  {Cox}(1963)}]{brenner1963resistance}%
  \BibitemOpen
  \bibfield  {author} {\bibinfo {author} {\bibfnamefont {H.}~\bibnamefont
  {Brenner}}\ and\ \bibinfo {author} {\bibfnamefont {R.~G.}\ \bibnamefont
  {Cox}},\ }\bibfield  {title} {\enquote {\bibinfo {title} {The resistance to a
  particle of arbitrary shape in translational motion at small {R}eynolds
  numbers},}\ }\href@noop {} {\bibfield  {journal} {\bibinfo  {journal} {J.
  Fluid Mech.}\ }\textbf {\bibinfo {volume} {17}},\ \bibinfo {pages}
  {561–595} (\bibinfo {year} {1963})}\BibitemShut {NoStop}%
\bibitem [{\citenamefont {Cox}(1965)}]{cox1965steady}%
  \BibitemOpen
  \bibfield  {author} {\bibinfo {author} {\bibfnamefont {R.G.}\ \bibnamefont
  {Cox}},\ }\bibfield  {title} {\enquote {\bibinfo {title} {The steady motion
  of a particle of arbitrary shape at small {R}eynolds numbers},}\ }\href
  {\doibase 10.1017/S0022112065001593} {\bibfield  {journal} {\bibinfo
  {journal} {J. Fluid Mech.}\ }\textbf {\bibinfo {volume} {23}},\ \bibinfo
  {pages} {625--643} (\bibinfo {year} {1965})}\BibitemShut {NoStop}%
\bibitem [{\citenamefont {Candelier}\ \emph {et~al.}(2015)\citenamefont
  {Candelier}, \citenamefont {Einarsson}, \citenamefont {Lundell},
  \citenamefont {Mehlig},\ and\ \citenamefont
  {Angilella}}]{candelier2015role_a}%
  \BibitemOpen
  \bibfield  {author} {\bibinfo {author} {\bibfnamefont {F.}~\bibnamefont
  {Candelier}}, \bibinfo {author} {\bibfnamefont {J.}~\bibnamefont
  {Einarsson}}, \bibinfo {author} {\bibfnamefont {F.}~\bibnamefont {Lundell}},
  \bibinfo {author} {\bibfnamefont {B.}~\bibnamefont {Mehlig}}, \ and\ \bibinfo
  {author} {\bibfnamefont {J.-R.}\ \bibnamefont {Angilella}},\ }\bibfield
  {title} {\enquote {\bibinfo {title} {Role of inertia for the rotation of a
  nearly spherical particle in a general linear flow},}\ }\href@noop {}
  {\bibfield  {journal} {\bibinfo  {journal} {Phys. Rev. E}\ }\textbf {\bibinfo
  {volume} {91}},\ \bibinfo {pages} {053023} (\bibinfo {year}
  {2015})}\BibitemShut {NoStop}%
\bibitem [{\citenamefont {Collis}\ \emph {et~al.}(2024)\citenamefont {Collis},
  \citenamefont {Nunn},\ and\ \citenamefont {Sader}}]{collis2024unsteady}%
  \BibitemOpen
  \bibfield  {author} {\bibinfo {author} {\bibfnamefont {J.~F.}\ \bibnamefont
  {Collis}}, \bibinfo {author} {\bibfnamefont {A.}~\bibnamefont {Nunn}}, \ and\
  \bibinfo {author} {\bibfnamefont {J.~E.}\ \bibnamefont {Sader}},\ }\bibfield
  {title} {\enquote {\bibinfo {title} {Unsteady motion of nearly spherical
  particles in viscous fluids: a second-order asymptotic theory},}\ }\href@noop
  {} {\bibfield  {journal} {\bibinfo  {journal} {Journal of Fluid Mechanics}\
  }\textbf {\bibinfo {volume} {1001}},\ \bibinfo {pages} {A23} (\bibinfo {year}
  {2024})}\BibitemShut {NoStop}%
\bibitem [{\citenamefont {Candelier}\ and\ \citenamefont
  {Mehlig}(2016)}]{Candelier_Mehlig_2016}%
  \BibitemOpen
  \bibfield  {author} {\bibinfo {author} {\bibfnamefont {F.}~\bibnamefont
  {Candelier}}\ and\ \bibinfo {author} {\bibfnamefont {B.}~\bibnamefont
  {Mehlig}},\ }\bibfield  {title} {\enquote {\bibinfo {title} {Settling of an
  asymmetric dumbbell in a quiescent fluid},}\ }\href {\doibase
  10.1017/jfm.2016.350} {\bibfield  {journal} {\bibinfo  {journal} {J. Fluid
  Mech.}\ }\textbf {\bibinfo {volume} {802}},\ \bibinfo {pages} {174--185}
  (\bibinfo {year} {2016})}\BibitemShut {NoStop}%
\bibitem [{\citenamefont {Roy}\ \emph {et~al.}(2019)\citenamefont {Roy},
  \citenamefont {Hamati}, \citenamefont {Tierney}, \citenamefont {Koch},\ and\
  \citenamefont {Voth}}]{Roy_2019}%
  \BibitemOpen
  \bibfield  {author} {\bibinfo {author} {\bibfnamefont {A.}~\bibnamefont
  {Roy}}, \bibinfo {author} {\bibfnamefont {R.J.}\ \bibnamefont {Hamati}},
  \bibinfo {author} {\bibfnamefont {L.}~\bibnamefont {Tierney}}, \bibinfo
  {author} {\bibfnamefont {D.L.}\ \bibnamefont {Koch}}, \ and\ \bibinfo
  {author} {\bibfnamefont {G.A.}\ \bibnamefont {Voth}},\ }\bibfield  {title}
  {\enquote {\bibinfo {title} {Inertial torques and a symmetry breaking
  orientational transition in the sedimentation of slender fibres},}\ }\href
  {\doibase 10.1017/jfm.2019.492} {\bibfield  {journal} {\bibinfo  {journal}
  {J. Fluid Mech.}\ }\textbf {\bibinfo {volume} {875}},\ \bibinfo {pages}
  {576--596} (\bibinfo {year} {2019})}\BibitemShut {NoStop}%
\bibitem [{\citenamefont {Ravichandran}\ and\ \citenamefont
  {Wettlaufer}(2023)}]{ravichandran2023orientation}%
  \BibitemOpen
  \bibfield  {author} {\bibinfo {author} {\bibfnamefont {S.}~\bibnamefont
  {Ravichandran}}\ and\ \bibinfo {author} {\bibfnamefont {J.~S.}\ \bibnamefont
  {Wettlaufer}},\ }\bibfield  {title} {\enquote {\bibinfo {title} {Orientation
  dynamics of two-dimensional concavo-convex bodies},}\ }\href@noop {}
  {\bibfield  {journal} {\bibinfo  {journal} {Phys. Rev. Fluids}\ }\textbf
  {\bibinfo {volume} {8}},\ \bibinfo {pages} {L062301} (\bibinfo {year}
  {2023})}\BibitemShut {NoStop}%
\bibitem [{\citenamefont {Tatsii}\ \emph {et~al.}(2024)\citenamefont {Tatsii},
  \citenamefont {Bucci}, \citenamefont {Bhowmick}, \citenamefont {Guettler},
  \citenamefont {Bakels}, \citenamefont {Bagheri},\ and\ \citenamefont
  {Stohl}}]{tatsii2024shape}%
  \BibitemOpen
  \bibfield  {author} {\bibinfo {author} {\bibfnamefont {D.}~\bibnamefont
  {Tatsii}}, \bibinfo {author} {\bibfnamefont {S.}~\bibnamefont {Bucci}},
  \bibinfo {author} {\bibfnamefont {T.}~\bibnamefont {Bhowmick}}, \bibinfo
  {author} {\bibfnamefont {J.}~\bibnamefont {Guettler}}, \bibinfo {author}
  {\bibfnamefont {L.}~\bibnamefont {Bakels}}, \bibinfo {author} {\bibfnamefont
  {G.}~\bibnamefont {Bagheri}}, \ and\ \bibinfo {author} {\bibfnamefont
  {A.}~\bibnamefont {Stohl}},\ }\bibfield  {title} {\enquote {\bibinfo {title}
  {Shape matters: Long-range transport of microplastic fibers in the
  atmosphere},}\ }\href@noop {} {\bibfield  {journal} {\bibinfo  {journal}
  {Environmental Science \& Technology}\ }\textbf {\bibinfo {volume} {58}},\
  \bibinfo {pages} {671--682} (\bibinfo {year} {2024})}\BibitemShut {NoStop}%
\bibitem [{\citenamefont {Bhowmick}\ \emph {et~al.}(2025)\citenamefont
  {Bhowmick}, \citenamefont {Wang}, \citenamefont {Latt},\ and\ \citenamefont
  {Bagheri}}]{bhowmick2024twist}%
  \BibitemOpen
  \bibfield  {author} {\bibinfo {author} {\bibfnamefont {T.}~\bibnamefont
  {Bhowmick}}, \bibinfo {author} {\bibfnamefont {Y.}~\bibnamefont {Wang}},
  \bibinfo {author} {\bibfnamefont {J.}~\bibnamefont {Latt}}, \ and\ \bibinfo
  {author} {\bibfnamefont {G}~\bibnamefont {Bagheri}},\ }\bibfield  {title}
  {\enquote {\bibinfo {title} {Twist, turn and encounter: the trajectories of
  small atmospheric particles unravelled},}\ }\href@noop {} {\bibfield
  {journal} {\bibinfo  {journal} {Journal of Fluid Mechanics}\ }\textbf
  {\bibinfo {volume} {1021}},\ \bibinfo {pages} {A27} (\bibinfo {year}
  {2025})}\BibitemShut {NoStop}%
\bibitem [{\citenamefont {Kim}\ and\ \citenamefont
  {Karrila}(1991)}]{Kim_Karrila_1991}%
  \BibitemOpen
  \bibfield  {author} {\bibinfo {author} {\bibfnamefont {S.}~\bibnamefont
  {Kim}}\ and\ \bibinfo {author} {\bibfnamefont {S.J.}\ \bibnamefont
  {Karrila}},\ }\href {\doibase 10.1016/C2013-0-04644-0} {\emph {\bibinfo
  {title} {Microhydrodynamics: principles and selected applications}}}\
  (\bibinfo  {publisher} {Butterworth-Heinemann},\ \bibinfo {address}
  {Boston},\ \bibinfo {year} {1991})\BibitemShut {NoStop}%
\bibitem [{\citenamefont {Brenner}(1964{\natexlab{b}})}]{brenner1963stokes}%
  \BibitemOpen
  \bibfield  {author} {\bibinfo {author} {\bibfnamefont {H.}~\bibnamefont
  {Brenner}},\ }\bibfield  {title} {\enquote {\bibinfo {title} {The {S}tokes
  resistance of an arbitrary particle—{III}: Shear fields},}\ }\href@noop {}
  {\bibfield  {journal} {\bibinfo  {journal} {Chem. Eng. Sci.}\ }\textbf
  {\bibinfo {volume} {19}},\ \bibinfo {pages} {631--651} (\bibinfo {year}
  {1964}{\natexlab{b}})}\BibitemShut {NoStop}%
\bibitem [{\citenamefont {Ishimoto}(2020)}]{ishimoto2020jeffery}%
  \BibitemOpen
  \bibfield  {author} {\bibinfo {author} {\bibfnamefont {K.}~\bibnamefont
  {Ishimoto}},\ }\bibfield  {title} {\enquote {\bibinfo {title} {Jeffery orbits
  for an object with discrete rotational symmetry},}\ }\href@noop {} {\bibfield
   {journal} {\bibinfo  {journal} {Phys. Fluids}\ }\textbf {\bibinfo {volume}
  {32}} (\bibinfo {year} {2020})}\BibitemShut {NoStop}%
\bibitem [{\citenamefont {Lovalenti}\ and\ \citenamefont
  {Brady}(1993)}]{lovalenti1993hydrodynamic}%
  \BibitemOpen
  \bibfield  {author} {\bibinfo {author} {\bibfnamefont {P.~M.}\ \bibnamefont
  {Lovalenti}}\ and\ \bibinfo {author} {\bibfnamefont {J.~F.}\ \bibnamefont
  {Brady}},\ }\bibfield  {title} {\enquote {\bibinfo {title} {The hydrodynamic
  force on a rigid particle undergoing arbitrary time-dependent motion at small
  {R}eynolds number},}\ }\href {\doibase 10.1017/S0022112093002885} {\bibfield
  {journal} {\bibinfo  {journal} {Journal of Fluid Mechanics}\ }\textbf
  {\bibinfo {volume} {256}},\ \bibinfo {pages} {561–605} (\bibinfo {year}
  {1993})}\BibitemShut {NoStop}%
\bibitem [{\citenamefont {Childress}(1964)}]{childress1964slow}%
  \BibitemOpen
  \bibfield  {author} {\bibinfo {author} {\bibfnamefont {S.}~\bibnamefont
  {Childress}},\ }\bibfield  {title} {\enquote {\bibinfo {title} {The slow
  motion of a sphere in a rotating, viscous fluid},}\ }\href@noop {} {\bibfield
   {journal} {\bibinfo  {journal} {J. Fluid Mech.}\ }\textbf {\bibinfo {volume}
  {20}},\ \bibinfo {pages} {305--314} (\bibinfo {year} {1964})}\BibitemShut
  {NoStop}%
\bibitem [{\citenamefont {Saffman}(1965)}]{saffman1965lift}%
  \BibitemOpen
  \bibfield  {author} {\bibinfo {author} {\bibfnamefont {P.~G.~T.}\
  \bibnamefont {Saffman}},\ }\bibfield  {title} {\enquote {\bibinfo {title}
  {The lift on a small sphere in a slow shear flow},}\ }\href@noop {}
  {\bibfield  {journal} {\bibinfo  {journal} {Journal of fluid mechanics}\
  }\textbf {\bibinfo {volume} {22}},\ \bibinfo {pages} {385--400} (\bibinfo
  {year} {1965})}\BibitemShut {NoStop}%
\bibitem [{\citenamefont {Meibohm}\ \emph {et~al.}(2016)\citenamefont
  {Meibohm}, \citenamefont {Candelier}, \citenamefont {Ros\'en}, \citenamefont
  {Einarsson}, \citenamefont {Lundell},\ and\ \citenamefont
  {Mehlig}}]{meibohm2016angular}%
  \BibitemOpen
  \bibfield  {author} {\bibinfo {author} {\bibfnamefont {J.}~\bibnamefont
  {Meibohm}}, \bibinfo {author} {\bibfnamefont {F.}~\bibnamefont {Candelier}},
  \bibinfo {author} {\bibfnamefont {T.}~\bibnamefont {Ros\'en}}, \bibinfo
  {author} {\bibfnamefont {J.}~\bibnamefont {Einarsson}}, \bibinfo {author}
  {\bibfnamefont {F.}~\bibnamefont {Lundell}}, \ and\ \bibinfo {author}
  {\bibfnamefont {B.}~\bibnamefont {Mehlig}},\ }\bibfield  {title} {\enquote
  {\bibinfo {title} {Angular velocity of a spheroid log rolling in a simple
  shear at small {R}eynolds number},}\ }\href@noop {} {\bibfield  {journal}
  {\bibinfo  {journal} {Phys. Rev. Fluids}\ }\textbf {\bibinfo {volume} {1}},\
  \bibinfo {pages} {084203} (\bibinfo {year} {2016})}\BibitemShut {NoStop}%
\bibitem [{\citenamefont {Redaelli}\ \emph {et~al.}(2023)\citenamefont
  {Redaelli}, \citenamefont {Candelier}, \citenamefont {Mehaddi}, \citenamefont
  {Eloy},\ and\ \citenamefont {Mehlig}}]{redaelli2023hydrodynamic}%
  \BibitemOpen
  \bibfield  {author} {\bibinfo {author} {\bibfnamefont {T.}~\bibnamefont
  {Redaelli}}, \bibinfo {author} {\bibfnamefont {F.}~\bibnamefont {Candelier}},
  \bibinfo {author} {\bibfnamefont {R.}~\bibnamefont {Mehaddi}}, \bibinfo
  {author} {\bibfnamefont {C.}~\bibnamefont {Eloy}}, \ and\ \bibinfo {author}
  {\bibfnamefont {B.}~\bibnamefont {Mehlig}},\ }\bibfield  {title} {\enquote
  {\bibinfo {title} {Hydrodynamic force on a small squirmer moving with a
  time-dependent velocity at small {R}eynolds numbers},}\ }\href@noop {}
  {\bibfield  {journal} {\bibinfo  {journal} {Journal of Fluid Mechanics}\
  }\textbf {\bibinfo {volume} {973}},\ \bibinfo {pages} {A11} (\bibinfo {year}
  {2023})}\BibitemShut {NoStop}%
\bibitem [{\citenamefont {Flapper}\ \emph {et~al.}(2025)\citenamefont
  {Flapper}, \citenamefont {Piumini}, \citenamefont {Verzicco}, \citenamefont
  {Huisman},\ and\ \citenamefont {Lohse}}]{flapper2025settling}%
  \BibitemOpen
  \bibfield  {author} {\bibinfo {author} {\bibfnamefont {M.~M.}\ \bibnamefont
  {Flapper}}, \bibinfo {author} {\bibfnamefont {G.}~\bibnamefont {Piumini}},
  \bibinfo {author} {\bibfnamefont {R.}~\bibnamefont {Verzicco}}, \bibinfo
  {author} {\bibfnamefont {S.~G.}\ \bibnamefont {Huisman}}, \ and\ \bibinfo
  {author} {\bibfnamefont {D.}~\bibnamefont {Lohse}},\ }\href
  {https://arxiv.org/abs/2511.05137} {\enquote {\bibinfo {title} {Settling
  dynamics of an oloid: experiments and simulations},}\ } (\bibinfo {year}
  {2025}),\ \Eprint {http://arxiv.org/abs/2511.05137} {arXiv:2511.05137
  [physics.flu-dyn]} \BibitemShut {NoStop}%
\bibitem [{\citenamefont {Kelvin}(1871)}]{kelvin1871hydrokinetic}%
  \BibitemOpen
  \bibfield  {author} {\bibinfo {author} {\bibfnamefont {{Lord}}\ \bibnamefont
  {Kelvin}},\ }\bibfield  {title} {\enquote {\bibinfo {title} {Hydrokinetic
  solutions and observations},}\ }\href@noop {} {\bibfield  {journal} {\bibinfo
   {journal} {Phil. Mag.}\ }\textbf {\bibinfo {volume} {42}},\ \bibinfo {pages}
  {362} (\bibinfo {year} {1871})}\BibitemShut {NoStop}%
\bibitem [{\citenamefont {Ashcroft}\ and\ \citenamefont
  {Mermin}(1976)}]{Ashcroft}%
  \BibitemOpen
  \bibfield  {author} {\bibinfo {author} {\bibfnamefont {N.~W.}\ \bibnamefont
  {Ashcroft}}\ and\ \bibinfo {author} {\bibfnamefont {N.~D.}\ \bibnamefont
  {Mermin}},\ }\href@noop {} {\emph {\bibinfo {title} {{S}olid {S}tate
  {P}hysics}}}\ (\bibinfo  {publisher} {Holt-Saunders},\ \bibinfo {year}
  {1976})\BibitemShut {NoStop}%
\bibitem [{\citenamefont {Nakaya}(1954)}]{nakaya1954snow}%
  \BibitemOpen
  \bibfield  {author} {\bibinfo {author} {\bibfnamefont {U.}~\bibnamefont
  {Nakaya}},\ }\href@noop {} {\emph {\bibinfo {title} {Snow Crystals: Natural
  and Artificial}}}\ (\bibinfo  {publisher} {Harvard University Press},\
  \bibinfo {year} {1954})\BibitemShut {NoStop}%
\bibitem [{\citenamefont {Furukawa}\ and\ \citenamefont
  {Wettlaufer}(2007)}]{furukawa2007snow}%
  \BibitemOpen
  \bibfield  {author} {\bibinfo {author} {\bibfnamefont {Y.}~\bibnamefont
  {Furukawa}}\ and\ \bibinfo {author} {\bibfnamefont {J.~S.}\ \bibnamefont
  {Wettlaufer}},\ }\bibfield  {title} {\enquote {\bibinfo {title} {Snow and ice
  crystals},}\ }\href@noop {} {\bibfield  {journal} {\bibinfo  {journal}
  {Physics Today}\ }\textbf {\bibinfo {volume} {60}},\ \bibinfo {pages}
  {70--71} (\bibinfo {year} {2007})}\BibitemShut {NoStop}%
\bibitem [{\citenamefont {Rubinow}\ and\ \citenamefont
  {Keller}(1961)}]{rubinow1961transverse}%
  \BibitemOpen
  \bibfield  {author} {\bibinfo {author} {\bibfnamefont {S.~I.}\ \bibnamefont
  {Rubinow}}\ and\ \bibinfo {author} {\bibfnamefont {Joseph~B.}\ \bibnamefont
  {Keller}},\ }\bibfield  {title} {\enquote {\bibinfo {title} {The transverse
  force on a spinning sphere moving in a viscous fluid},}\ }\href@noop {}
  {\bibfield  {journal} {\bibinfo  {journal} {Journal of Fluid Mechanics}\
  }\textbf {\bibinfo {volume} {11}},\ \bibinfo {pages} {447–459} (\bibinfo
  {year} {1961})}\BibitemShut {NoStop}%
\bibitem [{\citenamefont {Maches}\ \emph {et~al.}(2024)\citenamefont {Maches},
  \citenamefont {Houssais}, \citenamefont {Sauret},\ and\ \citenamefont
  {Meiburg}}]{maches2024settling}%
  \BibitemOpen
  \bibfield  {author} {\bibinfo {author} {\bibfnamefont {Z.}~\bibnamefont
  {Maches}}, \bibinfo {author} {\bibfnamefont {M.}~\bibnamefont {Houssais}},
  \bibinfo {author} {\bibfnamefont {A.}~\bibnamefont {Sauret}}, \ and\ \bibinfo
  {author} {\bibfnamefont {E.}~\bibnamefont {Meiburg}},\ }\href@noop {}
  {\enquote {\bibinfo {title} {Settling of two rigidly connected spheres},}\ }
  (\bibinfo {year} {2024}),\ \Eprint {http://arxiv.org/abs/2406.10381}
  {arXiv:2406.10381} \BibitemShut {NoStop}%
\bibitem [{\citenamefont {Cox}(1970)}]{cox1970motion}%
  \BibitemOpen
  \bibfield  {author} {\bibinfo {author} {\bibfnamefont {R.~G.}\ \bibnamefont
  {Cox}},\ }\bibfield  {title} {\enquote {\bibinfo {title} {The motion of long
  slender bodies in a viscous fluid {Part 1. G}eneral theory},}\ }\href@noop {}
  {\bibfield  {journal} {\bibinfo  {journal} {J. Fluid Mech.}\ }\textbf
  {\bibinfo {volume} {44}},\ \bibinfo {pages} {791--810} (\bibinfo {year}
  {1970})}\BibitemShut {NoStop}%
\bibitem [{\citenamefont {Keller}\ and\ \citenamefont
  {Rubinow}(1976)}]{keller1976slender}%
  \BibitemOpen
  \bibfield  {author} {\bibinfo {author} {\bibfnamefont {J.~B.}\ \bibnamefont
  {Keller}}\ and\ \bibinfo {author} {\bibfnamefont {S.~I.}\ \bibnamefont
  {Rubinow}},\ }\bibfield  {title} {\enquote {\bibinfo {title} {Slender-body
  theory for slow viscous flow},}\ }\href@noop {} {\bibfield  {journal}
  {\bibinfo  {journal} {Journal of Fluid Mechanics}\ }\textbf {\bibinfo
  {volume} {75}},\ \bibinfo {pages} {705--714} (\bibinfo {year}
  {1976})}\BibitemShut {NoStop}%
\bibitem [{\citenamefont {Andersen}\ \emph {et~al.}(2005)\citenamefont
  {Andersen}, \citenamefont {Pesavento},\ and\ \citenamefont
  {Wang}}]{Andersen2005-2}%
  \BibitemOpen
  \bibfield  {author} {\bibinfo {author} {\bibfnamefont {A.}~\bibnamefont
  {Andersen}}, \bibinfo {author} {\bibfnamefont {U.}~\bibnamefont {Pesavento}},
  \ and\ \bibinfo {author} {\bibfnamefont {Z.~Jane}\ \bibnamefont {Wang}},\
  }\bibfield  {title} {\enquote {\bibinfo {title} {Unsteady aerodynamics of
  fluttering and tumbling plates},}\ }\href@noop {} {\bibfield  {journal}
  {\bibinfo  {journal} {Journal of Fluid Mechanics}\ }\textbf {\bibinfo
  {volume} {541}},\ \bibinfo {pages} {65} (\bibinfo {year} {2005})}\BibitemShut
  {NoStop}%
\bibitem [{\citenamefont {Pierson}\ \emph {et~al.}(2021)\citenamefont
  {Pierson}, \citenamefont {Kharrouba},\ and\ \citenamefont
  {Magnaudet}}]{pierson2021hydrodynamic}%
  \BibitemOpen
  \bibfield  {author} {\bibinfo {author} {\bibfnamefont {J.-L.}\ \bibnamefont
  {Pierson}}, \bibinfo {author} {\bibfnamefont {M.}~\bibnamefont {Kharrouba}},
  \ and\ \bibinfo {author} {\bibfnamefont {J.}~\bibnamefont {Magnaudet}},\
  }\bibfield  {title} {\enquote {\bibinfo {title} {Hydrodynamic torque on a
  slender cylinder rotating perpendicularly to its symmetry axis},}\
  }\href@noop {} {\bibfield  {journal} {\bibinfo  {journal} {Phys. Rev.
  Fluids}\ }\textbf {\bibinfo {volume} {6}},\ \bibinfo {pages} {094303}
  (\bibinfo {year} {2021})}\BibitemShut {NoStop}%
\bibitem [{\citenamefont {Landau}\ and\ \citenamefont
  {Lifshitz}(1987)}]{landau1987hydrodynamics}%
  \BibitemOpen
  \bibfield  {author} {\bibinfo {author} {\bibfnamefont {L.~D.}\ \bibnamefont
  {Landau}}\ and\ \bibinfo {author} {\bibfnamefont {E.~M.}\ \bibnamefont
  {Lifshitz}},\ }\href@noop {} {\emph {\bibinfo {title} {Fluid mechanics}}},\
  \bibinfo {edition} {2nd}\ ed.\ (\bibinfo  {publisher} {Pergamon Press},\
  \bibinfo {year} {1987})\BibitemShut {NoStop}%
\bibitem [{\citenamefont {Candelier}\ \emph {et~al.}(2023)\citenamefont
  {Candelier}, \citenamefont {Mehaddi}, \citenamefont {Mehlig},\ and\
  \citenamefont {Magnaudet}}]{candelier2023second}%
  \BibitemOpen
  \bibfield  {author} {\bibinfo {author} {\bibfnamefont {F.}~\bibnamefont
  {Candelier}}, \bibinfo {author} {\bibfnamefont {R.}~\bibnamefont {Mehaddi}},
  \bibinfo {author} {\bibfnamefont {B.}~\bibnamefont {Mehlig}}, \ and\ \bibinfo
  {author} {\bibfnamefont {J.}~\bibnamefont {Magnaudet}},\ }\bibfield  {title}
  {\enquote {\bibinfo {title} {Second-order inertial forces and torques on a
  sphere in a viscous steady linear flow},}\ }\href@noop {} {\bibfield
  {journal} {\bibinfo  {journal} {J. Fluid Mech.}\ }\textbf {\bibinfo {volume}
  {954}},\ \bibinfo {pages} {A25} (\bibinfo {year} {2023})}\BibitemShut
  {NoStop}%
\bibitem [{\citenamefont {Jiang}\ \emph {et~al.}(2025)\citenamefont {Jiang},
  \citenamefont {Qiu}, \citenamefont {Gustavsson}, \citenamefont {Mehlig},\
  and\ \citenamefont {Zhao}}]{jiang2025smart}%
  \BibitemOpen
  \bibfield  {author} {\bibinfo {author} {\bibfnamefont {X.}~\bibnamefont
  {Jiang}}, \bibinfo {author} {\bibfnamefont {J.}~\bibnamefont {Qiu}}, \bibinfo
  {author} {\bibfnamefont {K.}~\bibnamefont {Gustavsson}}, \bibinfo {author}
  {\bibfnamefont {B.}~\bibnamefont {Mehlig}}, \ and\ \bibinfo {author}
  {\bibfnamefont {L.}~\bibnamefont {Zhao}},\ }\href@noop {} {\enquote {\bibinfo
  {title} {Smart navigation of a gravity-driven glider with adjustable
  centre-of-mass},}\ } (\bibinfo {year} {2025}),\ \Eprint
  {http://arxiv.org/abs/2510.09250} {arXiv:2510.09250} \BibitemShut {NoStop}%
\bibitem [{\citenamefont {Chester}\ \emph {et~al.}(1969)\citenamefont
  {Chester}, \citenamefont {Breach},\ and\ \citenamefont
  {Proudman}}]{chester1969flow}%
  \BibitemOpen
  \bibfield  {author} {\bibinfo {author} {\bibfnamefont {W.}~\bibnamefont
  {Chester}}, \bibinfo {author} {\bibfnamefont {D.R.}\ \bibnamefont {Breach}},
  \ and\ \bibinfo {author} {\bibfnamefont {I.}~\bibnamefont {Proudman}},\
  }\bibfield  {title} {\enquote {\bibinfo {title} {On the flow past a sphere at
  low {R}eynolds number},}\ }\href@noop {} {\bibfield  {journal} {\bibinfo
  {journal} {J. Fluid Mech.}\ }\textbf {\bibinfo {volume} {37}},\ \bibinfo
  {pages} {751--760} (\bibinfo {year} {1969})}\BibitemShut {NoStop}%
\bibitem [{\citenamefont {Oesterle}\ and\ \citenamefont
  {Dinh}(1998)}]{oesterle1998experiments}%
  \BibitemOpen
  \bibfield  {author} {\bibinfo {author} {\bibfnamefont {B.}~\bibnamefont
  {Oesterle}}\ and\ \bibinfo {author} {\bibfnamefont {T.~Bui}\ \bibnamefont
  {Dinh}},\ }\bibfield  {title} {\enquote {\bibinfo {title} {Experiments on the
  lift of a spinning sphere in a range of intermediate {R}eynolds numbers},}\
  }\href@noop {} {\bibfield  {journal} {\bibinfo  {journal} {Experiments in
  Fluids}\ }\textbf {\bibinfo {volume} {25}},\ \bibinfo {pages} {16} (\bibinfo
  {year} {1998})}\BibitemShut {NoStop}%
\bibitem [{\citenamefont {Fintzi}\ \emph {et~al.}(2023)\citenamefont {Fintzi},
  \citenamefont {Gamet},\ and\ \citenamefont {Pierson}}]{fintzi2023inertial}%
  \BibitemOpen
  \bibfield  {author} {\bibinfo {author} {\bibfnamefont {N.}~\bibnamefont
  {Fintzi}}, \bibinfo {author} {\bibfnamefont {L.}~\bibnamefont {Gamet}}, \
  and\ \bibinfo {author} {\bibfnamefont {J.-L.}\ \bibnamefont {Pierson}},\
  }\bibfield  {title} {\enquote {\bibinfo {title} {Inertial loads on a
  finite-length cylinder embedded in a steady uniform flow},}\ }\href@noop {}
  {\bibfield  {journal} {\bibinfo  {journal} {Phys. Rev. Fluids}\ }\textbf
  {\bibinfo {volume} {8}},\ \bibinfo {pages} {044302} (\bibinfo {year}
  {2023})}\BibitemShut {NoStop}%
\bibitem [{\citenamefont {Ouchene}\ \emph
  {et~al.}(2016{\natexlab{b}})\citenamefont {Ouchene}, \citenamefont {Khalij},
  \citenamefont {Arcen},\ and\ \citenamefont {{A.
  Tani{\`e}re}}}]{ouchene2016new}%
  \BibitemOpen
  \bibfield  {author} {\bibinfo {author} {\bibfnamefont {R.}~\bibnamefont
  {Ouchene}}, \bibinfo {author} {\bibfnamefont {M.}~\bibnamefont {Khalij}},
  \bibinfo {author} {\bibfnamefont {B.}~\bibnamefont {Arcen}}, \ and\ \bibinfo
  {author} {\bibnamefont {{A. Tani{\`e}re}}},\ }\bibfield  {title} {\enquote
  {\bibinfo {title} {A new set of correlations of drag, lift and torque
  coefficients for non-spherical particles at large {R}eynolds numbers},}\
  }\href {\doibase 10.1016/j.powtec.2016.07.067} {\bibfield  {journal}
  {\bibinfo  {journal} {Powder Technology}\ }\textbf {\bibinfo {volume}
  {303}},\ \bibinfo {pages} {33--43} (\bibinfo {year}
  {2016}{\natexlab{b}})}\BibitemShut {NoStop}%
\bibitem [{\citenamefont {Sanjeevi}\ and\ \citenamefont
  {Padding}(2017)}]{sanjeevi2017orientational}%
  \BibitemOpen
  \bibfield  {author} {\bibinfo {author} {\bibfnamefont {S.~K.~P.}\
  \bibnamefont {Sanjeevi}}\ and\ \bibinfo {author} {\bibfnamefont {J.~T.}\
  \bibnamefont {Padding}},\ }\bibfield  {title} {\enquote {\bibinfo {title} {On
  the orientational dependence of drag experienced by spheroids},}\ }\href@noop
  {} {\bibfield  {journal} {\bibinfo  {journal} {J. Fluid Mech.}\ }\textbf
  {\bibinfo {volume} {820}},\ \bibinfo {pages} {R1} (\bibinfo {year}
  {2017})}\BibitemShut {NoStop}%
\bibitem [{\citenamefont {Kharrouba}\ \emph {et~al.}(2021)\citenamefont
  {Kharrouba}, \citenamefont {Pierson},\ and\ \citenamefont
  {Magnaudet}}]{kharrouba2021flow}%
  \BibitemOpen
  \bibfield  {author} {\bibinfo {author} {\bibfnamefont {M.}~\bibnamefont
  {Kharrouba}}, \bibinfo {author} {\bibfnamefont {J.-L.}\ \bibnamefont
  {Pierson}}, \ and\ \bibinfo {author} {\bibfnamefont {J.}~\bibnamefont
  {Magnaudet}},\ }\bibfield  {title} {\enquote {\bibinfo {title} {Flow
  structure and loads over inclined cylindrical rodlike particles and
  fibers},}\ }\href {\doibase 10.1103/PhysRevFluids.6.044308} {\bibfield
  {journal} {\bibinfo  {journal} {Phys. Rev. Fluids}\ }\textbf {\bibinfo
  {volume} {6}},\ \bibinfo {pages} {044308} (\bibinfo {year}
  {2021})}\BibitemShut {NoStop}%
\bibitem [{\citenamefont {Gavze}(1990)}]{gavze1990accelerated}%
  \BibitemOpen
  \bibfield  {author} {\bibinfo {author} {\bibfnamefont {E.}~\bibnamefont
  {Gavze}},\ }\bibfield  {title} {\enquote {\bibinfo {title} {The accelerated
  motion of rigid bodies in non-steady stokes flow},}\ }\href@noop {}
  {\bibfield  {journal} {\bibinfo  {journal} {Int. J. Multiph. Flow}\ }\textbf
  {\bibinfo {volume} {16}},\ \bibinfo {pages} {153} (\bibinfo {year}
  {1990})}\BibitemShut {NoStop}%
\bibitem [{\citenamefont {Lamb}(1924)}]{lamb1924hydrodynamics}%
  \BibitemOpen
  \bibfield  {author} {\bibinfo {author} {\bibfnamefont {H.}~\bibnamefont
  {Lamb}},\ }\href@noop {} {\emph {\bibinfo {title} {Hydrodynamics}}}\
  (\bibinfo  {publisher} {University Press},\ \bibinfo {year}
  {1924})\BibitemShut {NoStop}%
\bibitem [{\citenamefont {Howe}(1995)}]{Howe_1995}%
  \BibitemOpen
  \bibfield  {author} {\bibinfo {author} {\bibfnamefont {M.~S.}\ \bibnamefont
  {Howe}},\ }\bibfield  {title} {\enquote {\bibinfo {title} {On the force and
  moment on a body in an incompressible fluid, with application to rigid bodies
  and bubbles at high and low {R}eynolds numbers},}\ }\href@noop {} {\bibfield
  {journal} {\bibinfo  {journal} {The Quarterly Journal of Mechanics and
  Applied Mathematics}\ }\textbf {\bibinfo {volume} {48}},\ \bibinfo {pages}
  {401--426} (\bibinfo {year} {1995})}\BibitemShut {NoStop}%
\bibitem [{\citenamefont {Legendre}\ and\ \citenamefont
  {Magnaudet}(1997)}]{legendre1997note}%
  \BibitemOpen
  \bibfield  {author} {\bibinfo {author} {\bibfnamefont {D.}~\bibnamefont
  {Legendre}}\ and\ \bibinfo {author} {\bibfnamefont {J.}~\bibnamefont
  {Magnaudet}},\ }\bibfield  {title} {\enquote {\bibinfo {title} {A note on the
  lift force on a spherical bubble or drop in a low-{R}eynolds-number shear
  flow},}\ }\href@noop {} {\bibfield  {journal} {\bibinfo  {journal} {Phys.
  Fluids}\ }\textbf {\bibinfo {volume} {9}},\ \bibinfo {pages} {3572--3574}
  (\bibinfo {year} {1997})}\BibitemShut {NoStop}%
\end{thebibliography}
\end{document}